\documentclass[letterpaper,titlepage,11pt]{article}

\usepackage[T1]{fontenc}
\usepackage[utf8]{inputenc}
\usepackage{lmodern}
\usepackage{enumerate}
\usepackage[english]{babel}
\usepackage{comment}
\usepackage[numbers,sort&compress]{natbib}
\usepackage{bm}
\usepackage[usenames,dvipsnames]{xcolor}
\usepackage[a4paper]{geometry}
\usepackage[normalem]{ulem}

\usepackage[colorlinks=true,urlcolor=blue,citecolor=magenta]{hyperref}
\usepackage{amsmath,bbm}
\usepackage{amsfonts}
\usepackage{amssymb}
\usepackage{graphicx}
\usepackage[dvipsnames]{xcolor}
\usepackage{mathtools}
\usepackage{simplewick}
\usepackage{hyperref}
\usepackage{multirow}
\usepackage{enumitem}
\usepackage{mathrsfs}
\usepackage{physics}
\usepackage[compat=1.1.0]{tikz-feynman}
\usepackage{subcaption}
\usepackage{booktabs}
\usepackage{xcolor}

\usepackage{lipsum}

\usepackage{amsthm}
\newtheorem*{remark}{Quantization prescription}

\allowdisplaybreaks[1]

\newcommand{\CJ}{\mathcal{J}}

\newcommand{\BCJ}{\bar{\mathcal{J}}}

\def\mQ{\mathcal{Q}}

\def\mN{\mathcal{N}}

\def\dalpha{\dot{\alpha}}
\def\dbeta{\dot{\beta}}

\newcommand{\bracenom}{\genfrac{\lbrace}{\rbrace}{0pt}{}}

\newcommand\blfootnote[1]{%
  \begingroup
  \renewcommand\thefootnote{}\footnote{#1}%
  \addtocounter{footnote}{-1}%
  \endgroup
}

\newcommand{\SU}{\text{SU}}

\newcommand{\ds}{\displaystyle}

\newcommand{\spa}{\ , \ \ }

\newcommand{\psd}{\psi^\dagger}

\def\mba{\mathbf{a}}
\def\mbb{\mathbf{b}}
\def\mbc{\mathbf{c}}
\usepackage{etoolbox}

\newcommand{\beq}{\begin{equation}}
\newcommand{\eeq}{\end{equation}}
\newcommand{\bea}{\begin{eqnarray}}
\newcommand{\eea}{\end{eqnarray}}

\hypersetup{   colorlinks=true,  citecolor=blue, linkcolor=blue, urlcolor=red }

\begin{document}

\numberwithin{equation}{section}

\begin{titlepage}
\rightline{\vbox{   \phantom{ghost} }}

 \vskip 1.8 cm
\begin{center}
{\LARGE \bf
Spin Matrix Theory in near $\frac{1}{8}$-BPS corners of 
\\[2mm]
  ${\cal N} = 4$ super-Yang-Mills}
\end{center}
\vskip 1 cm

\title{}
\date{\today}
\author{Stefano Baiguera}
\author{Troels Harmark}
\author{Yang Lei}

\centerline{\large {{\bf Stefano Baiguera$^1$, Troels Harmark$^2$, Yang Lei$^{3}$}}}

\vskip 1.0cm

\begin{center}
\sl ${}^1$Department of Physics, Ben-Gurion University of the Negev, \\ Beer Sheva 84105, Israel \\[1mm]
\sl ${}^2$The Niels Bohr Institute, University of Copenhagen,\\
Blegdamsvej 17, DK-2100 Copenhagen \O, Denmark\\[1mm]
\sl ${}^3$ Kavli Institute for Theoretical Sciences (KITS), \\
University of Chinese Academy of Sciences, 100190 Beijing, P.R.~China
\end{center}

\vskip 1.3cm \centerline{\bf Abstract} \vskip 0.2cm \noindent
We consider limits of ${\cal N} = 4$ super-Yang-Mills (SYM) theory that approach BPS bounds. 
These limits result in non-relativistic theories that describe the effective dynamics near the BPS bounds and upon quantization are known as Spin Matrix Theories. The near-BPS theories can be obtained by reducing ${\cal N}=4$ SYM on a three-sphere and integrating out the fields that become non-dynamical in the limits. 
In the previous works \cite{Harmark:2019zkn,Baiguera:2020jgy,Baiguera:2020mgk} we have considered various SU(1,1) and SU(1,2) types of subsectors in this limit. 
In the current work, we will construct the remaining Spin Matrix Theories defined near the $\frac{1}{8}$-BPS subsectors, which include the PSU(1,1|2) and SU(2|3) cases. 
%These sectors involve interactions between fields with different flavours.
We derive the Hamiltonians by applying the spherical reduction algorithm and show that they match with the spin chain result, coming from the loop corrections to the dilatation operator.
In the PSU(1,1|2) case, we prove the positivity of the spectrum by constructing cubic supercharges using the enhanced PSU$(1|1)^2$ symmetry and show that they close to the interacting Hamiltonian.
%is exactly the anticommutator of the cubic supercharges. 
%of the  PSU$(1,1|2)$ Hamiltonian.
We finally analyse the symmetry structure of the sectors in view of an interpretation of the interactions in terms of fundamental blocks.
% the symmetry analysis developped in SU$(1,2)$ subsectors can be applied to SU$(2|3)$ but failed in PSU$(1,1|2)$ subsector. 
% 

\blfootnote{ \scriptsize{ \texttt{baiguera@post.bgu.ac.il, harmark@nbi.ku.dk, leiyang@ucas.ac.cn}} }

\end{titlepage}
\newpage
\tableofcontents

%%%%%%%%%%%%%%%%%%%%%%%%%%%%%%%%%%%%

\section{Introduction}

During the last decades there have been several attempts to get a deeper understanding of quantum gravity via the AdS/CFT duality, whose most studied example is the correspondence between $\mathcal{N}=4$ super Yang-Mills (SYM) theory with gauge group $\SU(N)$ and type IIB string theory on $\mathrm{AdS}_5 \times S^5 $ \cite{Maldacena:1997re}.
Among the celebrated tests of the duality, the integrability structure arising in the large $N$ limit provides a playground where the matching can be performed explicitly.
Indeed, when $N \rightarrow \infty$ the interactions simplify because only single trace operators survive the limit, and the system can be interpreted as a periodic spin chain \cite{Minahan:2008hf, Beisert:2003tq}. 
Another relevant scenario where the matching between the two sides of the duality is possible consists in the BMN limit considered in \cite{Berenstein:2002jq}, which is a regime where the string along the equator of the five-sphere moves at high speed.
All these scenarios are well-suited for the weak coupling case, but they are not able to investigate the regime where extended objects like D-branes or black holes arise in the gravity side.

As it is often the case in theoretical physics, one can hope that restricting to particular limits can help to simplify the problem under consideration. 
The idea behind the so-called Spin Matrix Theories (SMT) \cite{Harmark:2014mpa} is to zoom in close to a BPS bound to find an effective theory describing the surviving degrees of freedom.
The main advantage is that there exists a unique extension from the $N= \infty$ scenario to the case at finite $N,$ therefore these quantum-mechanical models constitute a generalization of the spin chains where non-perturbative effects can be included.
Starting from the observation that non-abelian gauge theories on compact manifolds admit an Hagedorn/deconfinement phase transition  \cite{ATICK1988291,Aharony:2003sx,Sundborg:1999ue}, SMTs were found by considering zero-temperature critical points of the grand-canonical partition function of $\mathcal{N}=4$ SYM defined on $\mathbb{R} \times S^3 .$
This procedure defines a set of decoupling limits where only a subset of the original degrees of freedom of the theory survive and the action of the one-loop dilatation operator closes inside a given SMT \cite{Harmark:2007px}.
In particular, the fields surviving the limit transform under a certain representation  $R_s$ of the spin group which characterizes the symmetries of the model, and they transform under the adjoint representation of the colour group $\SU(N).$
Several aspects of SMTs were studied in \cite{Harmark:2006di,Harmark:2006ta,Harmark:2008gm}.
Another interesting feature of these models is that they are naturally non-relativistic: an emergent U(1) symmetry associated to particle number conservation arises.
For this reason, another motivation to study SMT limits is that they represent non-trivial realizations of the non-relativistic symmetries, which are known to have several applications in condensed matter systems.
Few examples are given by superconductors \cite{Grover:2013rc}, cold atoms \cite{Nishida:2010tm} and the quantum Hall effect \cite{Son:2013rqa,Geracie:2014nka}. 

\begin{figure}[ht]
	\centering
	\includegraphics[trim=5cm 24cm 4cm 2cm,width=0.7\linewidth]{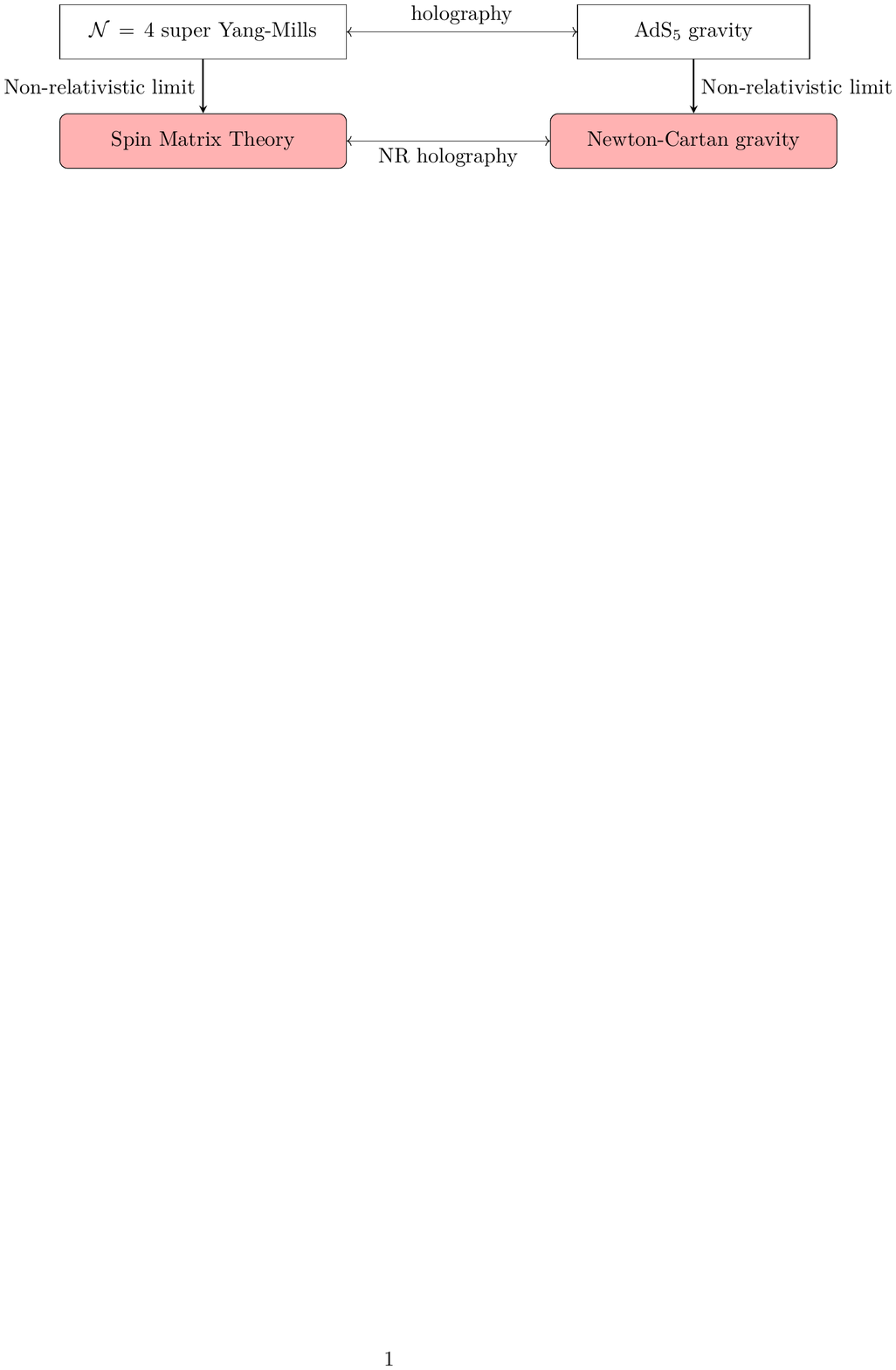}
	\caption{The holography between SMT and Newton-Cartan gravity can be viewed as the non-relativistic limit of the holography between $\mN=4$ SYM and AdS$_5$ gravity.}
	\label{commutative_diagram_0}
\end{figure}

The holographic duals of SMT are constructed by taking the same kind of decoupling limit at the level of string theory in the AdS$_5\times S^5$ background \cite{Harmark:2018cdl,roychowdhury2021decoding,Harmark:2020vll}. 
The resulting models admit a natural coupling to Newton-Cartan geometry in the target space, as shown in Figure~\ref{commutative_diagram_0}. 
More recent developments about the non-relativistic string theory can be found in \cite{Harmark:2017rpg,Bergshoeff:2018vfn,Bergshoeff:2018yvt,Gallegos:2019icg,Roychowdhury:2019sfo,Blair:2019qwi,Bergshoeff:2019pij,Yan:2019xsf,Gomis:2019zyu,Harmark:2019upf,Roychowdhury:2020yun,Kluson:2019xuo,Gomis:2020izd,Fontanella:2021hcb,Bidussi:2021ujm,Yan:2021lbe,Hartong:2021ekg}.

%\frac{For both SU$(2|3)$ and PSU$(1,1|2)$ sectors, the target spacetime of sigma model describing the string theory are six dimensional submanifolds inside the ten dimensional spacetime. 
%Studying the SMT in these subsectors thus provides simplified models to test the non-relativistic limits of holography. }{den}

The microcanonical formulation of the decoupling limits of $\mathcal{N}=4$ SYM defined on $\mathbb{R} \times S^3$ is based on considering BPS-bounds of the form
\beq
E \geq a_1 S_1 + a_2 S_2 + b_1 Q_1 + b_2 Q_2 + b_3 Q_3 \, ,
\eeq
where $a_i$ and $b_i$ are constant chemical potentials, $S_i$ are the Cartan generators of rotations and $Q_i$ the Cartan generators of R-symmetry.
Once a near-BPS limit is performed, there exist two
main techniques to obtain an Hamiltonian describing the effective interactions in the sector.
One possibility is compute the loop corrections to the dilatation operator of $\mathcal{N} = 4$ SYM, and then zoom in towards the unitarity bound of interest. This method has been extensively
applied in \cite{Beisert:2003tq,Minahan:2002ve,Beisert:2002ff}.
On the other hand, the opposite procedure can be implemented: one
starts from classical $\mathcal{N}=4$ SYM defined on $\mathbb{R} \times S^3$ and performs a dimensional reduction along the three-sphere. 
Afterwards, a quantization prescription is given. Non-trivially, it
turns out that these techniques are equivalent, \emph{i.e.} the quantization and the near-BPS limit
are commuting operations in all the cases considered. While the decoupling limits give rise to
quantum mechanical models after the sphere reduction, it was also observed that an effective
description in terms of semi-local fields exists, giving rise to lower-dimensional QFTs \cite{Harmark:2019zkn,Baiguera:2020jgy}.
When supersymmetry is preserved in the spin group, we also found a concrete realization
in terms of superfields. Given the non-relativistic traits of these theories, one can hope to
make contact with the superfield formulation of supersymmetric quantum field theories with
Lifshitz or Schroedinger invariance in the spirit of the models considered in \cite{Meyer:2017zfg,Auzzi:2019kdd,Arav:2019tqm}.

An important feature of the SMT Hamiltonians investigated so far is that their interactions
are positive definite. 
One of the main novelties of the present work is the presentation of a new
%another different 
method to derive the interacting Hamiltonian in the near-BPS limit when supersymmetry is preserved.
It is based on representing fermionic generators in terms of cubic supercharges whose anticommutator closes into the Hamiltonian of the sector.
This does not only provide another consistency check of the interactions derived from the dilatation operator and from the sphere reduction methods, but it also proves the positivity of the spectrum.

Analyzing the algebraic properties of the spin group, we also derived for several sectors a set of fundamental blocks to express the interactions in a manifest
positive-definite form \cite{Baiguera:2020mgk}. 
The interacting Hamiltonian is thus interpreted as a norm in the linear space of the representation.
This property is of great interest in view of the investigation of the theories at strong coupling, since the leading-order approximation simply consists in setting the corresponding blocks to zero. 
The interesting physics in the strongly coupled limit includes the dual black hole solutions  \cite{Gutowski:2004yv,Kunduri:2006ek} in the largest SMT sector - the PSU$(1,2|3)$. 
Furthermore the strongly-coupled SU$(2)$ SMT admits an effective description in terms of dual interacting giant gravitons obtained from the Dirac-Born-Infeld action  \cite{Harmark:2016cjq}. 
%Furthermore the largest sector, which has spin group PSU(1,2|3) and contains all the other near-BPS limits as subcases, admits dual black hole solutions \cite{Gutowski:2004yv,Kunduri:2006ek}.

The aim of the present paper is to continue the route towards the understanding of AdS/CFT duality through near-BPS limits of $\mathcal{N} = 4 $ SYM by constructing the SMTs corresponding to $\frac{1}{8}$-BPS subsectors \cite{Mandal:2006tk,Grant:2008sk}.
They are characterized by the unitarity bounds
\beq
E \geq  \alpha S_1 + Q_1 + Q_2 + \left( 1- \alpha \right) Q_3  \,.
\label{eq:near_BPS_limits_1over8BPS_subsectors}
\eeq
The parameter $\alpha$ runs over $\alpha=0,1$ and gives rise to the $\SU(2|3)$ or $\mathrm{PSU}(1,1|2)$ subsectors, respectively\footnote{The third possible $\frac{1}{8}$-BPS subsector is SU$(1,2|2)$, which was already reported in \cite{Baiguera:2020mgk}.}.
The works \cite{Kinney:2005ej,Choi:2018hmj} studied these subsectors in the exact BPS limit by means of the superconformal index.
In particular, our work focuses on the near-BPS limit given by
\beq
\lambda \rightarrow 0 \, , \qquad
\frac{1}{\lambda} \left( E - S_1 - Q_1 - Q_2 - \left( 1- \alpha \right) Q_3  \right) \,\,\, \mathrm{finite}  \, , \qquad
N \,\,\, \mathrm{finite} \, ,
\eeq
where $\lambda=g^2 N$ is the 't Hooft coupling.
It is evident that $N$ is kept constant but not necessarily large, which allows to investigate the non-perturbative regime, too.

The paper is organized as follows.
In Section \ref{sect-comparison_spin_chain} we derive the effective Hamiltonian near the BPS limits defined by Eq.~\eqref{eq:near_BPS_limits_1over8BPS_subsectors} starting from the action on spin chains derived from the one-loop corrections to the dilatation operator.
We apply the sphere reduction approach in Section \ref{sect-effective_ham_from_sphere} and we give in Section \ref{sect-quantization_sphere_red} a prescription to quantize the classical Hamiltonian obtained in this way, showing that the two procedures are equivalent.
We show in Section \ref{sect-positivity} that the effective Hamiltonian in the near-BPS limit is positive definite. 
We then explore in Section \ref{sec:symmetry} the algebraic structure of the interactions, which can potentially make the positiveness property manifest.
A discussion of the results and their applications for future investigations is given in Section \ref{sect-discussion}.
Technical details about the decomposition of fields into spherical harmonics, conventions and the relevant algebra are put in the Appendices.

\section{Hamiltonian from quantization of the dilatation operator}
\label{sect-comparison_spin_chain}

Starting from the one-loop corrections to the dilatation operator of $\mathcal{N}=4$ SYM theory \cite{Minahan:2002ve,Beisert:2003tq,Beisert:2002ff}, it is possible to derive an effective Hamiltonian involving the surviving degrees of freedom close 
to a BPS limit of interest.
In the cases described by Eq.~\eqref{eq:near_BPS_limits_1over8BPS_subsectors}, the quantization of the dilatation operator and the corresponding action of the Hamiltonian on a spin chain were considered in \cite{Beisert:2003ys,Beisert:2008qy} for the $\SU(2|3)$ sector, and in \cite{Bellucci:2006bv,Beisert:2007sk, Zwiebel:2007cpa} for the $\mathrm{PSU}(1,1|2)$ sector.

We review in Section \ref{sect-SMT-from_spin_chain} the definition of SMT and its relation to spin chains, following \cite{Harmark:2014mpa}.
Then we derive the effective SMT Hamiltonian for the near-BPS limits of interest in Sections \ref{sect-spin_chain_SU23_sector} and \ref{sect-spin_chain_PSU112_sector}.
It was shown for the  $\mathrm{SU}(1,1)$ sector and some of its extensions that the results coming from the quantization of the dilatation operator can be equivalently derived by performing the sphere reduction of classical $\mathcal{N}=4$ SYM, and then giving a prescription for the quantization \cite{Harmark:2019zkn,Baiguera:2020jgy}.
We will come back to this comparison and show the equivalence of the procedures for the $\mathrm{SU}(2|3)$ and $\mathrm{PSU}(1,1|2)$ sectors in Section \ref{sect-quantization_sphere_red}.

%
%It is possible to derive an effective Hamiltonian describing the degrees of freedom close to a BPS bound by computing the quantum corrections to the dilatation operator of $\mathcal{N}=4$ SYM theory \cite{Minahan:2002ve,Beisert:2003tq,Beisert:2002ff}. 
%The procedure is to zoom in towards the BPS limit of interest where only the one-loop contributions to the dilatation operator survive, and then apply the dictionary presented in \cite{Harmark:2014mpa} to define a SMT Hamiltonian.
%The Hilbert space reduces to states generated by the ladder operators  selected by the limit, in such a way to preserve a certain spin group.
%

\subsection{Spin Matrix Theory from spin chains}
\label{sect-SMT-from_spin_chain}

SMTs are quantum-mechanical models whose defining ingredients are the representation $R_s$ of a semi-simple Lie (super)-group $G_s$ and the adjoint representation of $\SU(N).$
Their definition does not require a priori any reference to a parent theory like $\mathcal{N}=4$ SYM; instead one builds an Hilbert space using ladder operators with a spin index $s \in R_s $ and a matrix structure due to the adjoint representation of the unitary group.
In the bosonic case, the vacuum state and the commutation relations are given by
\beq
(a_s)^i_{\,\, j} |0 \rangle \, , \qquad
\left[ (a^r)^i_{\,\, j} , (a^{\dagger}_s)^k_{\,\, l}  \right] = 
\delta^r_s \delta^k_j \delta^i_l \, , \qquad
\forall s,i,j \, ,
\eeq
and the Hilbert space is spanned by the states 
\beq
\tr \left( a^{\dagger}_{s_1} \dots a^{\dagger}_{s_l}  \right)
\tr \left( a^{\dagger}_{s_{l+1}} \dots \right) \dots
\tr \left( a^{\dagger}_{s_{k+1}} \dots a^{\dagger}_{s_L} \right) |0 \rangle  \, ,
\eeq
where $L$ is called the length and the traces are performed over the indices of the colour group.
This construction with multi-trace operators ensures that the singlet condition is satisfied, \emph{i.e.} physical states are annihilated by the operator
\beq
\sum_{s \in R_s} \left[ (a^{\dagger}_s)^i_{\,\,k} (a^s)^k_{\,\,j} 
- (a^{\dagger}_s)^k_{\,\,j} (a^s)^i_{\,\,k}  \right]  \, .
\eeq
One can include fermionic operators in a similar way, \emph{i.e.}
\beq
(b_s)^i_{\,\, j} |0 \rangle \, , \qquad
\left\lbrace (b^r)^i_{\,\, j} , (b^{\dagger}_s)^k_{\,\, l}  \right\rbrace = 
\delta^r_s \delta^k_j \delta^i_l \, , \qquad
\forall s,i,j \, .
\eeq
The bosonic raising operators commute among themselves and with the fermionic ones; furthermore the fermionic creation operators anticommute between themselves.
This is summarized by using graded commutators.

We consider quartic interactions which annihilate two excitations at a time and create two new ones; they are required to commute with the generators of the spin group. 
In the bosonic case with a single copy of the creation operator, the interacting Hamiltonian is given by
\beq
H_{\rm int}^{\rm bos} = \frac{1}{N} \, U^{s' r'}_{s r}  \sum_{\sigma \in S(4)} 
T_{\sigma} (a^{\dagger}_{s'})^{i_{\sigma(1)}}_{\,\,\, i_3}
(a^{\dagger}_{r'})^{i_{\sigma(2)}}_{\,\,\, i_4}
(a^{s})^{i_{\sigma(3)}}_{\,\,\, i_1}
(a^{r})^{i_{\sigma(4)}}_{\,\,\, i_2} \, , 
\eeq
where the coefficients of the permutation read
\beq
\sum_{\sigma \in S(4)} T_{\sigma} \sigma = (14) + (23) - (12) - (34) \, .
\eeq
The coefficients $U^{s' r'}_{s r}$ encode the spin structure of the theory; the reality and the symmetries of the Hamiltonian are guaranteed if they satisfy the relations
\beq
(U^{s' r'}_{s r})^* = U^{s r}_{s' r'} \, , \qquad
U^{s' r'}_{s r} = U^{r' s'}_{r s} \, .
\eeq
The previous structure can be written more compactly using the definition of normal ordering and the properties of commutators and cyclicity of the trace:
\beq
H_{\rm int}^{\rm bos} =  - \frac{1}{N} \, U^{s' r'}_{s r} : \tr \left( [a^{\dagger}_{s'}, a^s ] [a^{\dagger}_{r'}, a^{r}] \right) :  
\label{eq:dictionary_bos_term_SMT}
\eeq 
In the fermionic case, commutators are replaced by anti-commutators and there are additional minus signs when exchanging them. The corresponding term is
\beq
H_{\rm int}^{\rm ferm} = - \frac{1}{N} U^{s' r'}_{s r} : \tr \left( \lbrace b^{\dagger}_{s'}, b^s \rbrace \lbrace b^{\dagger}_{r'}, b^{r}\rbrace \right) :  
\label{eq:dictionary_ferm_term_SMT}
\eeq
One can similarly define a quartic interaction for mixed bosonic--fermionic terms.

All the definitions and the manipulations performed until now make sense for a generic quantum-mechanical model where we give a prescription for the coefficients $U^{s'r'}_{sr}.$
The link with $\mathcal{N}=4$ SYM and the quantization of the dilatation operator appear once we consider the large $N$ limit.
In this scenario, the Hilbert space collapses to states generated by single trace operators
\beq
| s_1 \dots s_L \rangle = \tr \left( a^{\dagger}_{s_1} \dots a^{\dagger}_{s_L}  \right) | 0 \rangle \, ,
\eeq
and the cyclicity property of the trace makes evident that this defines a spin chain with translation invariance.
In the limit $N \rightarrow \infty,$ the action of the interacting Hamiltonian on the singlet states reads
\beq
H_{\rm int}  | s_1 s_2 \dots s_L \rangle =
2 \sum_{k=1}^L U^{m n}_{s_k \, s_{k+1}} | s_1 \dots s_{k-1} \, m \, n \, s_{k+2} \dots s_L \rangle \, .
\eeq
This describes a nearest neighbour spin chain Hamiltonian, which is characterized precisely by the coefficients $U^{s'r'}_{sr}.$
They can be extracted by acting on two-particle states\footnote{With an abuse of notation, we refer here to the states as
\[
| k l \rangle \equiv | \dots k l \dots \rangle  \, ,
\]
where the dots denote other sites of the spin chain on which the Hamiltonian is not acting.
Therefore, the right-hand side of Eq.~\eqref{eq:spin_chain_part2_su23} does not vanish  due to the periodicity of the states on the spin chain, because it contains an arbitrary number of other excitations. }
\beq
H_{\rm int} | k \, l \rangle =
H_{\rm spin} | k \, l \rangle = 
 2 U^{m n}_{k l} | m \, n \rangle  \, .
 \label{eq:action_spin_chain_two_particle_states}
\eeq
Since the extension from large to finite $N$ is unique, the SMT in a given near-BPS limit is defined unambiguously from the action of the spin chain Hamiltonian on the letters describing the surviving degrees of freedom of the sector.
This action is in turn obtained by the one-loop corrections of the dilatation operator.
Therefore we will use the results from \cite{Beisert:2003ys,Bellucci:2006bv,Beisert:2007sk, Zwiebel:2007cpa} to derive the quantized SMT Hamiltonian in Sections \ref{sect-spin_chain_SU23_sector} and \ref{sect-spin_chain_PSU112_sector}.

\subsection{SU(2|3) sector}
\label{sect-spin_chain_SU23_sector}

We start from the spin chain result derived in \cite{Beisert:2003ys}.
In order to make the comparison simpler, we review the notation used in such reference. 
We call the fields of the $\SU(2|3)$ sector as
\beq
W_A = \lbrace a_a, b_{\alpha}  \rbrace 
\eeq
with $a=1,2,3$ and  $\alpha=1,2$ (consequently $A=1,2,3,4,5$) referring to the scalars in the triplet of $\SU(3)$ and to the fermions in the doublet of the $\SU(2)$ symmetry, respectively.
Defining the symbol
\beq
\bracenom{A_1 \dots A_n}{B_1 \dots B_m} \equiv
W_{B_1} \dots W_{B_n} \, \frac{\delta}{\delta W_{A_m}} \dots 
\frac{\delta}{\delta W_{A_1}} \, ,
\eeq 
which performs a permutation of the letters, the spin chain Hamiltonian can be written as
\beq
\frac{H_{\rm spin}}{c^2} =  \bracenom{a b}{a b} 
+ \left(  \bracenom{a \beta}{a \beta}  +  \bracenom{\alpha b}{\alpha b}  \right) +  \bracenom{\alpha \beta}{\alpha \beta} 
-  \bracenom{a b}{b a} - \left(  \bracenom{a \beta}{\beta a} + \bracenom{\alpha b}{b \alpha}  \right)
+ \bracenom{\alpha \beta}{\beta \alpha}  \, ,
\eeq
where $c$ is an overall normalization.
We also collectively denote the letters of the sector as $ |a \rangle \equiv \lbrace | Z \rangle , | X \rangle , | W \rangle  \rbrace $ for the bosonic excitations, and as $ | \alpha \rangle \equiv \lbrace  | \psi_+, \psi_- \rangle \rbrace $ for the fermionic modes\footnote{More details on the letters and the algebra of the $\SU(2|3)$ sector are given in Appendix \ref{app-review_SU23}.}.
Consequenty, the action of the Hamiltonian on the letters of the $\SU(2|3)$ sector is
\begin{align}
	\begin{split}
&  H_{\rm spin} | a b \rangle = 2 | a b \rangle - 2 | b a \rangle \, , \qquad
H_{\rm spin} | \alpha \beta \rangle = 2 | \alpha \beta \rangle 
+ 2 | \beta \alpha \rangle \, , 
\\
& H_{\rm spin} | a \beta \rangle = 2 | a \beta \rangle - 2 | \beta a \rangle \, , \qquad
  H_{\rm spin} | \alpha b  \rangle = 2 | \alpha b \rangle - 2 | b \alpha \rangle \, .
  \label{eq:spin_chain_part2_su23}
  \end{split}
\end{align}
The factors of 2 come from the combinations counted by the abovementioned symbols and from the periodicity of the spin chain.
Using the dictionary presented in Eq.~\eqref{eq:action_spin_chain_two_particle_states}, we translate the action of the spin chain Hamiltonian on the letters to a quartic expression containing two creation and two annihilation operators.
Omitting momentarily overall factors of $1/N$ included in the dictionary, the purely bosonic part reads
\beq
- \tr \left(  [a^{\dagger}_a, a_a] [a^{\dagger}_b, a_b]  \right)
+ \tr \left(  [a^{\dagger}_b , a_a] [a^{\dagger}_a , a_b]  \right) =
\tr \left( [a^{\dagger}_b , a^{\dagger}_a][a_a , a_b] \right) \, ,
\eeq
therefore we notice that it naturally reorganizes as an F-term after using the cyclicity properties of the trace\footnote{The terminology of D-term and F-term is commonly used in the supersymmetry literature to refer to expressions integrated over either the full superspace or half of it, respectively \cite{Weinberg:2000cr}.}.
The purely fermionic part is similar, but crucially there is a different relative sign which also allows to get an F-term using the corresponding identity for the trace of anticommuting quantities:
\beq \label{eq:Jacobi-identity-fourscalar}
- \tr \left( \lbrace b^{\dagger}_{\alpha}, b_{\alpha} \rbrace  \lbrace b^{\dagger}_{\beta}, b_{\beta} \rbrace \right)  
- \tr \left(  \lbrace b^{\dagger}_{\beta}, b_{\alpha} \rbrace  \lbrace b^{\dagger}_{\alpha}, b_{\beta} \rbrace \right) =
\tr \left( \lbrace b^{\dagger}_{\beta}, b^{\dagger}_{\alpha} \rbrace  \lbrace b_{\alpha}, b_{\beta} \rbrace  \right)   \, .
\eeq
The interactions are completed by the mixed bosonic--fermionic term, which combines into
\beq  \label{eq:Jacobi-identity-fourfermion}
- 2 \tr \left( [a^{\dagger}_a, a_a] \lbrace b^{\dagger}_{\beta}, b_{\beta} \rbrace \right)
+ 2 \tr \left( [b^{\dagger}_{\beta}, a_a] [a^{\dagger}_a, b_{\beta}] \right) =
- 2 \tr \left(  [b_{\beta}, a_a] [a^{\dagger}_a , b^{\dagger}_{\beta}]  \right) \, .
\eeq
Putting all the terms together and restoring factors in the numbers of colours, we find the SMT Hamiltonian of the sector:
\beq
\frac{H_{\rm SMT}}{c^2} = \frac{1}{2N} \tr \left( [a^{\dagger}_b , a^{\dagger}_a][a_a , a_b] \right) 
+ \frac{1}{2N} \tr \left( \lbrace b^{\dagger}_{\beta}, b^{\dagger}_{\alpha} \rbrace  \lbrace b_{\alpha}, b_{\beta} \rbrace  \right) 
 + \frac{1}{N} \tr \left(  [a^{\dagger}_a , b^{\dagger}_{\beta}] [b_{\beta}, a_a] \right) \, . 	
 \label{eq:spin_chain_Ham_su23}
\eeq
Written in this form, the Hamiltonian is manifestly positive definite and all the interactions naturally organize into an F-term structure.
We also observe that the interactions naturally appear in a normal-ordered form; this will not be the case in the $\mathrm{PSU}(1,1|2)$ sector treated in Section \ref{sect-spin_chain_PSU112_sector}.

\subsection{PSU(1,1|2) sector}
\label{sect-spin_chain_PSU112_sector}

The quantization of the dilatation operator in the $\mathrm{PSU}(1,1|2)$ sector was analyzed in \cite{Bellucci:2006bv,Beisert:2007sk, Zwiebel:2007cpa}.
In this case, the application of the dictionary between spin chains and SMT Hamiltonian involves summations over momenta, since the letters can contain an arbitrary number of covariant derivatives.
Following the notation in Appendix \ref{app-review_PSU112}, the bosonic letters are denoted as $|Z_n \rangle, |X_n \rangle $ and the fermionic ones as $ | \psi^1_n \rangle , | \psi^2_n \rangle , $ where the label $n$ refers to the number of covariant derivatives $d_1$ acting on their zero-modes.
These states are normalized to unity\footnote{Notice that instead reference \cite{Bellucci:2006bv} uses a different normalization for the fermions. We obtain our conventions by rescaling their fermionic letters as $| \chi_n \rangle \rightarrow \sqrt{n+1} \, | \chi_n \rangle $ and $| \zeta_n \rangle \rightarrow \sqrt{n+1} \, | \zeta_n \rangle .$}.

\subsubsection*{Warm-up: SU(1,1) bosonic subsector}

We show an explicit example of the manipulations to get the SMT Hamiltonian starting from the action on the states of the spin chain.
We restrict to the case where the only non-trivial letters are $|Z_n \rangle ,$ which corresponds to the $\SU(1,1)$ bosonic subsector.
We start from the result given \emph{e.g.} in \cite{Beisert:2004ry}
\beq
H_{\rm spin} | Z_m Z_n \rangle =
(h(m) + h(n)) | Z_m Z_n \rangle - \sum_{l=1}^m \frac{1}{l} | Z_{m-l} Z_{n+l} \rangle 
- \sum_{l=1}^n \frac{1}{l} | Z_{m+i} Z_{n-l} \rangle \, ,
\eeq
where $h(n) = \sum_{i=1}^n \frac{1}{i}$ are the harmonic numbers.
Using Eq.~\eqref{eq:action_spin_chain_two_particle_states}, we immediately conclude that the coefficients $U^{s'r'}_{sr}$ of the interactions written in SMT form are 
\begin{equation}
\begin{array}{c} \ds
-2 U^{mn}_{mn} = h(m) + h(n) \\[3mm] \ds
- 2 U^{m-l,n+l}_{mn} = - \frac{1}{l} \ \ \mbox{for} \ \ l = 1,...,m \\[3mm] \ds
- 2 U^{m+l,n-l}_{mn} = - \frac{1}{l} \ \ \mbox{for} \ \ l = 1,...,n\,.
\end{array}
\end{equation} 
Inserting these expressions inside Eq.~\eqref{eq:dictionary_bos_term_SMT} and exploiting the symmetry of the summation indices $m,n$ plus the cyclicity properties of the trace yields
\begin{equation}
H_{\rm SMT}  = - \frac{1}{2N} \sum_{m,n = 0}^\infty :\tr\bigg(h(m) \left[a_{m}^\dagger,a^m\right] \left[a_{n}^\dagger, a^n\right] - \sum_{l=1}^n \frac{1}{l} \left[a_{m+l}^\dagger,a^m\right] \left[a_{n-l}^\dagger, a^n\right] \bigg):  
\label{eq:first_step_SMT_Hamiltonian_su11}
\end{equation}
In this form, the interacting Hamiltonian is written in terms of a renormalized four-point interaction, which is the result of the one-loop correction to the dilatation operator in the $\SU(1,1)$ bosonic subsector.

For comparison with the results that will be presented in Section \ref{sect-effective_ham_from_sphere} and in view of a local description, it proves useful to rewrite the Hamiltonian obtained above in a form which is not normal ordered.
First of all, it is convenient to express the interactions in terms of the quantum-mechanical bosonic charge density defined as
\beq
q_l \equiv \sum_{n=0}^{\infty}  : [a^{\dagger}_n , a^{n+l}] : 
\eeq
We achieve this task by manipulating summations over momenta as follows
%\beq
%\sum_{m,n=0}^{\infty} \sum_{l=1}^n \rightarrow
%\sum_{l=1}^{\infty} \sum_{m=0}^{\infty}  \sum_{n=l}^{\infty} 
%\eeq
%and then we send $k \rightarrow k+i $ everywhere in order to obtain an expression containing only sums over positive integers.
%At the end of the day, we obtain
\beq
\sum_{m,n=0}^{\infty} \sum_{l=1}^n F(m,n,l) =
\sum_{l=1}^{\infty} \sum_{m=0}^{\infty}  \sum_{n=l}^{\infty} F(m,n,l) =
\sum_{m,n=0}^{\infty} \sum_{l=1}^{\infty} F(m,n+l,l) \, ,
\label{eq:manipulations_sums_momenta}
\eeq
for $F(m,n,l)$ a generic function of the three integer momenta.
Similar manipulations can be performed when the sum starts from $l=0,$
since it is sufficient to shift $l \rightarrow l -1$ to end up in the same setting as Eq.~\eqref{eq:manipulations_sums_momenta}.

At this point we can show by direct computation that
\beq
\sum_{l=1}^n\frac{1}{l}\tr(: q^{\dagger}_l q_l :) = \sum_{l=1}^n\frac{1}{l}\tr(q^{\dagger}_l q_l)
- 2 N \sum_{n = 0}^\infty h(n) \tr(a_{n}^\dagger a^{n}) + 2 \sum_{n = 0}^\infty h(n) \tr(a_{n}^\dagger) \tr(a^{n})\,.
\label{eq:normal_ordering_identity_su11bos}
\eeq
The story is not yet complete: the SMT Hamiltonian is defined on the Hilbert space of singlet states under the colour group $\SU(N),$ therefore we have the following constraint which applies to all physical states\footnote{Precisely, this constraint applies in the $\SU(1,1)$ bosonic subsector. In the full $\mathrm{PSU}(1,1|2)$ sector, Eq.~\eqref{eq:su11_bos_singlet_constraint} gets modified with a different definition of the charge density.}:
\beq
q_0 | \mathrm{phys} \rangle = 0 \, .
\label{eq:su11_bos_singlet_constraint}
\eeq
Similar manipulations with normal ordering give rise to the identity
\beq
\begin{aligned}
\sum_{m,n = 0}^\infty h(m) \tr(:\left[a_{m}^\dagger,a^m\right] \left[a_{n}^\dagger, a^n\right]:) = & \sum_{m=0}^{\infty} h(m) \tr \left( :[a^{\dagger}_m , a_m]: q_0 \right) 
 \\
& - 2 N \sum_{m=0}^\infty h(m) \tr(a_{m}^\dagger a^m) + 2 \sum_{m=0}^\infty h(m) \tr(a_{m}^\dagger)\tr(a^m)\,.
\end{aligned}
\label{eq:singlet_condition_boson_su111}
\eeq
Crucially, the self-energy contributions simplify with Eq.~\eqref{eq:normal_ordering_identity_su11bos}. 
Hence we find the interacting Hamiltonian
\begin{equation}
H_{\text{SMT}} = \frac{1}{2N} \sum_{l=1}^\infty  \frac{1}{l} \tr(q^{\dagger}_l q_l)\,,
\end{equation} 
which is precisely the same result obtained from sphere reduction combined with a quantization prescription, up to an overall normalization \cite{Baiguera:2020jgy}.

\subsubsection*{Extending to PSU(1,1|2) sector: double trace terms}

Moving to the full $\mathrm{PSU}(1,1|2)$ sector, there is a richer structure of interactions.
In the following we will distinguish these terms into single and double trace contributions, referring to the sum over the indices under the residual $\SU(2)$ internal symmetry of the model.
This terminology should not be confused with the sum over the indices of the $\SU(N)$ colour group; from that perspective, the interacting Hamiltonian will only contain single trace structures.

The SMT Hamiltonian of the $\mathrm{PSU}(1,1|2)$ sector contains several interactions which can be combined into a double trace term under the residual $\SU(2)$ symmetry, built with the total charge density of the system.
It is straightforward to generalize the steps undertaken for the $\SU(1,1)$ bosonic subsector to include the fermionic terms.
We define the total charge density of the sector as
\beq
q_l \equiv \sum_{n=0}^{\infty} \sum_{a=1,2} :[(a_a)^\dagger_n , (a_a)_{n+l}]: \, , \quad
\tilde{q}_l = \sum_{n=0}^{\infty} \sum_{a=1,2} \sqrt{\frac{n+1}{n+l+1}} :\lbrace (b^{\dagger}_a)_n , (b_a)_{n+l} \rbrace: \, , \quad
\mathbf{Q}_{l} = q_l + \tilde{q}_l \, .
\label{eq:definition_charge_densities_SMT_Hamiltonian_su111}
\eeq
One can show that the following identities involving fermionic ladder operators hold:
\beq
\sum_{l=1}^{\infty} \frac{1}{s} \tr \left( : \tilde{q}^{\dagger}_s \tilde{q}_s: \right) =
\sum_{s=1}^{\infty} \frac{1}{s} \tr \left( \tilde{q}^{\dagger}_s \tilde{q}_s \right) 
- 2 N \sum_{s=0}^{\infty} h(s+1) \tr ( b^{\dagger}_s b_s ) 
- 2 \sum_{s=0}^{\infty} h(s+1) \tr ( b^{\dagger}_s ) \tr (b_s) \, ,
\label{eq:self_energy_corrections_fermions}
\eeq
and
\begin{multline}
 \sum_{m,n=0}^{\infty} h(m+1) \tr(:\left\{b_{m}^\dagger,b^m\right\}\left\{b_{n}^\dagger, b^n\right\}:) = \sum_{m=0}^{\infty} h(m+1) \tr(:\left\{b_{m}^\dagger,b^m\right\}: \tilde{q}_0 ) \\ 
 - 2 N \sum_{m=0}^{\infty} h(m+1) \tr(b_m^\dagger b^m) - 2 \sum_{m=0}^{\infty} h(m+1) \tr(b_m^\dagger) \tr(b^m)\,.
 \label{eq:singlet_condition_fermion_su111}
\end{multline}
Using these identities together with Eqs.~\eqref{eq:normal_ordering_identity_su11bos} and \eqref{eq:singlet_condition_boson_su111}, plus the action of the spin chain Hamiltonian on the letters of the $\mathrm{PSU}(1,1|2)$ sector, we find the following double trace contribution to the SMT Hamiltonian
\beq
\frac{1}{2N} \sum_{l=1}^{\infty} \frac{1}{l} \tr \left( \mathbf{Q}_l^{\dagger} \mathbf{Q}_l  \right) \, , 
\label{eq:double_trace_terms_SMT_Ham_PSU112}
\eeq
where the $\SU(N)$ singlet condition has already been used to get rid of some terms.
This result was already derived from the sphere reduction perspective in \cite{Baiguera:2020jgy}.

\subsubsection*{Extending to PSU(1,1|2) sector: single trace terms}

The interacting Hamiltonian obtained in Eq.~\eqref{eq:double_trace_terms_SMT_Ham_PSU112} is not complete: there are other single trace terms which conclude the description of the SMT.
We present a sample of such calculations involving the purely fermionic part, and then we will state the full result (which can be derived in a similar way).
The spin chain Hamiltonian acting on the fermionic letters gives rise to \cite{Bellucci:2006bv,Beisert:2007sk}
\begin{align}
	\begin{split}
& H_{\rm spin} | \psi^1_m \psi^2_n \rangle  \supset \\
& - \sum_{l=1}^m \sqrt{\frac{(n+1)(m-l+1)}{(m+1)(n+l+1)}}  \, 
\frac{| \psi^1_{m-l} \psi^2_{n+l} \rangle}{m+n+2} 
- \sum_{l=1}^n \sqrt{\frac{(m+1)(n-l+1)}{(n+1)(m+l+1)}} \,
  \frac{ | \psi^1_{m+l} \psi^2_{n-l} \rangle}{m+n+2} 
 \\
 &    + \sum_{l=1}^m \sqrt{\frac{(n+1)(m-l+1)}{(m+1)(n+l+1)}} \, \frac{| \psi^2_{m-l} \psi^1_{n+l} \rangle}{m+n+2} 
 + \sum_{l=1}^n \sqrt{\frac{(m+1)(n-l+1)}{(n+1)(m+l+1)}} \,
  \frac{ | \psi^2_{m+l} \psi^1_{n-l} \rangle}{m+n+2}  \, ,
\\
& H_{\rm spin} | \psi^2_m \psi^1_n \rangle \supset \\
 & - \sum_{l=1}^m \sqrt{\frac{(n+1)(m-l+1)}{(m+1)(n+l+1)}} \,
  \frac{| \psi^2_{m-l} \psi^1_{n+l} \rangle}{m+n+2} 
  - \sum_{l=1}^n \sqrt{\frac{(m+1)(n-l+1)}{(n+1)(m+l+1)}} \,
   \frac{| \psi^2_{m+l} \psi^1_{n-l} \rangle}{m+n+2}  
 \\
 &  + \sum_{l=1}^m \sqrt{\frac{(n+1)(m-l+1)}{(m+1)(n+l+1)}} \,
  \frac{| \psi^1_{m-l} \psi^2_{n+l} \rangle}{m+n+2} 
 + \sum_{l=1}^n \sqrt{\frac{(k+1)(n-l+1)}{(n+1)(k+l+1)}} \,
  \frac{| \psi^1_{m+l} \psi^2_{n-l} \rangle}{m+n+2}   \, .
\end{split}
\end{align}
Here we are taking into account only the terms contributing to the single trace part. 
Using the dictionary in Eq~\eqref{eq:dictionary_ferm_term_SMT} with two copies of fermions, we find
\beq
\begin{aligned}
& - 2 U^{(m-l)_1,(n+l)_2}_{m_1,n_2} =
2 U^{(m-l)_2,(n+l)_1}_{m_1,n_2} = 
- 2 U^{(m-l)_2,(n+l)_1}_{m_2,n_1} =  \\
& = 2 U^{(m-l)_1,(n+l)_2}_{m_2,n_1}
 =  \sqrt{\frac{(n+1)(m-l+1)}{(m+1)(n+l+1)}}  \frac{1}{m+n+2} \, , \\[3mm]
& - 2 U^{(m+l)_1,(n-l)_2}_{m_1,n_2} =
2 U^{(m+l)_2,(n-l)_1}_{m_1,n_2} = 
- 2 U^{(m+l)_2,(n-l)_1}_{m_2,n_1} =  \\
& = 2 U^{(m+l)_1,(n-l)_2}_{m_2,n_1}  =
  \sqrt{\frac{(m+1)(n-l+1)}{(n+1)(m+l+1)}}   \frac{1}{m+n+2}  \, .
\end{aligned}
\label{eq:translation_spin_chain_SMT_quartic_fermion_term}
\eeq
In the previous notation, the subscipt $j$ in the labels $m_j$ of momenta refers to the flavor of the spinor, to distinguish the two copies of fermions surviving the near-BPS limit.
Now one makes use of the identity \eqref{eq:manipulations_sums_momenta} and of the cyclicity properties of the trace to express the interactions in the form
\beq
\begin{aligned}
& \sum_{l=0}^{\infty} \sum_{m,n=0}^{\infty} \frac{\sqrt{(m+1)(n+1)}}{\sqrt{(m+l+1)(n+l+1)}} 
\frac{\tr \left(  \epsilon^{ac}  \epsilon^{bd} \lbrace (b_a^{\dagger})_m , (b_b)_{m+l} \rbrace \lbrace (b_c^{\dagger})_{n+l} , (b_d)_n \rbrace \right)}{m+n+l+2} \\
& - \frac{1}{2} \sum_{m,n=0}^{\infty} \frac{1}{m+n+2}  \tr \left( \epsilon^{ac} \epsilon^{bd} \lbrace (b_a^{\dagger})_m , (b_b)_{m} \rbrace \lbrace (b_c^{\dagger})_{n} , (b_d)_n \rbrace \right)  \, .
\end{aligned}
\eeq
Notice that the second line is a boundary term, arising from evaluating half of the terms of the first line in $l=0.$
It will play a crucial role in the check of the symmetries of the SMT.

Working in a similar way, we obtain the full interacting Hamiltonian of the model
\beq
\begin{aligned}
H_{\rm SMT} & = H_B +  \frac{1}{2N} \sum_{l=1}^{\infty}  \frac{1}{l}  \tr \left( \mathbf{Q}_l^{\dagger} \, \mathbf{Q}_l  \right) 
	+ \frac{1}{2N} \sum_{l = 0}^{\infty}   \tr ( (F_{ab})_l^\dagger (F_{ab})_l )  \\
&  - \frac{1}{2N} \sum_{l=0}^{\infty} \sum_{m,n=0}^{\infty} \frac{1}{m+n+l+1} \tr \left(  \epsilon^{ac}  \epsilon^{bd} [(a_a^{\dagger})_m , (a_b)_{m+l}] [(a_c^{\dagger})_{n+l} , (a_d)_n] \right)    \\
& +  \frac{1}{2N} \sum_{l=0}^{\infty} \sum_{m,n=0}^{\infty} \frac{\sqrt{(m+1)(n+1)}}{\sqrt{(m+l+1)(n+l+1)}} 
\frac{ \tr \left(  \epsilon^{ac}  \epsilon^{bd} \lbrace (b_a^{\dagger})_m , (b_b)_{m+l} \rbrace \lbrace (b_c^{\dagger})_{n+l} , (b_d)_n \rbrace \right) }{m+n+l+2}
 \\
& + \frac{1}{2N} \sum_{l=0}^{\infty} \sum_{m,n=0}^{\infty} \sqrt{\frac{m+1}{n+l+1}} \frac{\epsilon^{ac} \epsilon^{bd}}{m+n+l+2}
 \tr \left( [(b_a^{\dagger})_m , (a_b)_{m+l+1}] [(b_c^{\dagger})_{n+l} , (a_d)_n ] \right)  \\
& -\frac{1}{2N} \sum_{l=0}^{\infty} \sum_{m,n=0}^{\infty} \sqrt{\frac{m+1}{n+l+1}}  \frac{\epsilon^{ac} \epsilon^{bd}}{m+n+l+2}  \tr \left( [(a_a^{\dagger})_{m+l+1} , (b_b)_{m}] [(a_c^{\dagger})_{n} , (b_d)_{n+l} ] \right)  \, ,
\end{aligned} 
\label{eq:total_spin_chain_Ham_psu112}
\eeq
where we defined a boundary Hamiltonian in the following way
\beq
\begin{aligned}
H_B & \equiv \frac{1}{4N}  \sum_{m,n=0}^{\infty} \frac{1}{m+n+1} 
\tr \left( \epsilon^{ac} \epsilon^{bd} [(a_a^{\dagger})_m , (a_b)_{m}] [(a_c^{\dagger})_{n} , (a_d)_n] \right)   \\
& - \frac{1}{4N}  \sum_{m,n=0}^{\infty} \frac{1}{m+n+2}  \tr \left( \epsilon^{ac} \epsilon^{bd} \lbrace (b_a^{\dagger})_m , (b_b)_{m} \rbrace \lbrace (b_c^{\dagger})_{n} , (b_d)_n \rbrace \right) \, ,
\end{aligned} 
\label{eq:boundary_Hamiltonian_spin_chain}
\eeq
and the block
\beq
(F_{ab})_l \equiv - \sum_{m=0}^\infty  \frac{ : [(a_b)_m^\dagger, (b_a)_{m+l}] :}{\sqrt{m+l+1}} \, .
\eeq
This concludes the derivation of the SMT Hamiltonian in the $\mathrm{PSU}(1,1|2)$ near-BPS limit.
We will derive the classical expression from sphere reduction in Section \ref{sect-sphere_reduction_PSU112} and define a quantization prescription in Section \ref{sect-quantization_sphere_red} to show that the two procedures are equivalent.

%%%%%%%%%%%%%%%%%%%%

\section{Hamiltonian from sphere reduction}
\label{sect-effective_ham_from_sphere}

In this Section we derive the classical effective Hamiltonian describing $\mathcal{N}=4$ SYM theory close to the BPS limits defined in Eq.~\eqref{eq:near_BPS_limits_1over8BPS_subsectors}, which give rise to $\SU(2|3)$ or $\mathrm{PSU}(1,1|2)$ when $\alpha=0,1$ respectively.
The general procedure is based on the sphere reduction technique used in \cite{Harmark:2019zkn, Baiguera:2020mgk, Baiguera:2020jgy}. 
We review this method and specify the features that characterize the near-BPS limits of interest in Section \ref{sect-general_procedure_sphere_reduction}.
Then we derive the effective Hamiltonian describing the degrees of freedom of the $\SU(2|3)$ sector in Section \ref{sect-sphere_reduction_SU23} and of the $\mathrm{PSU}(1,1|2)$ sector in Section \ref{sect-sphere_reduction_PSU112}.
We present a quantization prescription and compare the result with the alternative method based on the loop corrections of the dilatation operator in Section \ref{sect-quantization_sphere_red}.

\subsection{The sphere reduction method}
\label{sect-general_procedure_sphere_reduction}

We consider the classical action of $\mathcal{N}=4$ SYM on $\mathbb{R} \times S^3$ normalized as
\begin{multline}
S  = \int_{\mathbb{R} \times S^3} \sqrt{-\mathrm{det} \, g_{\mu\nu}} \, \tr  \left\lbrace - \frac{1}{4} F_{\mu\nu}^2 - |D_{\mu} \Phi_a|^2 - |\Phi_a|^2 - i \psi^{\dagger}_a \bar{\sigma}^{\mu} D_{\mu} \psi^A + g \sum_{A,B,a} C_{AB}^{a} \psi^A [\Phi_a, \psi^B]  \right. \\
 \left. + g \sum_{A,B,a} \bar{C}^{aAB} \psi^{\dagger}_A [\Phi^{\dagger}_a, \psi^{\dagger}_B]
- \frac{g^2}{2} \sum_{a,b} \left( |[\Phi_a, \Phi_b]|^2 + |[\Phi_a, \Phi^{\dagger}_b]|^2 \right)
 \right\rbrace \, .
\label{eq:theS}
\end{multline}
The conventions are summarized as follows:
\begin{itemize}
\item The Yang-Mills coupling constant is denoted with $g.$ 
\item The metric on $\mathbb{R} \times S^3$ is $g_{\mu\nu}$ and the radius of the three-sphere is set to unity.
\item The complex scalar fields transform in the $\mathbf{6}$ representation of the R-symmetry group $\mathrm{SO}(6) \simeq \SU(4)$ and are defined in terms of the real scalars as $\Phi_a = \frac{1}{\sqrt{2}} \left(\phi_{2a-1} + i \phi_{2a} \right)$ with $a \in \lbrace 1,2,3 \rbrace .$
\item The Weyl fermions $\psi^A$ transform in the fundamental representation $\mathbf{4}$ of $\SU(4),$ with index $A \in \lbrace 1,2,3,4 \rbrace .$
\item The field strength is defined as
\beq
F_{\mu\nu} = \partial_{\mu} A_{\nu} - \partial_{\nu} A_{\mu} + i g [A_{\mu} , A_{\nu}] \, .
\eeq
\item The covariant derivatives read
\bea
& D_{\mu} \Phi_a = \partial_{\mu} \Phi_a + i g [A_{\mu}, \Phi_a] \, , & \\
& D_{\mu} \psi^A = \nabla_{\mu} \psi^A + i g [A_{\mu}, \psi^A] \, , &
\eea
where $\nabla_{\mu}$ is the part of the connection containing only the geometrical contribution due to the curved space where the theory lives.
\item The cubic interactions involving two fermions and one scalar are mediated by the Clebsch-Gordan coefficients $C^a_{AB}$ coupling two $\mathbf{4}$  representations and one $\mathbf{6}$ representation of the R-symmetry group $\SU(4).$
\item All the fields transform in the adjoint representation of the gauge group $\SU(N)$.
\end{itemize}
Starting from the action \eqref{eq:theS}, it is straightforward to obtain the classical Hamiltonian $H$ of $\mathcal{N}=4$ SYM on $\mathbb{R} \times S^3$ by means of a Legendre transform.
We consider near-BPS limits of the form
\begin{equation}
\label{eq:newnearBPSlimit}
g \rightarrow 0 \quad \mbox{with} \quad \frac{H - \alpha S_1 - Q_1 - Q_2 - (1-\alpha) Q_3}{g^2} \quad \mbox{fixed} \,,
\end{equation}
where $\alpha = 0,1.$ We denoted the Cartan generators associated to rotational symmetry on the three-sphere as $S_1, S_2$ and the R-charges as $Q_i$ with $i \in \lbrace 1,2,3 \rbrace .$
In this procedure, we keep $N$ fixed while $g \rightarrow 0.$
The surviving degrees of freedom of the sector are described in terms of the interacting Hamiltonian defined by
\begin{equation}
\label{eq:full_ham_gen}
 H_{\rm int} = \lim_{g \rightarrow 0} \frac{H - \alpha S_1 -Q_1 - Q_2- (1- \alpha) Q_3}{g^2 N} \,.
\end{equation}
The general procedure to extract the effective Hamiltonian is the following:
\begin{enumerate}
\item Isolate the propagating modes in a given near-BPS limit from the quadratic classical Hamiltonian.
\item Derive the form of the currents that couple to the gauge fields.
\item Integrate out additional non-dynamical modes that give rise to effective interactions at order $g^2.$ 
\item Compute the interacting Hamiltonian by taking the limit \eqref{eq:full_ham_gen}.
\end{enumerate}
Dynamical components of the gauge field only survive in sectors where both the rotation generators $S_1,S_2$ are turned on \cite{Baiguera:2020mgk}; instead in the near-BPS limits considered here, we turn off $S_2$ or both of them.
However, the gauge field still plays an important role to determine the effective interactions in the 1/8 near-BPS limits \eqref{eq:newnearBPSlimit}, because it always mediates effective interactions at order $g^2$ even when all its degrees of freedom decouple on-shell.
In order to determine such contributions in the interacting Hamiltonian, we need to perform the Dirac analysis of constraints.
We work in the Coulomb gauge
\beq
\nabla_i A^i = 0 \, ,
\label{eq:Coulomb_gauge}
\eeq
which gets rid of the temporal and longitudinal components of the gauge field.
The quadratic part of the Yang-Mills action plus the source term is given by
\begin{equation}
S_A = \int_{\mathbb{R}\times S^3} \sqrt{-\mathrm{det} \, g_{\mu\nu}} \, \tr(-\frac{1}{4} F_{\mu\nu}^2 - A^\mu j_\mu)
\,.
\end{equation}
The corresponding Hamiltonian is obtained via Legendre transform with the inclusion of a Lagrange multiplier to impose the Coulomb gauge.
After expanding the fields into spherical harmonics according to the conventions in Appendix \ref{app-notations_conventions}, the constraints become algebraic and can be solved explicitly to get rid of the unphysical modes.
The explicit intermediate steps can be found in Section 2.1 of \cite{Baiguera:2020jgy}.
We report the result for the unconstrained Hamiltonian density
\begin{equation}
\label{eq:Ham_freeYM}
H_A = \tr\sum_{J,m,\tilde{m}} \left[ \sum_{\rho = \pm1}\left( \frac{1}{2} |\Pi_{(\rho)}^{Jm\tilde{m}}|^2 + \frac{1}{2} \omega_{A,J}^2 |A_{(\rho)}^{Jm\tilde{m}}|^2+ A_{(\rho)}^{Jm\tilde{m}} j_{(\rho)}^{\dagger\,Jm\tilde{m}}\right) + \frac{1}{8J(J+1)} |j_0^{Jm\tilde{m}}|^2 \right] \,.
\end{equation}
We apply the near-BPS limit \eqref{eq:newnearBPSlimit} at quadratic level. Since the gauge field is neutral under R-symmetry, we find
\begin{equation}
H_0 - \alpha  S_1  = 
\sum_{J,M}  \sum_{\rho=-1,1} \frac{1}{2} \left(|\Pi_{(\rho)}^{JM} - i \alpha (m- \tilde{m}) A_{(\rho)}^{\dagger\,JM}|^2+ (\omega_{A,J}^2 - \alpha^2 (m-\tilde{m})^2) |A_{(\rho)}^{JM}|^2\right)
\, ,
\label{eq:quadratic_gauge_Hamiltonian_su12sectors}
\end{equation}
where $H_0$ is the free Hamiltonian.
According to the general procedure listed above, we need to impose that this combination vanishes.
Since $|m- \tilde{m}| \leq 2J +1,$ there is no possibility to make the prefactor of the second term to vanish; therefore the constraints read
\beq
A_{(\rho)}^{J M} = 0 \, , \qquad
\Pi^{JM}_{(\rho)} - i \alpha (m -\tilde{m}) A^{\dagger \, JM}_{(\rho)} = 0 \ . 
\label{eq:general_constraints_gauge_fields}
\eeq
It is clear that each of the above constraints eliminates a dynamical degree of freedom
from the theory, as one forfeits the choice of freely choosing initial conditions. Instead,
the corresponding fields are entirely determined by the remaining degrees of the freedom,
as encoded by the right hand sides of the above relations. We can make these explicit
by demanding compatibility with Hamiltonian evolution.

The first constraint in Eq.~\eqref{eq:general_constraints_gauge_fields} does not give rise to new relations; the second one entails instead a new consistency condition, coming from restoring the sources as in Eq.~\eqref{eq:Ham_freeYM} inside the Hamiltonian \eqref{eq:quadratic_gauge_Hamiltonian_su12sectors}. We find
\beq
\lbrace H, \Pi^{JM}_{(\rho)} - i  \alpha (m-\tilde{m}) A^{\dagger \, JM}_{(\rho)} \rbrace =
(\omega_{A,J}^2 - \alpha^2 (m- \tilde{m})^2) A_{(\rho)}^{\dagger\,Jm\tilde{m}} + j_{(\rho)}^{\dagger\,Jm\tilde{m}} 
= 0 \, ,
\eeq
or analogously
\beq
A_{(\rho)}^{Jm\tilde{m}} = -\frac{j^{ Jm\tilde{m}}_{(\rho)} }{\omega_{A,J}^2  - \alpha^2 (m-\tilde{m})^2} \, .
\label{eq:general_constraints_gauge_field}
\eeq
Plugging this relation in Eq.~\eqref{eq:quadratic_gauge_Hamiltonian_su12sectors} with the sources restored, we obtain
\begin{equation}
H - \alpha S_1  = \tr \left( \sum_{J,m,\tilde{m}} \frac{1}{8J(J+1)} |j_0^{Jm\tilde{m}}|^2 - \frac{1}{2} \sum_{\rho=\pm1} \sum_{J,m,\tilde{m}}\frac{1}{\omega_{A,J}^2 - \alpha^2 (m-\tilde{m})^2} |j_{(\rho)}^{Jm\tilde{m}}|^2\right)  \, .
\label{eq:general_gauge_mediated_interaction_su12limits}
\end{equation}
Once the currents $j_0^{JM}, j^{JM}_{(\rho)}$ are extracted from the interacting Hamiltonian of $\mathcal{N}=4$ SYM, this expression must be used to compute the  gauge-mediated interactions.
At this point, we are ready to study the two near-BPS limits of interest explicitly.

\subsection{SU(2|3) sector}
\label{sect-sphere_reduction_SU23}

Turning off both angular momenta gives rise to a theory without remnants of the gluon in the original $\mathcal{N}=4$ SYM action, but containing three dynamical scalars and two fermions with the same chirality.
The BPS bound \eqref{eq:near_BPS_limits_1over8BPS_subsectors} with $\alpha=0$ is $H \geq Q_1 + Q_2 + Q_3, $ giving rise to interactions invariant under the $\SU(2|3)$ group.
The corresponding theory is supersymmetric and contains residues of the original R-symmetry group: there are three scalars transforming as a triplet under the $\SU(3)$ subgroup, and two fermions with the same chirality transforming as a doublet under another residual $\SU(2)$ subgroup.

\subsubsection*{Quadratic Hamiltonian and physical degrees of freedom}

At quadratic order, the combination of the free Hamiltonian and Cartan charges reads
\begin{multline}
 H_0 - Q_1 - Q_2 - Q_3  = \\
  \tr \sum_{JM} \Bigg\{ 
\sum_{a = 1,2,3}|\Pi_a^{JM} + i \Phi^{a \, \dagger}_{JM} |^2 + (\omega_J^2 - 1 ) |\Phi_a^{JM}|^2  
+  \sum_{\rho=-1,1} \frac{1}{2} \left(|\Pi_{(\rho)}^{Jm\tilde{m}} |^2+ \omega_{A,J}^2  |A_{(\rho)}^{Jm\tilde{m}}|^2\right)  \\
+ \sum_{\kappa=\pm 1}  \left[ \sum_{A=1}^4  \left( \kappa \omega^{\psi}_J - \frac{3}{2}  \right) (\psi^1_{JM,\kappa})^\dagger \psi^1_{JM,\kappa} + \sum_{A=2,3,4} \left(  \kappa \omega_{J}^{\psi} + \frac{1}{2} \right) (\psi^{A}_{JM,\kappa})^\dagger \psi^{A}_{JM,\kappa}  \right]  \Bigg\} \, .
\end{multline}
We impose that such expression vanishes.
By direct inspection, it is evident that all the modes satisfy $J=0,$ which agrees with the fact that the effective interactions are encoded by a theory of quantum mechanics.
We list the surviving degrees of freedom:
\begin{itemize}
\item \textbf{Scalars.}
All the scalar fields are dynamical with $J=m=\tilde{m}=0$ and satisfy the constraint
\beq
\Pi^a_{(0,0,0)} = - i \Phi^{a \, \dagger}_{(0,0,0)} \, ,
\label{eq:constraints_scalars_SU23}
\eeq
where we denoted the momenta with the subscript $(J,m,\tilde{m}).$
\item \textbf{Gauge fields.} They are not allowed to have any dynamical mode, as discussed in Section \ref{sect-general_procedure_sphere_reduction}.
\item \textbf{Fermions.} There are two fermions surviving the limit, and they are both characterized by the chirality $\kappa=1,$ with momentum $J=0$ and eigenvalues $m= \pm 1/2.$
We define them as spin up or down using the notation
\beq
\psi_+ \equiv \psi^1_{(0,\frac{1}{2},0)} \, , \qquad
\psi_- \equiv \psi^1_{(0,-\frac{1}{2},0)} \, .
\eeq
Their momenta are denoted with the subscript $(J,m,\tilde{m}),$ as we did for the scalars.
In the following, we also adopt the compact notation $\psi_{\alpha} \equiv (\psi_+, \psi_-).$
\end{itemize}
In order to get the standard normalization for the brackets, we redefine the scalars as
\beq
\Phi_a \equiv \sqrt{2} \Phi^a_{(0,0,0)} \, .
\label{eq:rescaling_scalars_SU_2_3}
\eeq
Thus all the fields have canonical Dirac brackets, \emph{i.e.} (no sum over $a, \alpha$)
\beq
\lbrace \Phi_a, \Phi^{\dagger}_a \rbrace_D = i \, , \qquad
\lbrace \psi_{\alpha} , \psi_{\alpha}^{\dagger}  \rbrace_D 
= i \, . 
\label{eq:Dirac_brackets_fields_SU23}
\eeq
The free Hamiltonian is obtained by using the constraints \eqref{eq:general_constraints_gauge_field} and \eqref{eq:constraints_scalars_SU23} inside Eq.~\eqref{eq:general_gauge_mediated_interaction_su12limits}.
The result is given by
\beq
H_0 =  \tr |\Phi_a |^2 + \frac{3}{2} \tr |\psi_{\alpha}|^2 \, .
\label{eq:free_Hamiltonian_SU23}
\eeq
The coefficients of the free Hamiltonian reflect the different scaling dimensions of the fields.

\subsubsection*{Interacting Hamiltonian}

Since by construction $H_0 - Q_1 - Q_2 - Q_3=0 , $ the interacting part of the Hamiltonian is defined via the limit \eqref{eq:quadratic_gauge_Hamiltonian_su12sectors} with $\alpha=0,$ \emph{i.e.} 
\beq
H_{\rm int} = \lim_{g \rightarrow 0} \frac{H - Q_1 - Q_2 - Q_3}{g^2 N} \, .
\eeq
We split the computation of such expression into three cathegories:
\begin{itemize}
\item Terms mediated by the gauge fields.
\item Quartic scalar self-interaction.
\item Yukawa term.
\end{itemize}
The expression of the Hamiltonian of $\mathcal{N}=4$ SYM decomposed into spherical harmonics from where we pick the terms is given in Eq.~\eqref{eq:app_full_interacting_N=4SYM_Hamiltonian}.

\vskip 4mm

\noindent
\textbf{Terms mediated by non-dynamical gauge fields.}

An immediate consequence of the absence of rotation charges in this near-BPS limit is that the total momentum of both dynamical fields and modes mediating the interactions is vanishing.
This implies that the first contribution in Eq.~\eqref{eq:general_gauge_mediated_interaction_su12limits}, arising from the current $j_0,$ identically vanishes because $J=0$ is excluded in the range of summation\footnote{More precisely, the constraint imposed by Coulomb gauge reads (after the decomposition of fields into spherical harmonics):
\[
2 i \sqrt{J(J+1)} \Pi_{(0)}^{J m \tilde{m}} 
+ j_0^{\dagger \, J m \tilde{m}} = 0  \, .
\] 
This implies $  j_0^{ J m \tilde{m}} = 0 $ when $J=0.$}.
Therefore the contribution to the effective Hamiltonian coming from the mediation of gauge fields is determined by the formula
\beq
- \frac{1}{8} \sum_{\rho=\pm1} \sum_m \tr |j^m_{(\rho)}|^2  \, ,
\label{eq:formula_terms_mediated_gauge_fields_SU23}
\eeq
where we used the fact that $J=0,$ and we explicitly denoted that the sums over free momenta only involve the spin $m,$ as we will see below.

The current $j^m_{(\rho)}$ contains a priori contributions from both the scalar and the fermion fields.
However, a direct inspection of the Clebsch-Gordan coefficient $\mathcal{D}$ defined in Eq.~\eqref{eq:app_definitionClebsch_D}  shows that the scalar part of the current identically vanishes.
%Indeed, given the assignment of momenta allowed to find interactions at order $g^2,$ the following Wigner 9--j symbol is involved (with $\rho = \pm 1$):
%\beq
%\begin{Bmatrix}
%0 & 0 & 1 \\
%\frac{\rho(\rho+1)}{2} & - \frac{\rho(1-\rho)}{2} & 1 \\
%0 & 0 & 0 
%\end{Bmatrix} = 0 \, .
%\eeq
The fermionic contribution to the current reads
\beq
j^{\dagger \, J M}_{(\rho)} = 
g  \, \mathcal{G}^{J_1 M_1 \kappa_1}_{J_2 M_2 \kappa_2; J M \rho} 
\lbrace  \psi^{\dagger}_{J_1 M_1 \kappa_1, A} , \psi^A_{J_2 M_2 \kappa_2} \rbrace \, ,
\label{eq:fermion_current_general_SU23}
\eeq
where $\mathcal{G}$ is another Clebsch-Gordan coefficient defined in Eq.~\eqref{eq:app_definitionClebsch_G}.
Imposing conservation of momentum and considering non-vanishing contributions to the effective Hamiltonian, we find
\begin{itemize}
\item $ \tilde{m} = \tilde{m}_1 = \tilde{m}_2=0.$ 
\item $J=J_1=J_2=0.$ 
\item $A=1$ and $\kappa_1=\kappa_2=1.$  
\end{itemize}
This confirms that the only relevant labels distinguishing the various cases are the eigenvalue $m$ of the momentum and the label $\rho$ describing the component of the gauge field.
Therefore we denote the current\footnote{Notice that we exchanged the role of the fermionic field with their hermitian conjugate compared to Eq.~\eqref{eq:fermion_current_general_SU23}. The reason is technical, arising from the comparison with the conventions adopted in \cite{Ishiki:2006rt}. The reader interested to this point is referred to Appendix B in \cite{Baiguera:2020jgy}.} as
\beq
j^m_{(\rho)} = g \, \mathcal{G}^{0,(m_1,0),\kappa_1=1}_{0,(m_2,0),\kappa_2=1;0,(m,0),\rho} 
\lbrace \psi_{m_1}, \psi^{\dagger}_{m_2}  \rbrace \, ,
\eeq
with the conventions $\psi_{m=1/2}=\psi_+ $ and $\psi_{m=-1/2}=\psi_-,$ and where we denoted $M=(m, \tilde{m}).$
There are four possible combinations of fermionic modes, because both the fermions $\psi_+, \psi_-$ are originally coming from the fermion field $\psi^A$ with $A=1$ in the $\mathcal{N}=4$ SYM action.
For this reason, the fermionic current is 
\beq
j_{(\rho)}^m = i g \, \delta_{\rho,1} \left( \delta_{m,0} \lbrace \psi_+^{\dagger} , \psi_+ \rbrace - \delta_{m,0}   \lbrace \psi_-^{\dagger} , \psi_- \rbrace 
-  \sqrt{2} \delta_{m,-1}  \lbrace \psi_-^{\dagger} , \psi_+ \rbrace + \sqrt{2} \delta_{m,1}  \lbrace \psi_+^{\dagger} , \psi_- \rbrace  \right) 
 \, .
\eeq
Using the following Jacobi identity coming from the application of the cyclicity properties of the trace
\beq\label{eq:Jacobi-identity-fermion}
\tr \left( \lbrace \psi_{\alpha}, \psi_{\beta} \rbrace \lbrace \psi^{\dagger}_{\beta}, \psi^{\dagger}_{\alpha} \rbrace \right) =
 - \tr \left( \lbrace \psi_{\alpha}, \psi^{\dagger}_{\beta} \rbrace \lbrace \psi_{\beta}, \psi^{\dagger}_{\alpha} \rbrace \right) 
 - \tr \left( \lbrace \psi^{\dagger}_{\alpha}, \psi_{\alpha} \rbrace \lbrace \psi^{\dagger}_{\beta}, \psi_{\beta} \rbrace \right) \, ,
\eeq
together with Eq.~\eqref{eq:formula_terms_mediated_gauge_fields_SU23}, we find that the contribution to the effective Hamiltonian mediated by the non-dynamical gauge field reads
\beq
  \frac{3}{8N}  \tr \left( \lbrace \psi^{\dagger}_{\alpha}, \psi_{\alpha} \rbrace  \lbrace \psi^{\dagger}_{\beta}, \psi_{\beta} \rbrace  \right) 
+ \frac{1}{4N} \tr \left( \lbrace \psi^{\dagger}_{\beta}, \psi^{\dagger}_{\alpha} \rbrace  \lbrace \psi_{\alpha}, \psi_{\beta} \rbrace  \right) 
\, .
\label{eq:purely_fermionic_term_su23}
\eeq
This is the purely fermionic part of the interactions.

\vskip 4mm

\noindent
\textbf{Quartic scalar self-interaction.}

Since all the scalars are dynamical, we simply take the quartic scalar self-interaction from the $\mathcal{N}=4$ SYM Hamiltonian and we manipulate the trace structure to get
\beq
 \frac{1}{8 N} \tr \left( [\Phi^{\dagger}_a, \Phi_a] [\Phi^{\dagger}_b, \Phi_b] \right)
+ \frac{1}{4 N} \tr \left(  [\Phi^{\dagger}_b, \Phi_a]
[\Phi_a, \Phi_b] \right) \, .
 \label{eq:purely_bosonic_term_su23}
\eeq
where we have rescaled the scalars according to the normalization introduced in Eq.~\eqref{eq:rescaling_scalars_SU_2_3}.
We notice the appearance of double and single trace terms, respectively.

\vskip 4mm

\noindent
\textbf{Yukawa term.}
According to the analysis of the physical degrees of freedom in this sector, we have at disposal three scalars $\Phi_a$ and two fermions $\psi_{\alpha}$ to build terms at order $g^2$ starting from the cubic Yukawa term.
However, the antisymmetry of the fermionic fields is responsible for the vanishing of all the terms mediated by non-dynamical scalar fields.
Instead it is allowed to have effective interactions mediated by non-auxiliary modes of the fermions. Such terms can be written as
\beq
 - \sqrt{2} i g \sum_{J_i m_i \tilde{m}_i} \sum_{J M \kappa} 
(-1)^{-m+\tilde{m}+\frac{\kappa}{2}} \mathcal{F}^{J,-M,\kappa}_{J_1, M_1, \kappa_1; J_2, M_2}
\tr \left( \psi^4_{J M \kappa} [(\Phi_a)^{J_2 M_2}, (\psi_1)^{J_1 M_1} ]  \right) 
+ \mathrm{h.c.}
\eeq
A direct evaluation of the Clebsch-Gordan coefficient $\mathcal{F}$ defined in Eq.~\eqref{eq:app_definitionClebsch_F} shows that there are some assignments of momenta such that the Yukawa term is non-vanishing.
The non-dynamical modes are integrated out by using the equations of motion
\beq
(\psi_4)_{J m \tilde{m} \kappa}  = \frac{i g}{2}  \delta_{J,0} 
\delta_{\tilde{m}, 0} \delta_{\kappa,1} \,
 \left( \delta_{m,- \frac{1}{2}} \, [\Phi_a^{\dagger}, \psi_+^{\dagger}]-  \delta_{m,\frac{1}{2}} \, [\Phi_a^{\dagger}, \psi_-^{\dagger}] \right)  \, ,
\eeq
and the similar one obtained by hermitian conjugation.
After plugging in the Hamiltonian, we get
\beq
- \frac{1}{2N} \tr \left( [\Phi_a^{\dagger}, \psi_{\alpha}] [\psi_{\alpha}^{\dagger}, \Phi_a] \right)  \, .
\eeq
Using the following Jacobi identity coming from the cyclicity properties of the trace
\beq
 \tr \left( [\Phi^{\dagger}_a, \Phi_a] \lbrace \psi^{\dagger}_{\beta}, \psi_{\beta} \rbrace \right)
-  \tr \left( [\psi^{\dagger}_{\beta}, \Phi_a] [\Phi^{\dagger}_a, \psi_{\beta}] \right) =
 \tr \left(  [\psi_{\beta}, \Phi_a] [\Phi^{\dagger}_a , \psi^{\dagger}_{\beta}]  \right) \, ,
\eeq
we obtain 
\beq
 \frac{1}{2N} \tr \left(  [\Phi^{\dagger}_a , \psi^{\dagger}_{\beta}] [\psi_{\beta}, \Phi_a] \right) 
 + \frac{1}{2N} \tr \left( [\Phi_a^{\dagger}, \Phi_{a}] \lbrace \psi^{\dagger}_{\beta}, \psi_{\beta} \rbrace \right)   \, .
\label{eq:mixed_terms_sphere_red_su23}
\eeq

\vskip 4mm

\noindent
\textbf{Total interacting Hamiltonian.}
We get the interacting Hamiltonian by summing the terms in Eqs.~\eqref{eq:purely_fermionic_term_su23}, \eqref{eq:purely_bosonic_term_su23} and \eqref{eq:mixed_terms_sphere_red_su23}.
Some terms simplify by means of the Gauss' law
\beq
[\Phi^{\dagger}_a, \Phi_a] + \lbrace \psi^{\dagger}_{\alpha}, \psi_{\alpha} \rbrace   = 0  \, ,
\label{eq:Gauss_law_su23}
\eeq
and we find
\beq
H_{\rm int}  =  \frac{1}{4 N} \tr \left( [\Phi^{\dagger}_b , \Phi^{\dagger}_a ] [\Phi_a, \Phi_b] \right)
+ \frac{1}{4N} \tr \left( \lbrace \psi^{\dagger}_{\beta}, \psi^{\dagger}_{\alpha} \rbrace  \lbrace \psi_{\alpha}, \psi_{\beta} \rbrace  \right)
+ \frac{1}{2N} \tr \left(  [\Phi^{\dagger}_a , \psi^{\dagger}_{\beta}] [\psi_{\beta}, \Phi_a] \right)  \, .
\label{eq:sphere_reduction_su23}
\eeq
We notice that the Gauss' law is responsible for making the D-terms vanish, while only the F-term interactions survive.
Written in this form, the Hamiltonian is manifestly positive definite.
The reason is that it is composed by a sum of F-terms containing the product of a quadratic (anti)commutator times its hermitian conjugate.
One may wonder if this structure gives only apparently a positive-definite expression, since for a single fermionic field $\psi$ we can find the relation
\beq
\tr \left( \lbrace  \psi^{\dagger}, \psi^{\dagger} \rbrace
\lbrace  \psi, \psi \rbrace \right) = 
- 2 \tr \left( \lbrace  \psi^{\dagger}, \psi \rbrace
\lbrace  \psi^{\dagger}, \psi \rbrace  \right) \, ,
\eeq
which seems to contradict the previous statement because of the minus sign in front of the D-term structure.

However, it is simple to show in a specific example that the interaction is positive definite.
Consider the case of the colour group with $N=2$ and parametrize the fermionic field as
\beq
\psi = \begin{pmatrix}
	0 & \zeta_1  \\
	\zeta_2 & 0 
\end{pmatrix} \, , \qquad
\psi^{\dagger} = 
\begin{pmatrix}
	0 & \zeta_2^{\dagger} \\
	\zeta_1^{\dagger} & 0 
\end{pmatrix} \, ,
\eeq
where the components are Grassmann-valued fields.
Since the fermion is carrying matrix indices, it is clear that the anticommutator with itself is in general non-vanishing, and we obtain
\beq
\tr \left( \lbrace  \psi^{\dagger}, \psi^{\dagger} \rbrace
\lbrace  \psi, \psi \rbrace \right) = 
8 (\zeta_1 \zeta_2)^{\dagger} (\zeta_1 \zeta_2) \geq 0 \, .
\eeq
Alternatively, one can compute 
\beq
\lbrace \psi^{\dagger}, \psi \rbrace = 
\begin{pmatrix}
	\zeta_2^{\dagger} \zeta_2 - \zeta_1^{\dagger} \zeta_1 & 0 \\
	0 &  -\zeta_2^{\dagger} \zeta_2 + \zeta_1^{\dagger} \zeta_1 
\end{pmatrix} \, ,
\eeq
which yields
\beq
- 2 \tr \left( \lbrace  \psi^{\dagger}, \psi \rbrace
\lbrace  \psi^{\dagger}, \psi \rbrace  \right)  = 
8 (\zeta_1 \zeta_2)^{\dagger} (\zeta_1 \zeta_2) \geq 0 \, .
\eeq
The results are consistent, and show the positivity of the fermionic F-term in a simple case.
This statement holds for any rank of the colour group.

\subsection{PSU(1,1|2) sector}
\label{sect-sphere_reduction_PSU112}

The sphere reduction technique was applied in the near-BPS limit giving rise to the $\mathrm{PSU}(1,1|2)$ sector in Section 2.5 of \cite{Baiguera:2020jgy}, where we derived almost entirely the effective Hamiltonian.
However, we point out few missing terms which slightly modify the fermionic part of the interactions.
In the following, we summarize the main steps of the computation and we point out the additional terms, referring the reader to \cite{Baiguera:2020jgy} for other technical details.

The BPS bound \eqref{eq:near_BPS_limits_1over8BPS_subsectors} with $\alpha=1$ reads $H \geq S_1 + Q_1 + Q_2 . $
The effective theory, emerging when zooming in towards this limit, has invariance under the $\mathrm{PSU}(1,1|2) \times \mathrm{U}(1)$ group.  
The physical degrees of freedom of the system are encoded by two scalars and two fermions with opposite chirality, transforming as a doublet under two different $\SU(2)$ subgroups: the first one is a remnant of the original R-symmetry of $\mathcal{N}=4$ SYM theory, while the second one is an enhanced symmetry of this subsector.

\subsubsection*{Quadratic Hamiltonian and physical degrees of freedom}

We impose at the quadratic level $H_0 - S_1 -Q_1 - Q_2 = 0 $ to determine the surviving degrees of freedom of the effective theory. We summarize the results:
\begin{itemize}
\item There are two scalar fields satisfying the non-relativistic constraint 
\beq
\Pi_a^{J,-J,J} + i \omega_J \Phi^{\dagger \, J, -J, J}_a = \mathcal{O} (g) \, \qquad (a=1,2) \, .
\label{eq:constraints_scalar_field}
\eeq
\item There are two unconstrained fermionic fields characterized by the  momenta and chiralities
\bea
\label{eq:dynamical_ferm_su112_1}
& A=1 \, , \qquad  \kappa=1 \, , \qquad m=-J-\frac{1}{2} \, , \qquad  \tilde{m} = J \, , & \\
& A=2 \, , \qquad  \kappa=-1 \, , \qquad m=-J \, , \qquad  \tilde{m} = J + \frac{1}{2} \, . &
\label{eq:dynamical_ferm_su112_2}
\eea
\item All the other bosonic and fermionic modes (including the gauge fields) decouple in the limit under consideration.
\end{itemize}
We canonically normalize the Dirac brackets of the fields and we introduce a short-hand notation as follows:
\beq
 \begin{aligned}
& \Phi_a^{2J} \equiv \Big( \sqrt{2\omega_J}\Phi_1^{J,-J,J} , \sqrt{2\omega_J}\Phi_2^{J,-J,J} \Big) \, ,  & \\
& \psi^a_{2J} \equiv  \Big( \psi^{A=1}_{J,-J-\tfrac{1}{2},J,\kappa = 1} , \psi^{A=2}_{J,-J,J+\tfrac{1}{2},\kappa = -1} \Big)  \, . & 
\label{eq:dynamical_modes_su_1_1_2} 
\end{aligned} 
\eeq
The free Hamiltonian of the system is given by
\beq
H_0 = \tr \sum_{n=0}^{\infty} \sum_{a=1,2} \left[\left( n+ 1 \right) |\Phi^a_n|^2 + \left( n+\frac{3}{2} \right) |\psi^a_n|^2 \right] \, .
\eeq
Comparing with $H_0$ in the $\SU(2|3)$ sector, we notice again that the scaling dimensions of the fields are explicitly entering the coefficients of the kinetic term.
However, the presence of a non-trivial rotation generator $S_1$ defining the BPS bound gives rise to an additional integer label creating an infinite tower of modes.

\subsubsection*{Interacting Hamiltonian}

The interacting Hamiltonian is defined via Eq.~\eqref{eq:quadratic_gauge_Hamiltonian_su12sectors}  with $\alpha=1,$ \emph{i.e.}
\beq
H_{\rm int} = \lim_{g \rightarrow 0} \frac{H- S_1 - Q_1 - Q_2}{g^2 N} \, .
\eeq
We split again the analysis of the Hamiltonian as: terms mediated by gauge fields, quartic scalar self-interaction, Yukawa term and we take as a starting point the expression collected in Eq.~\eqref{eq:app_full_interacting_N=4SYM_Hamiltonian}.  

\vskip 4mm

\noindent
\textbf{Terms mediated by non-dynamical gauge fields.}
The currents are singlet under the $\SU(2)$ subgroup and are given by
\beq
\label{eq:current_j0_su112}
\begin{aligned}
\hspace{-2mm} j_0^{\dagger\,Jm\tilde{m}}  = & g \,  \frac{J_1+J_2+1}{\sqrt{\omega_{J_1} \omega_{J_2}}} {\cal C}^{\CJ_2}_{\CJ_1,JM} [\Phi^a_{2J_1},(\Phi_a)^\dagger_{2J_2}] \\
& + g \mathcal{F}^{\BCJ_1}_{\BCJ_2, JM}  
\lbrace \psi^1_{2J_1} , (\psi_1)^{\dagger}_{2J_2} \rbrace 
 + g \mathcal{F}^{\BCJ_2}_{\BCJ_1, JM}  
\lbrace \psi^2_{2J_1} , (\psi_2)^{\dagger}_{2J_2} \rbrace  \, , 
\end{aligned}
\eeq
\beq
\label{eq:current_jrho_su112}
\begin{aligned}
\hspace{-2mm} j_{(\rho)}^{\dagger\,Jm\tilde{m}}  = &- 4g \sqrt{\frac{J_1(J_1+1)}{\omega_{J_1} \omega_{J_2}}} {\cal D}^{\CJ_2}_{\CJ_1, JM\rho}[\Phi^a_{2J_1},(\Phi_a)^\dagger_{2J_2}] \\
&  + g \mathcal{G}^{\BCJ_1}_{\BCJ_2, JM \rho}
\lbrace \psi^1_{2J_1} , (\psi_1)^{\dagger}_{2J_2} \rbrace 
- g \mathcal{G}^{\BCJ_2}_{\BCJ_1, JM, -\rho}
\lbrace \psi^2_{2J_1} , (\psi_2)^{\dagger}_{2J_2} \rbrace  \, ,
\end{aligned}
\eeq
where the Clebsch-Gordan coefficients are defined in Eqs.~\eqref{eq:app_definitionClebsch_C}, \eqref{eq:app_definitionClebsch_D}, \eqref{eq:app_definitionClebsch_F}, \eqref{eq:app_definitionClebsch_G} and we are using the short-hand notation
\beq
\CJ = (J,-J,J) \,,  \qquad
\BCJ = ( J,J+\frac{1}{2},-J,\kappa=1) \, .
\label{eq:short-hand_notation_momenta}
\eeq
Plugging these expressions inside Eq.~\eqref{eq:general_gauge_mediated_interaction_su12limits} and supplementing the scalar part with its double trace structure, we obtain the full contribution to the Hamiltonian mediated by non-dynamical modes of the gauge fields.
It reads
\beq
\frac{1}{2N}  \sum_{l=1}^{\infty} \frac{1}{l} \tr \left( \mathbf{Q}_l^{\dagger} \mathbf{Q}_l \right) \, , 
\label{eq:sphere_reduction_su112_term_mediated_gluon}
\eeq
where the charge densities are defined as $\mathbf{Q}_l = q_l + \tilde{q}_l$ with
\beq
 q_l \equiv \sum_{n=0}^{\infty} \sum_{a=1,2} [(\Phi_a^{\dagger})_n , (\Phi_a)^{n+l}] \, , \qquad
\tilde{q}_l \equiv \sum_{n=0}^{\infty} \sum_{a=1,2} \frac{\sqrt{n+1}}{\sqrt{n+l+1}}  \lbrace (\psi_a^{\dagger})_n , (\psi_a)^{n+l} \rbrace  \, .
\label{eq:charge_densities_psu112}
\eeq
In order to derive the result \eqref{eq:sphere_reduction_su112_term_mediated_gluon}, we use the Gauss' law $\mathbf{Q}_0 = 0 .$

The previous computation is explained in details in Section 2.5 of \cite{Baiguera:2020jgy}.
Here we point out that a careful analysis of the summation over Clebsch-Gordan coefficients reveals that there is an additional term in the purely fermionic interaction which was not spotted before, and which is given by
\beq
\begin{aligned}
& \frac{1}{2N} \sum_{l=1} \sum_{m,n=0}^{\infty} 
\sqrt{\frac{(m+1)(n+1)}{(m+l+1)(n+l+1)}}  \frac{1}{m+n+l+2} 
\tr \left( \lbrace (\psi^{\dagger}_1)_{m} , (\psi_1)_{m+l}  \rbrace  \lbrace (\psi^{\dagger}_2)_{n+l} , (\psi_2)_{n}  \rbrace \right)  \\
& + \frac{1}{2N} \sum_{l,m,n=0}^{\infty} 
\sqrt{\frac{(m+1)(n+1)}{(m+l+1)(n+l+1)}} \frac{1}{m+n+l+2} 
\tr \left( \lbrace (\psi^{\dagger}_2)_{m} , (\psi_2)_{m+l}  \rbrace  \lbrace (\psi^{\dagger}_1)_{n+l} , (\psi_1)_{n}  \rbrace \right) \, .
\end{aligned}
\label{eq:coefficients_mixed_fermionic_blocks_sphere_reduction}
\eeq
This part will combine with the Yukawa term to give rise to a different fermionic structure. 
It is important to observe that the summation range of $l$ starts from 1 for the terms in the first line and from 0 for the terms in the second line: this mismatch introduces a boundary term in the Hamiltonian which is also crucial to guarantee that the Hamiltonian is invariant under the symmetries of this sector.

\vskip 4mm

\noindent
\textbf{Quartic scalar self-interaction.}
The computation was performed in Section 2.5 of \cite{Baiguera:2020jgy}.
Here we only report the result without the details of the intermediate steps of the computation:
\beq
\begin{aligned}
 & \frac{1}{2N} \sum_{l,m,n=0}^{\infty} \frac{1}{m+n+l+1}\tr([\Phi_a^{m+l},\Phi_b^{n}][(\Phi_b^\dagger)^{n+l},(\Phi_a^\dagger)^{m}]) \\
&  - \frac{1}{4N} \sum_{m,n=0}^{\infty} \frac{1}{m+n+1}\tr([\Phi_a^{m},\Phi_b^{n}][(\Phi_b^\dagger)^{n},(\Phi_a^\dagger)^{m}])
  \, .
  \end{aligned}
 \label{eq:purely_bosonic_term_su_1_1_2}
\eeq 
It is worth noticing that the scalar interactions combine into an F-term structure, which is manifestly invariant under the $\SU(2)$ subgroup.
Furthermore, there is a term in the second line which counts correctly the boundary term $l=0.$ This plays a crucial role to determine the invariance under the $\SU(1,1)$ subgroup and under supersymmetry.

\vskip 4mm

\noindent
\textbf{Yukawa term.}
In the $\mathrm{PSU}(1,1|2)$ sector the Yukawa term shows a new feature compared to the $\SU(2|3)$ case: both non-dynamical  fermionic and bosonic modes can mediate an interaction.
The terms mediated by the fermions have been investigated extensively in \cite{Baiguera:2020jgy}, giving rise to the interactions
\beq
\begin{aligned}
 &   \frac{1}{2N} \sum_{m, n, l = 0}^{\infty} \frac{\tr \left( [(\Phi^\dagger_a)_{m},(\psi_b)_{m+l}] [(\psi_b)^{\dagger}_{n+l},(\Phi_a)_{n}] \right) }{\sqrt{(m+l+1)(n+l+1)}} \\
& -\frac{1}{2N} \sum_{m,n, l = 0}^{\infty} \sqrt{\frac{m +1}{n + l+1}} 
\frac{ \epsilon_{ac}\epsilon_{bd}\tr \left( [(\psi_a^{\dagger})_{m} , (\Phi_b)_{m+ l + 1}] [(\psi_c^{\dagger})_{n+ l} , (\Phi_d)_{n} ] \right) }{m+n+ l +2}  \\
& -\frac{1}{2N} \sum_{m,n,l=0}^{\infty} \sqrt{\frac{m+1}{n+ l+1}}  
\frac{ \epsilon_{ac}\epsilon_{bd} \tr \left( [(\Phi_a^{\dagger})_{m+ l + 1} , (\psi_b)_{m}] [(\Phi_c^{\dagger})_{n} , (\psi_d)_{n+l} ] \right) }{m+n+ l+2} \, .
\end{aligned} 
\label{eq:final_mixed_interaction_su_1_1_2}
\eeq
However a careful treatment of the terms mediated by the non-dynamical scalar shows that they contribute to the following effective Hamiltonian
\beq
\begin{aligned}
 & - \frac{1}{2N} \sum_{l=1} \sum_{m, n= 0}^{\infty}  \sqrt{\frac{(m +1)(n +1)}{(m  + l +1)(n  +l +1)}}
\frac{\tr \left( \lbrace (\psi_1)_{m+ l}, (\psd_2)_{m}   \rbrace 
		\lbrace (\psd_1)_{n + l} , (\psi_2)_{n}  \rbrace  \right)}{m + n + l +2} \\
		& - \frac{1}{2N} \sum_{l=0} \sum_{m, n=0}^{\infty}  \sqrt{\frac{(m +1)(n +1)}{(m  + l +1)(n  +l +1)}}
\frac{\tr \left( \lbrace (\psi_2)_{m+ l}, (\psd_1)_{m}   \rbrace 
		\lbrace (\psd_2)_{n + l} , (\psi_1)_{n}  \rbrace  \right)}{m + n + l +2}  \, .
		\label{eq:4point_fermionic_term_sphere_reduction_1}
		\end{aligned}
\eeq 
It is important to observe that the missing terms to build an $\SU(2)-$invariant structure in Eq.~\eqref{eq:4point_fermionic_term_sphere_reduction_1} are precisely contained inside Eq.~\eqref{eq:coefficients_mixed_fermionic_blocks_sphere_reduction}.
Moreover, we observe again that the first line contains a summation over $l$ starting from 1, while the second line starts from 0.

\vskip 4mm

\noindent
\textbf{Total interacting Hamiltonian.}
Summing all the expressions derived from sphere reduction, we obtain the effective Hamiltonian describing the $\mathrm{PSU}(1,1|2)$ near-BPS limit:
\beq
\begin{aligned}
H_{\rm int} & =  \frac{1}{2N} \sum_{l=1}^{\infty}  \frac{1}{l} \tr \left( \mathbf{Q}_l^{\dagger} \, \mathbf{Q}_l  \right)
	+ \frac{1}{2N} \sum_{l = 0}^{\infty}  \tr ( (F_{ab})_l^\dagger (F_{ab})_l ) \\
& +	\frac{1}{2N} \sum_{l,m,n=0}^{\infty} \frac{1}{m+n+l+1}\tr([\Phi_a^{m+l},\Phi_b^{n}][(\Phi_b^\dagger)^{n+l},(\Phi_a^\dagger)^{m}]) \\
& +  \frac{1}{2N} \sum_{l=0}^{\infty} \sum_{m,n=0}^{\infty} \frac{\sqrt{(m+1)(n+1)}}{\sqrt{(m+l+1)(n+l+1)}} 
\frac{ \tr \left(  \epsilon^{ac}  \epsilon^{bd} \lbrace (\psi_a^{\dagger})_m , (\psi_b)_{m+l} \rbrace \lbrace (\psi_c^{\dagger})_{n+l} , (\psi_d)_n \rbrace \right)  }{m+n+l+2}
 \\
& -\frac{1}{2N} \sum_{m,n,l= 0}^{\infty} \sqrt{\frac{m +1}{n + l+1}} 
\frac{ \epsilon_{ac}\epsilon_{bd}\tr \left( [(\psi_a^{\dagger})_{m} , (\Phi_b)_{m+ l + 1}] [(\psi_c^{\dagger})_{n+ l} , (\Phi_d)_{n} ] \right) }{m+n+ l +2}  \\
& -\frac{1}{2N} \sum_{m,n,l=0}^{\infty} \sqrt{\frac{m+1}{n+ l+1}}  
\frac{ \epsilon_{ac}\epsilon_{bd} \tr \left( [(\Phi_a^{\dagger})_{m+ l + 1} , (\psi_b)_{m}] [(\Phi_c^{\dagger})_{n} , (\psi_d)_{n+l} ] \right) }{m+n+ l+2} + H_B \, ,
\end{aligned} 
\label{eq:final_result_sphere_reduction}
\eeq
where we defined the blocks
\beq
\begin{aligned}
& \mathbf{Q}_l \equiv  \sum_{n=0}^{\infty} \sum_{a=1,2} \left(  [(\Phi_a^{\dagger})_n , (\Phi_a)^{n+l}] +  \frac{\sqrt{n+1}}{\sqrt{n+l+1}}  \lbrace (\psi_a^{\dagger})_n , (\psi_a)^{n+l} \rbrace \right) \, , & \\ 
& (F_{ab})_l \equiv \sum_{m=0}^\infty \frac{ [(\psi_a)_{m+l},(\Phi_b)_m^\dagger] }{\sqrt{m+l+1}} \, . &
\end{aligned}
\label{eq:Q_F_blocks_psu112}
\eeq 
The Hamiltonian contains a boundary term which reads
\beq
\begin{aligned}
 H_B = &  - \frac{1}{4N} \sum_{m,n=0}^{\infty} \frac{1}{m+n+1}\tr([\Phi_a^{m},\Phi_b^{n}][(\Phi_b^\dagger)^{n},(\Phi_a^\dagger)^{m}]) \\
& - \frac{1}{4N}  \sum_{m,n=0}^{\infty} \frac{1}{m+n+2}  \tr \left( \epsilon^{ac} \epsilon^{bd} \lbrace (\psi_a^{\dagger})_m , (\psi_b)_{m} \rbrace \lbrace (\psi_c^{\dagger})_{n} , (\psi_d)_n \rbrace \right)  \, .
\label{eq:boundary_term_sphere_reduction}
\end{aligned}
\eeq
Contrarily to the $\SU(2|3)$ case considered in Eq.~\eqref{eq:sphere_reduction_su23}, this Hamiltonian is not manifestly positive definite; in particular the single trace fermionic interactions do not combine into a pure F-term, but they have a more involved structure.
We will present in Section \ref{sect-positivity} a method based on the supersymmetry invariance of the sector to show that the Hamiltonian is indeed positive definite.
Then we will explore in Section \ref{sec:symmetry} the algebraic structure of the interactions, in view of the understanding of positivity from a block structure.

%%%%%%%%%%%%

\subsection{Quantization}
\label{sect-quantization_sphere_red}

\begin{figure}[ht]
\centering
\includegraphics[trim=4cm 24cm 4cm 2cm,width=0.7\linewidth]{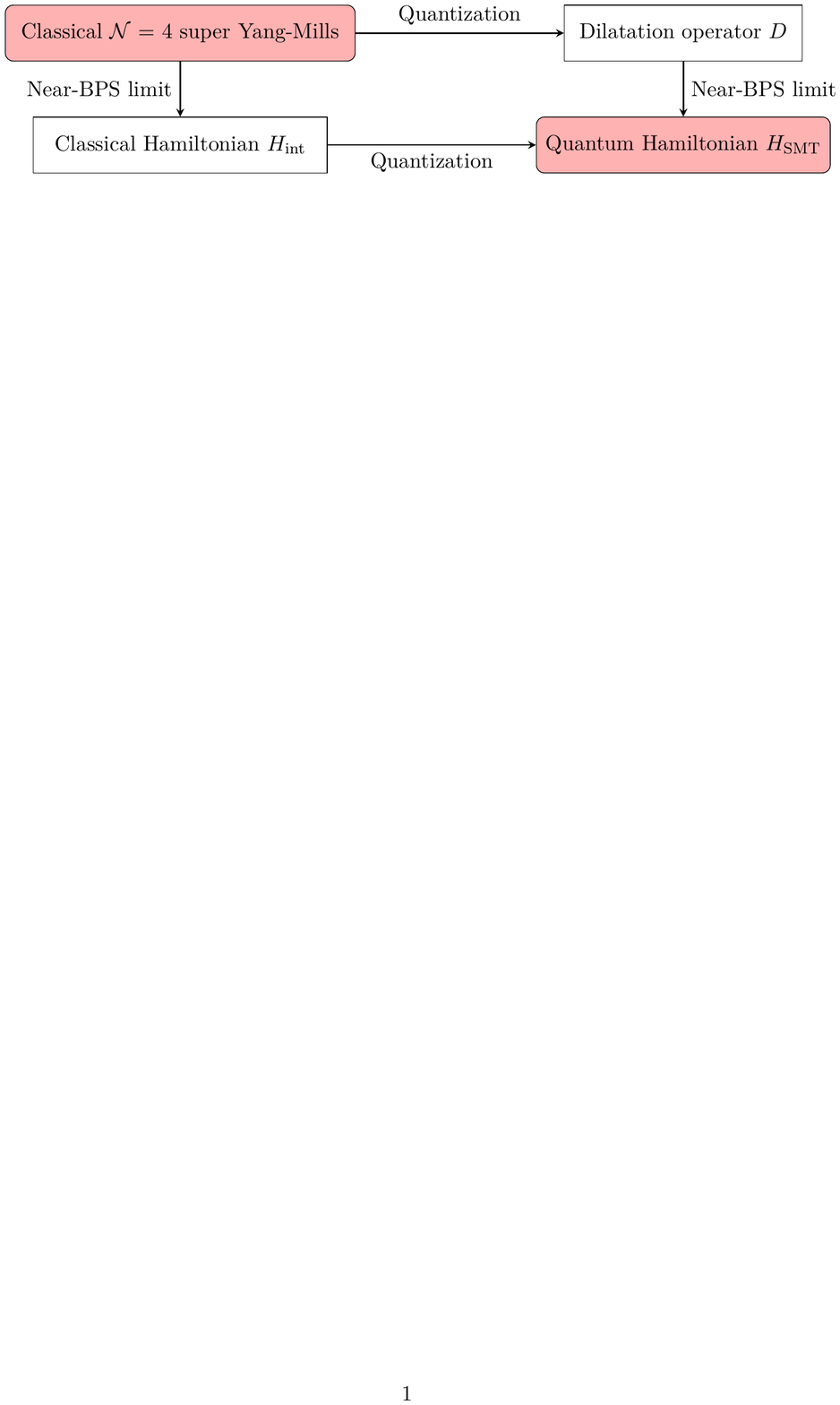}
\caption{\small Commutative diagram describing the two procedures to get a SMT Hamiltonian in a given near-BPS limit.
One can either compute one-loop corrections to the dilatation operator and then restrict to a certain limit (right-down), or perform the sphere reduction and give a recipe to quantize the theory (down-right).}
\label{fig:commutative_diagram}
\end{figure}

In Section \ref{sect-comparison_spin_chain} we started from the one-loop corrections to the dilatation operator and we employed a dictionary to derive a SMT Hamiltonian $H_{\rm SMT}$ describing the interactions of the system in the vicinity of a BPS bound.
In Section \ref{sect-effective_ham_from_sphere} we obtained another interacting Hamiltonian by performing the sphere reduction of $\mathcal{N}=4$ SYM theory in the near-BPS limit.
This gives a classical Hamiltonian $H_{\rm int}$ which we need to quantize.
We will show that there is a simple prescription such that the results obtained by the two methods are equivalent, \emph{i.e.} the diagram depicted in Fig.~\ref{fig:commutative_diagram} is commutative.

The common procedure to quantize a theory is to substitute the Dirac brackets with graded commutators
\beq
\lbrace \cdot, \cdot \rbrace_D \rightarrow i [ \cdot , \cdot \rbrace  \, . 
\eeq
Then we require that the fields are promoted to ladder operators satisying the commutation relations 
\beq
[ (a_r)^i_{\,\, j} , (a^{\dagger}_s)^{k}_{\,\, l}  ] = \delta^i_{\,\, l} \delta^k_{\,\, j} \delta_{rs} \, , \qquad
\lbrace (b_r)^i_{\,\, j} , (b^{\dagger}_s)^{k}_{\,\, l}  \rbrace = \delta^i_{\,\, l} \delta^k_{\,\, j} \delta_{rs} \, , 
\label{eq:commutation_relations_ladder_operators_SMT}
\eeq
where $ a_s \equiv \Phi_s , \, a^{\dagger}_s \equiv \Phi^{\dagger}_s $ are bosonic, and $ b_s \equiv \psi_s , \, b^{\dagger}_s \equiv \psi^{\dagger}_s  $ are fermionic.
This holds for all the copies of scalars or spinors that survive the near-BPS limit.
The set of these creation and annihilation operators corresponds precisely to the building blocks defining a SMT as described in Section \ref{sect-SMT-from_spin_chain}.
Notice that the commutation relations have a canonical normalization; this is naturally achieved for the fields obtained from sphere reduction thanks to the rescalings \eqref{eq:rescaling_scalars_SU_2_3} and \eqref{eq:dynamical_modes_su_1_1_2}  introduced for the scalars.

\begin{remark}
We directly promote the classical Hamiltonian obtained from sphere reduction at quantum level, without changes of ordering.
The Gauss law gets promoted to the $\SU(N)$ singlet condition.
\end{remark}

We proceed to show that this prescription gives consistent results with the spin chain Hamiltonian.
This is immediately true for the $\SU(2|3)$ sector, since the Hamiltonians \eqref{eq:spin_chain_Ham_su23} and \eqref{eq:sphere_reduction_su23} coincide once we choose the normalization to be $c^2=2.$
It is less trivial to show the equivalence in the $\mathrm{PSU}(1,1|2)$ case.
Manipulations involving the normal ordering have been performed in Section \ref{sect-spin_chain_PSU112_sector}, leading to the equivalence of most terms in the Hamiltonians \eqref{eq:total_spin_chain_Ham_psu112} and \eqref{eq:final_result_sphere_reduction}.

In order to complete the comparison, we need to show that the F-term and D-term structures for the single trace quartic scalar interaction are the same.
We start by defining the blocks
\begin{equation}
	J_L = \sum_{n=0}^L [\Phi^1_{L-n},\Phi_n^2], \qquad J_L^\dagger = \sum_{m=0}^L [(\Phi^{\dagger}_2)_{L-m},(\Phi^{\dagger}_1)_m] \, .
\end{equation}
It is straighforward to verify that the quartic scalar interaction in the Hamiltonian \eqref{eq:final_result_sphere_reduction} can be recast in the following form by defining $L=m+n+l$ and working on the summation ranges:
\begin{equation}\label{eq:Hscalar}
 \sum_{L=0}^\infty \frac{1}{L+1} \text{tr} (J_L^\dagger J_L ) \, .
\end{equation}
It is instead non-trivial to show that the quartic scalar term in Eq.~\eqref{eq:total_spin_chain_Ham_psu112} can be rewritten in this form.
By splitting the sum over the $\SU(2)$ indices, we notice that the SMT Hamiltonian derived from the action on spin chains is given by
\beq
\begin{aligned}
&	H_{\rm SMT}^{\rm bos}  =	\sum_{m,n=0}^\infty   \text{tr}\Big(- \sum_{l=0}^{\infty} [(\Phi_1^{\dagger})_m,(\Phi_1)_{m+l}] [(\Phi_2^{\dagger})_{n+l},(\Phi_2)_{n}]  
	+ \sum_{l=0}^{\infty} [(\Phi_1^{\dagger})_{m}, (\Phi_2)_{m+l}]  [(\Phi_2^{\dagger})_{n+l},(\Phi_1)_{n}]  \\
& 	+ \sum_{l=1}^{\infty} [(\Phi_2^{\dagger})_{m},(\Phi_1)_{m+l}] [(\Phi_1^{\dagger})_{n+l},(\Phi_2)_{n}]   
 - \sum_{l=1}^{\infty} [(\Phi_2^{\dagger})_{m},(\Phi_2)_{m+l}] [(\Phi_1^{\dagger})_{n+l}, (\Phi_1)_{n}] \Big)  \times \frac{1}{m+n+l+1}  \, ,
\end{aligned}
\label{eq:spinchainscalar-2}
\eeq
where we included the boundary terms in Eq.~\eqref{eq:boundary_Hamiltonian_spin_chain} by making half of the sums to start from $l=1.$
We call the four terms (following the order) composing the expression as $H_1, \dots , H_4$ such that
\beq
H_{\rm SMT}^{\rm bos} = -H_1 + H_2 + H_3- H_4 \, .
\label{eq:bosonic_quartic_term_SMT_single_trace}
\eeq
After defining $L=m+n+l$ and manipulating the sums, one can show that they can be rewritten as
\begin{align} \label{eq:L-notation-purescalar}
	\begin{split}
H_1 &= \sum_{m,n=0}^L \sum_{m+n\le L} \frac{1}{L+1} \text{tr} \, [(\Phi_1^{\dagger})_m,(\Phi^1)_{L-n}] [(\Phi_2^{\dagger})_{L-m},(\Phi_2)_n] \, , \\
H_2 &= \sum_{m,n=0}^L \sum_{m+n\le L} \frac{1}{L+1} \text{tr} \Big( 
[(\Phi_1)_n,(\Phi_2)_{L-n}] [(\Phi_2^{\dagger})_{L-m},(\Phi_1^{\dagger})_m] + 
[(\Phi_1^{\dagger})_m,(\Phi_1)_n] [(\Phi_2^{\dagger})_{L-m},(\Phi_2)_{L-n}]
\Big) \, , \\
H_3 &= \sum_{m,n=0}^L \sum_{m+n< L} \frac{1}{L+1} \text{tr} \Big( 
[(\Phi_2)_n,(\Phi_1)_{L-n}] [(\Phi_1^{\dagger})_{L-m},(\Phi_2^{\dagger})_m] + 
[(\Phi_2^{\dagger})_m,(\Phi_2)_n] [(\Phi_1^{\dagger})_{L-m},(\Phi_1)_{L-n}]
\Big) \, , \\
H_4 &=\sum_{m,n=0}^L \sum_{m+n< L} \frac{1}{L+1} \text{tr} \, [(\Phi_2^{\dagger})_m,(\Phi_2)_{L-n}] [(\Phi_1^{\dagger})_{L-m},(\Phi_1)_n] \, .
	\end{split}
\end{align}
The key simplification happens if we redefine the modes such that they all have the same labels.
This can be done with the following operations:
\begin{itemize}
\item Send $n\to L-n$ inside $H_1.$
\item Send $(m,n)\to (L-m,L-n)$ inside $H_3.$
\item Send $m\to L-m$ inside $H_4.$ 
\end{itemize}
Therefore we obtain
\begin{align} 
	\begin{split}
	H_1 &= \sum_{m,n=0}^L \sum_{m\le n} \frac{1}{L+1} \text{tr} \, [(\Phi_1^{\dagger})_m,(\Phi_1)_{n}] [(\Phi_2^{\dagger})_{L-m},(\Phi_2)_{L-n}] \, , \\
	H_2 &= \sum_{m,n=0}^L \sum_{m+n\le L} \frac{1}{L+1} \text{tr} \Big( 
		[(\Phi_1)_n,(\Phi_2)_{L-n}] [(\Phi_2^{\dagger})_{L-m},(\Phi_1^{\dagger})_m] + 
		[(\Phi_1^{\dagger})_m,(\Phi_1)_n] [(\Phi_2^{\dagger})_{L-m},(\Phi_2)_{L-n}]
		\Big) \, , \\
		H_3 &= \sum_{m,n=0}^L \sum_{m+n> L} \frac{1}{L+1} \text{tr} \Big( 
		[(\Phi_2)_{L-n},(\Phi_1)_{n}] [(\Phi_1^{\dagger})_{m},(\Phi_2^{\dagger})_{L-m}] + 
		[(\Phi_2^{\dagger})_{L-m}, (\Phi_2)_{L-n}] [(\Phi_1^{\dagger})_{m}, (\Phi_1)_{n}]
		\Big) \, , \\
		H_4 &=\sum_{m,n=0}^L \sum_{n< m} \frac{1}{L+1} \text{tr} \, [(\Phi_2^{\dagger})_{L-m}, (\Phi_2)_{L-n}] [(\Phi_1^{\dagger})_{m},(\Phi_1)_n] \, .
	\end{split}
\end{align}
Combining the terms as in Eq.~\eqref{eq:bosonic_quartic_term_SMT_single_trace} and using the cyclicity properties of the trace, we obtain precisely Eq~\eqref{eq:Hscalar}.
This concludes the comparison between the two procedures in the $\mathrm{PSU}(1,1|2)$ sector: the results agree.

%%%%%%%%%%%%%%%

\section{Positivity of the PSU(1,1|2) Hamiltonian}
\label{sect-positivity}

The SMT Hamiltonian represents an effective description of the degrees of freedom close to a BPS bound. It is then expected to be positive definite and to describe the reaction of a physical system to move away from the point in the space of parameters where the saturation occurs.
We have given a natural interpretation of this phenomenon in \cite{Baiguera:2020mgk} as a distance in the linear space of the representation identified by a set of fundamental blocks composing the Hamiltonian and classified by group theory.
Furthermore, we wrote the interactions of several sectors of $\mathcal{N}=4$ SYM in a manifest positive-definite form.

Since the effective theory in the $\mathrm{PSU}(1,1|2)$ near-BPS limit preserves some of the supercharges of the original $\mathcal{N}=4$ SYM, there is an alternative method to show that the interactions are positive definite, based on the supersymmetry algebra.
Indeed, it is possible to compute the one-loop corrections to the dilatation operator in the $\mathrm{PSU}(1,1|2)$ sector by using a representation of 
the fermionic generators composing the enhanced
$\mathrm{psu}(1|1)^2$ subalgebra \cite{Zwiebel:2005er,Beisert:2007sk,Zwiebel:2007cpa}.
There are in total four supercharges, denoted\footnote{In reference \cite{Zwiebel:2005er} the fermionic generators are denoted instead as $\overleftrightarrow{\mathfrak{T}}^{\pm},$ where the horizontal arrows correspond to the hypercharge. The precise dictionary is given by
\[
 \hat{\mathfrak{Q}}^{<} = \sqrt{2} \, \overrightarrow{\mathfrak{T}}^+ \, , \qquad
 \hat{\mathfrak{Q}}^{>} = \sqrt{2} \, \overrightarrow{\mathfrak{T}}^- \, , \qquad
  \hat{\mathfrak{S}}^{<} = \sqrt{2} \, \overleftarrow{\mathfrak{T}}^+ \, , \qquad
   \hat{\mathfrak{S}}^{>} = - \sqrt{2} \, \overleftarrow{\mathfrak{T}}^- \, .
\]} as $\hat{\mathfrak{Q}}^{(<,>)}$ and $\hat{\mathfrak{S}}^{(<,>)}$ in references \cite{Beisert:2007sk,Zwiebel:2007cpa}, where the symbols $(<,>)$ refer to their hypercharge under the emergent $\mathrm{SU}(2)$ automorphism symmetry.
In particular, it can be shown at the level of spin chains that the following relations hold:
\beq
\delta D = 2 \lbrace  \hat{\mathfrak{Q}}^{<}, \hat{\mathfrak{S}}^{>} \rbrace =
-  2 \lbrace  \hat{\mathfrak{Q}}^{<}, \hat{\mathfrak{S}}^{>} \rbrace \, , \qquad
\left( \hat{\mathfrak{Q}}^{>} \right)^{\dagger} = \hat{\mathfrak{S}}^{<}  \, , \qquad
\left( \hat{\mathfrak{Q}}^{<} \right)^{\dagger} = - \hat{\mathfrak{S}}^{>}  \, , 
\label{eq:relation_dilatation_operator_supercharges}
\eeq
where $\delta D$ is the one-loop correction to the dilatation operator.
These identities entail the positivity of the dilatation operator.

Our strategy is the following:
\begin{enumerate}
\item We define a couple of fermionic generators $\hat{\mathcal{Q}}, \hat{\mathcal{Q}}^{\dagger}$ cubic in the fields, and we show that they are the hermitian conjugate of each other.
They are the analog of the supercharges $\hat{\mathfrak{Q}}^{<}, \hat{\mathfrak{S}}^{>} $ introduced in Eq.~\eqref{eq:relation_dilatation_operator_supercharges}, written as generators in SMT language.
\item We show that these supercharges close into the interacting SMT Hamiltonian of the model, i.e.
\beq\label{eq:QQtoH}
\lbrace \hat{\mathcal{Q}}, \hat{\mathcal{Q}}^{\dagger} \rbrace_D = H_{\rm int} \, .
\eeq
\end{enumerate}
These conditions are sufficient to prove that the spectrum of the Hamiltonian is positive definite, as follows from standard manipulations in a supersymmetric-invariant theory \cite{Bilal:2001nv}.
In addition, this procedure is an alternative method to derive the effective Hamiltonian in the near-BPS limit, and provides a further check of the result \eqref{eq:final_result_sphere_reduction}.

\subsection{SU(1,1|1) subsector}

We start applying the technique in the $\SU(1,1|1)$ subsector, which has been shown with various techniques to have a positive definite spectrum \cite{Baiguera:2020jgy,Baiguera:2020mgk}.
It is possible to infer the action of the fermionic generators on the states of a spin chain from references \cite{Beisert:2007sk,Zwiebel:2007cpa}:
\beq
\begin{aligned}
& \hat{\mathcal{Q}} | Z_n \rangle = 
\sum_{k=0}^{n-1} \frac{1}{2\sqrt{k+1}} | \psi^1_{k} Z_{n-1-k} \rangle 
- \sum_{k=0}^{n-1} \frac{1}{2\sqrt{n-k}} | Z_k \psi^1_{n-1-k} \rangle \, , & \\
& \hat{\mathcal{Q}} | \psi^1_n \rangle = \frac{1}{2}
\sum_{k=0}^{n-1} \sqrt{\frac{n+1}{(k+1)(n-k)}} | \psi^1_{k} \psi^1_{n-1-k} \rangle  \, , & \\
&  \hat{\mathcal{Q}}^{\dagger}  | Z_{m} \psi^1_{n} \rangle = - \frac{1}{2\sqrt{n+1}} | Z_{m+n+1} \rangle \, ,  \qquad
 \hat{\mathcal{Q}}^{\dagger}  | \psi^1_{m} Z_{n} \rangle = 
 \frac{1}{2 \sqrt{m+1}} | Z_{m+n+1} \rangle \, , &  \\
& \hat{\mathcal{Q}}^{\dagger}  | \psi^1_{m} \psi^1_{n} \rangle = \frac{1}{2}
 \sqrt{\frac{m+n+2}{(m+1)(n+1)}} | \psi^1_{m+n+1} \rangle \, , 
\qquad
 \hat{\mathcal{Q}}^{\dagger}  | Z_{m} Z_{n} \rangle = 0  \, ,  &
\end{aligned}
\label{eq:cubic_supercharge_spin_chains_su111}
\eeq
where the letters of the sector are explicitly given in Appendix \ref{app-review_PSU112}.
Similarly to the case of the spin chain Hamiltonian described in Section \ref{sect-SMT-from_spin_chain}, there is a dictionary which allows to find a corresponding representation in terms of fields. 
We define the fermionic generators as
\begin{align}
\begin{split}
&\hat{\mathcal{Q}}  = \sum_{n=0}^{\infty} \sum_{k=0}^{n-1} 
 \tr \left(  \frac{1}{\sqrt{k+1}} \psi^{\dagger}_k [\Phi^{\dagger}_{n-1-k}, \Phi_n]
+\frac{1}{2} \sqrt{\frac{n+1}{(k+1)(n-k)}} \psi^{\dagger}_k \lbrace \psi^{\dagger}_{n-1-k}, \psi_n \rbrace
  \right) \, ,
  \label{eq:cubic_supercharge_Q1S2_su111}
\\
&\hat{\mathcal{Q}}^{\dagger}  = \sum_{m,n=0}^{\infty}  \tr \left( 
 \frac{1}{\sqrt{n+1}} [\Phi^{\dagger}_{m+n+1}, \Phi_{m}] \psi_n
+\frac{1}{2} \sqrt{\frac{m+n+2}{(m+1)(n+1)}} \lbrace \psi^{\dagger}_{m+n+1}, \psi_{m} \rbrace \psi_n   \right) \, . 
  \end{split}
\end{align}
There is no ambiguity in going from the action on spin chains in Eq.~\eqref{eq:cubic_supercharge_spin_chains_su111} to this representation as SMT generators, since they are all equivalently related by the cyclicity property of the trace.
In this form, the supercharges do not seem to be the hermitian conjugate of each others.
However, one can show that this requirement is indeed satisfied by performing the manipulations \eqref{eq:manipulations_sums_momenta} on the summations over momenta.
This allows to put the first supercharge into the form
\beq
\hat{\mathcal{Q}} = \sum_{n,m=0}^{\infty}  \tr \left(  \frac{1}{\sqrt{m+1}}
\psi^{\dagger}_m [\Phi^{\dagger}_n, \Phi_{n+m+1}] +
 \frac{1}{2}\sqrt{\frac{n+m+2}{(m+1)(n+1)}} \psi^{\dagger}_m \lbrace \psi^{\dagger}_n, \psi_{n+m+1} \rbrace
\right) \, ,
 \label{eq:cubic_supercharge_Q1_su111_form2}
\eeq
which is now evidently the hermitian conjugate of $\hat{\mathcal{Q}}^{\dagger}$ defined in Eq.~\eqref{eq:cubic_supercharge_Q1S2_su111}.

The Hamiltonian in the SU$(1,1|1)$ sector includes  tr$(\mathbf{Q}^\dagger \mathbf{Q})$ and one copy of tr$(F^\dagger F)$ term contained in Eq.~\eqref{eq:final_result_sphere_reduction}. 
We compute $\{\hat{\mathcal{Q}}^\dagger, \hat{\mathcal{Q}} \}$ in the following way. 
First of all, it is straightforward to acquire the pure scalar term by contracting the fermions in $\psi^{\dagger}_m [\Phi^{\dagger}_n, \Phi_{n+m+1}]$ through the Poisson bracket. 
This results in 
\begin{equation}
	\sum_{m,l,n=0 } \frac{1}{l+1} \text{tr} \left( [\Phi_{m+l+1}^\dagger,\Phi_m] [\Phi^\dagger_n,\Phi_{n+l+1}]  \right) \, .
\end{equation}
This is exactly the pure scalar interaction $\sum \frac{1}{l} \text{tr }(q_l^\dagger q_l)$ after shifting $l\to l-1$. 

We then move on to the purely fermionic contributions, obtained by contracting the fermions in the $ \psi^{\dagger}_m \lbrace \psi^{\dagger}_n, \psi_{n+m+1} \rbrace$ term. 
This will result in four D-terms and one F-term. 
By relabelling the momenta, the four D-terms can be shown to be identical. We thus find
\begin{align}\label{eq:D+Fterminfermion}
	\begin{split}
& \sum_{m,n=0}^\infty \sum_{l=1}^\infty \frac{1}{l} \sqrt{\frac{(m+l+1)(n+l+1)}{(m+1)(n+1)}} \text{tr} \, \left( \{ \psi^\dagger_{m+l},\psi_m \} \{\psi^\dagger_n,\psi_{n+l} \} \right)
\\
+& \sum_{m,k=0}^\infty \sum_{n=0}^{m+k}\frac{1}{4} \frac{m+k+2}{\sqrt{(m+1)(n+1)(k+1)(m+k-n+1)}}
\text{tr} \, \left( \{ \psi_{k},\psi_m \} \{\psi^\dagger_n,\psi_{m+k-n}^\dagger \}  \right) \, .
	\end{split}
\end{align}
We can then apply the Jacobi identity~\eqref{eq:Jacobi-identity-fermion} to the F-term 
\begin{equation}\label{eq:after-Jacobi-fourfermion}
\text{tr} \, \{ \psi_{k},\psi_m \} \{\psi^\dagger_n,\psi_{m+k-n}^\dagger \}  = - \text{tr} \, \left( \{ \psi_{k}, \psi^\dagger_n \} \{\psi_m,\psi_{m+k-n}^\dagger \}  - \text{tr} \, \{ \psi_{k}, \psi^\dagger_{m+k-n} \} \{\psi_m,\psi_{n}^\dagger \} \right) \, .
\end{equation}
The two D-terms in \eqref{eq:after-Jacobi-fourfermion} can be simplified as follows:

\begin{itemize}
	\item If $k-n=l\ge 0$,
	\begin{equation}\label{eq:fourcases-1}
		\text{tr} \, \{ \psi_{k}, \psi^\dagger_n \} \{\psi_m,\psi_{m+k-n}^\dagger \} = \text{tr} \, \left( \{ \psi_{n+l}, \psi^\dagger_n \} \{\psi_m,\psi_{m+l}^\dagger \} \right) \, .
	\end{equation}
\item If $n-k=l \ge 0$, as $m+k-n\ge0$, we can define $m-l= r \ge 0$, 
	\begin{equation}
	\text{tr} \, \{ \psi_{k}, \psi^\dagger_n \} \{\psi_m,\psi_{m+k-n}^\dagger \} = \text{tr} \, \left( \{ \psi_{k}, \psi^\dagger_{k+l} \} \{\psi_{l+r},\psi_{r}^\dagger \} \right) \, .
\end{equation}
\item If $l=m-n \ge 0$ 
\begin{equation}
	\text{tr} \, \{ \psi_{k}, \psi^\dagger_{m+k-n} \} \{\psi_m,\psi_{n}^\dagger \}  = 	\text{tr} \, \left( \{ \psi_{k}, \psi^\dagger_{k+l} \} \{\psi_{n+l},\psi_{n}^\dagger \} \right) \, .
\end{equation}
\item If $n-m=l\ge 0$, we can define $r=k-l \ge 0$, then 
\begin{equation}\label{eq:fourcases-4}
	\text{tr} \, \left( \{ \psi_{k}, \psi^\dagger_{m+k-n} \} \{\psi_m,\psi_{n}^\dagger \}  = 	\text{tr} \, \{ \psi_{l+r}, \psi^\dagger_{r} \} \{\psi_{m},\psi_{m+l}^\dagger \} \right) \, .
\end{equation}
\end{itemize}
Collecting the results from the relabelling procedure from \eqref{eq:fourcases-1} to \eqref{eq:fourcases-4}, the F-term in \eqref{eq:D+Fterminfermion} can rewritten in the unified form
\begin{equation}
-	\sum_{m,n=0}^\infty \sum_{l=1}^\infty \frac{m+n+l+2}{\sqrt{(m+1)(n+1)(m+l+1)(n+l+1)}} \text{tr}\, \left( \{\psi_{m+l}^\dagger,\psi_m \} \{\psi^\dagger_n,\psi_{n+l} \} \right)
\end{equation}
The final result for the purely fermionic term is then
\begin{align}
	\begin{split}
		&\sum_{m,n=0}^\infty \sum_{l=1}^\infty \text{tr}\, \left( \{\psi_{m+l}^\dagger,\psi_m \} \{\psi^\dagger_n,\psi_{n+l} \} \right) \\
		&\left(	\frac{1}{l}\sqrt{\frac{(m+l+1)(n+l+1)}{(m+1)(n+1)}}  -
		\frac{m+n+l+2}{\sqrt{(m+1)(n+1)(m+l+1)(n+l+1)}}
		\right) \\
		= &\sum_{m,n=0}^\infty \sum_{l=1}^\infty 
		\frac{1}{l} \sqrt{\frac{(m+1)(n+1)}{(m+l+1)(n+l+1)}}
\,	\text{tr}\, \left(	\{\psi_{m+l}^\dagger,\psi_m \} \{\psi^\dagger_n,\psi_{n+l} \} \right)
	\end{split}
\end{align}
which is exactly the one obtained from spherical reduction \eqref{eq:sphere_reduction_su112_term_mediated_gluon}. 

The last piece in the SU$(1,1|1)$ near-BPS theory is the mixed term between scalar and fermions. 
The contraction of \eqref{eq:cubic_supercharge_Q1S2_su111} will again result in both D-terms and F-terms.
By using Jacobi identity to eliminate F-terms, we can exactly reproduce the full Hamiltonian in SU$(1,1|1)$ case. 
The procedure is similar to the one discussed above.

\subsection{General PSU(1,1|2) sector}

We extend the definition of the supercharges to the full $\mathrm{PSU}(1,1|2)$ sector, including both copies of the scalars and fermions surviving the near-BPS limit.
Based on the spin chain calculation in \cite{Zwiebel:2005er,Beisert:2007sk,Zwiebel:2007cpa}, we infer the expression of these fermionic generators:
\begin{align}
	\begin{split}
 \hat{\mathcal{Q}} & =  \sum_{n=0}^{\infty} \sum_{k=0}^{n-1} \left[  
\frac{1}{\sqrt{k+1}} \tr \left( (\psi_2^{\dagger})_k [(\Phi^{\dagger}_a)_{n-k-1} , (\Phi_a)_n \right) \right. \\  & \left.
+  \sqrt{\frac{n-k}{(k+1)(n+1)}} \tr \left(  (\psi_2^{\dagger})_k \lbrace (\psi^{\dagger}_1)_{n-k-1}, (\psi_1)_n \rbrace  \right)
 +\frac{1}{2} \sqrt{\frac{n+1}{(k+1)(n-k)}} \tr \left(  (\psi_2^{\dagger})_k \lbrace (\psi^{\dagger}_2)_{n-k-1}, (\psi_2)_n \rbrace  \right)  \right] \\
 &   - \sum_{n=0}^{\infty} \sum_{k=0}^n \frac{1}{2\sqrt{n+1}} \, \epsilon^{ab}
  \tr \left( [ (\Phi^{\dagger}_a)_k , (\Phi^{\dagger}_b)_{n-k} ] (\psi_1)_n]  \right) \, ,
\\
 \hat{\mathcal{Q}}^{\dagger} &  =  \sum_{m,n=0}^{\infty} \left[  
 \frac{1}{\sqrt{n+1}}  \tr \left( [ (\Phi^{\dagger}_a)_{m+n+1}, (\Phi_a)_{m}] (\psi_2)_n \right)   +  \sqrt{\frac{m+1}{(n+1)(m+n+2)}}  \tr \left( \lbrace (\psi^{\dagger}_1)_{m+n+1}, (\psi_1)_{m} \rbrace (\psi_2)_n   \right)  \right.  \\
& \left.
+\frac{1}{2} \sqrt{\frac{m+n+2}{(m+1)(n+1)}}  \tr \left( \lbrace (\psi^{\dagger}_2)_{m+n+1}, (\psi_2)_{m} \rbrace (\psi_2)_n   \right)  - \frac{1}{2\sqrt{m+n+1}} \, \epsilon^{ab}
\tr \left( (\psi^{\dagger}_1)_{m+n}  [(\Phi_a)_m, (\Phi_b)_n]  \right)
 \right] \, .\label{eq:cubic_supercharge_Q1S2_psu112}
 \end{split}
\end{align} 
Using the same manipulations explained in Eq.~\eqref{eq:manipulations_sums_momenta}, we find that the generators are indeed the hermitian conjugate of each other.

As our primary goal is to show the positivity structure of the PSU$(1,1|2)$ Hamiltonian, we can again take the purely fermionic part as the example to illustrate the calculation. 
First of all it is very obvious that the tr$ \left( \{ (\psi^{\dagger}_1)_{m+l},(\psi_1)_m\} \{ (\psi_1^{\dagger})_n,(\psi_1)_{n+l}\} \right)$ can only be acquired from $ \lbrace (\psi^{\dagger}_1)_{m+n+1}, (\psi_1)_{m} \rbrace (\psi_2)_n   $ by contracting the $\psi_2$ field. 
The Hamiltonian with  four $\psi_2$ fermions receives contributions 
from $ \lbrace (\psi^{\dagger}_2)_{m+n+1}, (\psi_2)_{m} \rbrace (\psi_2)_n  $. As we can see, the calculations will exactly follow what we discussed in \eqref{eq:D+Fterminfermion} for SU$(1,1|1)$ Hamiltonian. 
The result matches with the spherical reduction computation and gives the  tr$\sum \frac{1}{l}\tilde{q}_l^{1\dagger} \tilde{q}_l^1$ term.

The remaining purely fermionic terms involve two copies of $\psi_1$ and two copies of the $\psi_2$ fields. 
For simplicity, we define
\begin{align}
	\begin{split}
\mathtt{Q}_1^\dagger &= \sum_{m,l=0}^\infty   \sqrt{\frac{m+1}{(l+1)(m+l+2)}}  \text{tr} \left( \lbrace (\psi^{\dagger}_1)_{m+l+1}, (\psi_1)_{m} \rbrace (\psi_2)_l   \right) \, , \\
\mathtt{Q}_2^\dagger &=\frac{1}{2} \sum_{m,l=0}^\infty \sqrt{\frac{m+l+2}{(m+1)(l+1)}}  \text{tr} \left( \lbrace (\psi^{\dagger}_2)_{m+l+1}, (\psi_2)_{m} \rbrace (\psi_2)_l   \right) \, , \\
\mathtt{Q}_1&= \sum_{m,l=0}^\infty   \sqrt{\frac{m+1}{(l+1)(m+l+2)}}  \text{tr} \left( \lbrace (\psi_1)_{m+l+1}, (\psi_1^\dagger)_{m} \rbrace (\psi_2^\dagger)_l   \right) \, , \\
\mathtt{Q}_2 &=\frac{1}{2} \sum_{m,l=0}^\infty \sqrt{\frac{m+l+2}{(l+1)(m+1)}}  \text{tr} \left( \lbrace (\psi^{\dagger}_2)_{m}, (\psi_{2})_{m+l+1} \rbrace (\psi_2^\dagger)_l   \right) \, .
	\end{split}
\end{align}
The terms involving two fermions in each flavor can be collected from the contraction $\{\mathtt{Q}_1^\dagger+\mathtt{Q}_2^\dagger, \mathtt{Q}_1+\mathtt{Q}_2\}$. 
These are
\begin{align}
	\begin{split}\label{eq:F-term-cubiccharges}
\{\mathtt{Q}_1^\dagger, \mathtt{Q}_1\} \supset		&  \sum_{m,n,k=0}^\infty\sqrt{\frac{(m+1)^2}{(n+1)(m+n+2)(k+1)(m+k+2)}} \text{tr} \, \left( \{ (\psi^{\dagger}_1)_{m+n+1},(\psi_2)_n \} \{(\psi^{\dagger}_2)_k, (\psi_1)_{m+k+1} \} \right) \\
		+& \sum_{m,n=0}^\infty \sum_{k<m+n} \sqrt{\frac{(m+1)(m+n-k+1)}{(k+1)(n+1)(m+n+2)^2}} 
		\text{tr} \, \left( \{(\psi_1)_{m},(\psi_2)_n \} \{(\psi^{\dagger}_2)_k, (\psi^{\dagger}_1)_{m+n-k} \} \right) \, ,  \\
\{\mathtt{Q}_2^\dagger,\mathtt{Q}_1 \} &= \sum_{m,n,l=0}^\infty \sqrt{\frac{(m+n+2)(l+1)}{(m+1)(n+l+2)(n+1)^2}} \text{tr} \, \left( \{(\psi^{\dagger}_1)_{l},(\psi_1)_{n+l+1} \} \{(\psi^{\dagger}_2)_{m+n+1}, (\psi_{2})_{m} \} \right) \, , \\
\{\mathtt{Q}_1^\dagger,\mathtt{Q}_2 \} &= \sum_{m,n,l=0}^\infty \sqrt{\frac{(l+n+2)(m+1)}{(l+1)(n+m+2)(n+1)^2}} \text{tr} \, \left( \{(\psi^{\dagger}_1)_{m+n+1},(\psi_1)_{m} \} \{(\psi^{\dagger}_2)_{l}, (\psi_{2})_{n+l+1} \} 	\right)	\, .
	\end{split}
\end{align}
We then use the Jacobi identity to eliminate the F-term in \eqref{eq:F-term-cubiccharges}. 
The coefficient for \newline
$ \text{tr} \, \left( \{(\psi^{\dagger}_1)_{n+l},(\psi_2)_{n} \} \{(\psi^{\dagger}_2)_{m}, (\psi_{1})_{m+l} \} \right) $
is 
\begin{align}\label{eq:coefficient-1221-fermion}
	\begin{split}
		&\frac{1}{\sqrt{(n+1)(m+1)(n+l+1)(m+l+1)}} \left(l- \frac{(m+l+1)(n+l+1)}{m+n+l+2} \right) \\
		=& - \frac{1}{m+n+l+2} \sqrt{\frac{(m+1)(n+1)}{(m+l+1)(n+l+1)}} \, .
	\end{split}
\end{align}
Similarly, the coefficient for 
tr$ \left( \{(\psi_1)_m,(\psi^{\dagger}_1)_{m+l} \} \{(\psi_2)_{n+l},(\psi^{\dagger}_2)_{n}  \} \right)$ can be calculated as follows:
\begin{align}\label{eq:coefficient-1122-fermion}
	\begin{split}
		&\sqrt{\frac{(n+l+1)(m+1)}{(n+1)l^2(m+l+1)}} - \sqrt{\frac{(m+1)(m+l+1)}{(n+l+1)(n+1)(m+n+l+2)^2}}  \\
		=& \sqrt{\frac{(m+1)(n+1)}{(m+l+1)(n+l+1)}} \left[\frac{1}{l} + \frac{1}{m+n+l+2} \right]
	\end{split}
\end{align}
The part with the $1/l$ coefficient works as the cross term which should be combined with tr$\left( \tilde{q}_l^1 \tilde{q}_l^1 \right)$ and tr$\left( \tilde{q}_l^2 \tilde{q}_l^2 \right)$
to form tr$(\tilde{q}_l^1 +\tilde{q}_l^2) (\tilde{q}_l^1 +\tilde{q}_l^2)$. 
This is exactly the fermionic part in the $\mathbf{Q}$ blocks obtained from spherical reduction computation \eqref{eq:sphere_reduction_su112_term_mediated_gluon}. 
The part with $1/(m+n+l+2)$ coefficient needs to be combined with \eqref{eq:coefficient-1221-fermion}
to get exactly the fermionic interaction term between two flavors~\eqref{eq:final_result_sphere_reduction}. 
This calculation confirms that all the purely fermionic terms in the PSU$(1,1|2)$ Hamiltonian can be acquired from the anticommutator of the supercharges originated from the enhanced psu$(1|1)^2$ symmetry. 

This computation can be completed by analysing the rest of the mixed terms in the overall Hamiltonian~\eqref{eq:final_result_sphere_reduction}. 
The explicit calculation shows exactly 
\begin{equation}
\{ \hat{\mathcal{Q}}
,  \hat{\mathcal{Q}}^\dagger
\}	= H_{\rm int}
\end{equation}
This proves that the Hamiltonian in PSU$(1,1|2)$ is positive definite.

%%%%%%%%%%%%%%%

\section{Symmetry structures}\label{sec:symmetry}

The positivity structure discussed in the Section \ref{sect-positivity} can be made manifest by writing the  Hamiltonian in terms of fundamental blocks
\begin{equation}\label{eq:H-blockform}
	H_{
	\text{int}}= \sum_{l,a} \text{tr} ( B_l^{a\dagger} B_l^a ) \, ,
\end{equation}
where $l$ stands for all the allowed descendants (including the $n,k$ labels in SU$(1,2)$ cases) while $a$ is the spin group index. 
In the previous works \cite{Baiguera:2020jgy,Baiguera:2020mgk}, we have shown that the near-BPS Hamiltonian of the sectors with SU$(1,1)$ and SU$(1,2)$ symmetry (including their supersymmetric extensions) has a structure quadratic in the blocks as shown in Eq.~\eqref{eq:H-blockform}.
The form~\eqref{eq:H-blockform} brings us following benefits:
\begin{enumerate}
	\item 
	It is manifestly positive-definite. Since Spin Matrix theory is defined as an effective theory in the near-BPS decoupling limit, the Hamiltonian is believed to be positive definite as a response when we move away from the saturation of the bound.
	\item The blocks are simultaneously transforming in the specific representations of spin group and in the adjoint representation of the SU$(N)$ gauge group. 
	This form has been shown \cite{Baiguera:2020jgy,Baiguera:2020mgk} to be useful for proving the invariance of the Hamiltonian under the actions of the generators. For example, in the SU$(1,1|1)$ case the blocks simply organize into an $\mN=1$ chiral multiplets, while in the SU$(1,2|2)$ case the blocks of the Hamiltonian form the $\mN=2$ vector multiplets. These facts make the spin group invariance manifest. 
	\item The ground state of SMT is determined by an infinite number of constraints\footnote{We have shown in \cite{Baiguera:2020mgk} that the constraints in SU$(1,2|2)$ subsector are both of bosonic and fermionic kind, as a complete generalization of the constraints studied in \cite{Grant:2008sk}. }
	\begin{equation}
		B_l^{a\dagger} =0
	\end{equation}
The states satisfying these constraints are the only configurations surviving in the strong coupling. 
Thus the block structure \eqref{eq:H-blockform} effectively provides a probe into the strong coupling dynamics of  SMT.
\end{enumerate} 
We will show that the Hamiltonian in the SU$(2|3)$ case can also be organized nicely into the norm of the blocks. It is not clear that the Hamiltonian in the PSU$(1,1|2)$ sector has this structure; however, we will show its invariance pattern by direct calculations.

\subsection{Symmetry structure of SU$(2|3)$ sector}

%\begin{itemize}
%	\item How blocks transform under SU$(2|3)$: use Cartan generators to classify
%	\item Emphasize the F-term structure of chiral multiplet
%\end{itemize}

The  SU$(2|3)$ Hamiltonian \eqref{eq:mixed_terms_sphere_red_su23} is manifestly written as a sum of F-terms.  
Its equivalent D-term form  is  (subject to Gauss constraint \eqref{eq:Gauss_law_su23})
\begin{align}\label{eq:D-term-SU23-Hamiltonian}
	\begin{split}
H&= \sum_{a,b=1}^3 \text{tr} \Big([\Phi_a^\dagger,\Phi_b] [\Phi_b^\dagger,\Phi_a]  - [\Phi_a^\dagger,\Phi_a] [\Phi_b^\dagger,\Phi_b] \Big)  \\
&+ \sum_{a=1}^3 \sum_{\alpha=1,2} 2\,\text{tr} \Big([\psi_\alpha^\dagger,\Phi_a] [\Phi_a^\dagger,\psi_\alpha]] \Big)  + \sum_{\alpha,\beta=1}^2 \text{tr} \Big(
\{\psi_\alpha^\dagger,\psi_\alpha \} \{\psi^\dagger_\beta,\psi_\beta \} - \{\psi_\alpha^\dagger,\psi_\beta \} \{\psi^\dagger_\beta,\psi_\alpha \} 
\Big) \, ,
	\end{split}
\end{align}
acquired by applying the Jacobi identities \eqref{eq:Jacobi-identity-fourscalar}, \eqref{eq:Jacobi-identity-fourfermion} and \eqref{eq:spin_chain_Ham_su23}.

However, the blocks entering the F-term or D-term form of the Hamiltonian are not transforming under the same representations.
Let's focus on the blocks of the F-term type in \eqref{eq:mixed_terms_sphere_red_su23} first. We introduce the following notation for the blocks
\begin{equation}
	B_{(j;m)}^a, \qquad a=1\,,2\,,3\,, \quad j=0\,,\frac{1}{2}\,,1, \quad |m|\le j \, .
\end{equation}
The indices $a$ label the following weights of the Cartan generators presented in table \ref{table:dimension-su23-letter} of Appendix \ref{app-algebra_SU23_PSU112}:
\begin{align}
	\begin{split}
		a=1 & \quad \leftrightarrow  \quad (R_0,\mathbf{R}_0)= \pm \left(\frac{1}{2},\frac{1}{2}\right) \, , \\
		a=2 & \quad \leftrightarrow  \quad (R_0,\mathbf{R}_0)= \pm \left(-\frac{1}{2}, \frac{1}{2}\right) \, ,  \\
		a=3 & \quad \leftrightarrow  \quad (R_0,\mathbf{R}_0)= \pm \left(0, -1\right) \, ,
	\end{split}
\end{align}
where the $\pm$ signs are related by complex conjugation and $R_0,\mathbf{R}_0$ are the Cartan generators of the SU$(3)$ algebra shown in \eqref{eq:SU3algebra}.
The $j$ parametrizes the representations of SU$(2)$. 

The scalar blocks transform nontrivially under the su$(3)$ algebra but as a singlet under the su$(2)$ subalgebra.
\begin{align}
	\begin{split}
		(R_0)_D [\Phi_1,\Phi_2] &= 0, \quad (\mathbf{R}_0)_D  [\Phi_1,\Phi_2]= - [\Phi_1,\Phi_2] \, , \\
		(R_0)_D [\Phi_2,\Phi_3] &= \frac{1}{2} [\Phi_2,\Phi_3], \quad (\mathbf{R}_0)_D  [\Phi_2,\Phi_3] = \frac{1}{2} [\Phi_2,\Phi_3] \, , \\
		(R_0)_D [\Phi_3,\Phi_1] &= -\frac{1}{2} [\Phi_3,\Phi_1], \quad (\mathbf{R}_0)_D  [\Phi_3,\Phi_1]  = \frac{1}{2} [\Phi_3,\Phi_1]  \, .
	\end{split}
\end{align}
These equations indicate that the blocks
$[\Phi_a,\Phi_b]$ transform in the exact same way as the letter $\epsilon_{abc}\Phi_c^\dagger$ (see table \ref{table:dimension-su23-letter} in Appendix \ref{app-algebra_SU23_PSU112}), which is again the fundamental representation $\mathbf{3}$. 
This fact is closely related to the representation theory of SU$(3)$. The interpretation is as follows. 
The block $[\Phi_a^\dagger,\Phi_b^\dagger] = if_{ABC}\Phi_a^{A\dagger} \Phi_b^{B\dagger} T^C$ can be considered
as the tensor product of two $\Phi_a^\dagger$ fields  transforming in the fundamental representation. 
It obeys the following rules: 
\begin{equation}
	(1,0)\otimes(1,0) =(2,0) \oplus (0,1) \quad \leftrightarrow  \quad \mathbf{3} \times \mathbf{3} =\mathbf{6} + \bar{\mathbf{3}}
	\quad 
\end{equation}
Therefore, an anti-fundamental triplet sits inside the tensor product of two triplets transforming in the fundamental representation. Then its complex conjugate $[\Phi_a,\Phi_b]$ transforms in the fundamental representation, the same as $\Phi_c^\dagger$. 
We thus identify the $\epsilon_{abc}[\Phi_b,\Phi_c]$ block as $B_0^a$. 

The mixed block $[\Phi_a^\dagger,\psi_\alpha^\dagger]$ transforms in the same way as $\Phi_a^\dagger$ under the SU$(3)$ R-symmetry action. Similarly, it transforms in the same way as $\psi_\alpha^\dagger$ under the SU$(2)$  action. Therefore, these blocks will be denoted as $B_{(\frac{1}{2},\frac{1}{2})}^{a\dagger}$ when $\alpha=1$  and 
$B_{(\frac{1}{2},-\frac{1}{2})}^{a\dagger}$ when $\alpha=2$. They transform in the $j=\frac{1}{2}$ representation of SU$(2)$ and in fundamental representation of SU$(3)$. 

The last part of the Hamiltonian is composed by the fermionic blocks. They are obviously singlets under the SU$(3)$ action, therefore we suppress the $a$ indices in the definition of the blocks. 
The action under the SU$(2)$ symmetry is more subtle. 
We can show that
\begin{align}
	\begin{split}
		(I_0)_D  	\{\psi_+^\dagger,\psi_+^\dagger \} & = 	\{\psi_+^\dagger,\psi_+^\dagger \} \, , \\
		(I_0)_D  	\{\psi_+^\dagger,\psi_-^\dagger \} & = 0 \, , \\
		(I_0)_D  	\{\psi_-^\dagger,\psi_-^\dagger \} & = - \{\psi_-^\dagger,\psi_-^\dagger \} \, .
	\end{split}
\end{align}
Therefore instead of the $j=\frac{1}{2}$ fundamental representation of SU$(2)$, the fermionic blocks 
transform in the adjoint representation $j=1$ representation of SU$(2)$. 
We thus denote these blocks as $B_{(1,m)}^\dagger$ for $m=\pm 1,0$. 
\begin{equation}
	B_{(1,1)}^\dagger=	\{\psi_+^\dagger,\psi_+^\dagger \}, \qquad  B_{(1,0)}^\dagger=	\{\psi_+^\dagger,\psi_-^\dagger \}, \qquad 
	B_{(1,-1)}^\dagger=	\{\psi_-^\dagger,\psi_-^\dagger \} \, .
\end{equation}
After combinining all of blocks, the final Hamiltonian can be better organized as
\begin{align}\label{eq:H-su23-block}
	\begin{split}
		H = \sum_{a=1}^3 \sum_{j=0,\frac{1}{2}} \sum_{m=-j}^j B_{(j;m)}^{a\dagger} B_{(j;m)}^a +  \sum_{m=-1}^1 \left(1-\frac{|m|}{2}\right) B_{(1;m)}^{\dagger} B_{(1;m)} \, .
	\end{split}
\end{align}
We thus interpret the Hamiltonian \eqref{eq:H-su23-block} as the norm of states generated by $B_{(j;m)}^a$ blocks.

To see how supersymmetry acts on the Hamiltonian, let's take $Q_{a1} = \text{tr}(\Phi_a^\dagger \psi_+)$ as an example. 
The supercharge action on the blocks in terms of $B_{(j;m)}^a$ is given by
\begin{align}\label{eq:supercharge-actionon-SU23-BLOCKS}
	\begin{split}
		(Q_{c1})_D B_0^a & =\epsilon_{abc} B^b_{(\frac{1}{2},\frac{1}{2})},\qquad 	(Q_{c1})_D  B_{(1,m)}^\dagger = (m+1) B^{a\dagger}_{(\frac{1}{2},m-\frac{1}{2})} \, , \\
		(Q_{c1})_D B_{(\frac{1}{2},\frac{1}{2})}^a&= \delta_{ac} B_{(1,1)},\quad
		(Q_{c1})_D  B_{(\frac{1}{2},\frac{1}{2})}^{a\dagger} =  \epsilon_{abc} B^{b\dagger}_0,\quad
		(Q_{c1})_D  B^a_{(\frac{1}{2},-\frac{1}{2})} = \delta_{ac} B_{(1,0)} \, .
	\end{split}
\end{align}
The other terms are vanishing. 
Then the supersymmetry action on Hamiltonian is easily seen to vanish.
It can be shown that with respect to the $Q_{c1}$ supercharge, the three $\mN=1$ supermultiplets are 
\begin{equation}
	\Big(B_0^a, \epsilon_{abc} B^b_{(\frac{1}{2},\frac{1}{2})} \Big),\qquad 
	\Big(B_{(\frac{1}{2},\frac{1}{2})}^a,\delta_{ac}B_{(1,1)}
	\Big),\quad
	\text{and} \qquad \Big(B^a_{(\frac{1}{2},-\frac{1}{2})}, \delta_{ac}B_{(1,0)} \Big) \, .
\end{equation}
The norms of these supermultiplets are invariants under the action of $Q_{c1}$.

The D-term form of Hamiltonian \eqref{eq:D-term-SU23-Hamiltonian} is not manifestly positive.  However, there is a neat interpretation of the Hamiltonian from the D-term perspective, too. 
Let's restrict to the SU$(3)$ scalars and decouple all the other fermionic modes. There are eight different and independent blocks
\begin{equation}\label{eq:adjointblocks}	
[\Phi_a^\dagger,\Phi_b],  \quad [\Phi_a^\dagger,\Phi_a]-[\Phi_b^\dagger,\Phi_b] \, .
\end{equation} 
These are sitting in the $(1,1)\equiv \mathbf{8}$ which is the adjoint  representation of SU$(3)$. 
This structure appears from the tensor product between the F-term block and its conjugate, resulting in the $\mathbf{8}$ representation coming from $\mathbf{3}\times \bar{\mathbf{3}}$.
The invariant Hamiltonian is  thus nothing but the Casimir operator of SU$(3)$ subject to the Gauss constraint. 
Such interpretation does not apply to the F-term structure straightforwardly. 
However, we can still regard the norm structure of the Hamiltonian as the tensor product between conjugate representations where the blocks transform. 
The singlet state from the tensor product is automatically the Hamiltonian, invariant under the global symmetry. 
We can summarize the following facts about the Hamiltonian: 
\begin{itemize}
	\item The blocks are transforming in some representation of the spin group.
	\item The Hamiltonian is made by singlets under the tensor product of the blocks in their representation and the conjugate one.
\end{itemize}

\subsection{Symmetry structure of PSU$(1,1|2)$ sector}
\label{ssec:symmetry-psu112}

\subsection*{Symmetry action}
The global symmetry of PSU$(1,1|2)$ subsector includes an SU$(2)_A$ automorphism \cite{Beisert:2007sk}, 
which is not a subalgebra of the global symmetry of $\mN=4$ SYM.
The two scalars $\Phi_{1,2}$ transform as a doublet under the action of SU$(2)$ R-symmetry, while it does not act on fermions. 
As a requirement of supersymmetry, the emergent SU$(2)_A$ automorphism transforms the fermions as a doublet. 
In terms of the mode expansion, the symmetry generators are explicitly given in Eq.\eqref{eq:SU2-automorphism}.
As a result, the overall Hamiltonian \eqref{eq:final_result_sphere_reduction} should be invariant under both PSU$(1,1|2)$ and SU$(2)_A$ automorphism symmetry. 

Let's first focus on the bosonic part of the global symmetry. 
This includes SU$(1,1) \times \SU(2)_R\times \SU(2)_A$. 
The general spin $j$ representation of SU$(1,1)$ satisfies 
\begin{equation}
	\begin{array}{l}
		\ds L_0 | j,j+n\rangle = (j+n)  | j,j+n\rangle\,, \\
		\ds L_+ | j,j+n\rangle = \sqrt{(n+1)(n+2j)} | j,j+n+1\rangle\,, \\
		\ds L_- | j,j+n\rangle = \sqrt{n(n+2j-1)} | j,j+n-1\rangle\,.
	\end{array}
\end{equation}
The $\text{tr}(\mathbf{Q}^\dagger_l \mathbf{Q}_l)$ and $\tr  (F_{ab})_l^\dagger (F_{ab})_l $ parts in~\eqref{eq:final_result_sphere_reduction} are similar to the SU$(1,1|1)$ Hamiltonian whose symmetry structure was analyzed in~\cite{Baiguera:2020mgk}. 
We simply find 
\begin{align}
	\begin{split}
(L_0)_D \mathbf{Q}^\dagger_l &= l \mathbf{Q}^\dagger_l\,, \quad (R_0)_D \mathbf{Q}^\dagger_l=0\,,\quad (\mathfrak{R}_0 )_D \mathbf{Q}^\dagger_l=0\,, \\
  (L_0)_D (F_{ab}^\dagger)_l &= \left(l+\frac{1}{2} \right) (F_{ab}^\dagger)_l , \\
  (R_0)_D (F_{ab}^\dagger)_l &=\left(\frac{1}{2}-\delta_{b2}\right) (F_{ab}^\dagger)_l \,, \quad 
(\mathfrak{R}_0 )_D (F_{ab}^\dagger)_l = \left(\frac{1}{2}-\delta_{a2}\right) (F_{ab}^\dagger)_l\,.
	\end{split}
\end{align}
The invariance of $\text{tr}(\mathbf{Q}^\dagger_l \mathbf{Q}_l)$ and $\tr  (F_{ab})_l^\dagger (F_{ab})_l $ terms under $L_+$ actions follows similarly from the SU$(1,1|1)$ Hamiltonian, \emph{i.e.}
\begin{eqnarray}\label{eq:invariance-QQ-FF}
(L_+)_D \text{tr} \sum_{l=1}^{\infty} \frac{1}{l }(\mathbf{Q}^\dagger_l \mathbf{Q}_l) & =& \sum_{l=1}^{\infty} \text{tr} (\mathbf{Q}^\dagger_{l+1} \mathbf{Q}_l) -
\sum_{l=1}^{\infty} \text{tr} (\mathbf{Q}^\dagger_{l} \mathbf{Q}_{l-1})  = -\text{tr} (\mathbf{Q}^\dagger_{1} \mathbf{Q}_{0}) =0\,, \\  \nonumber
(L_+)_D \text{tr} \sum_{l=0}^{\infty} (F_{ab})_l^\dagger (F_{ab})_l  & =& \sum_{l=0}^\infty (l+1) \text{tr}(F_{ab})_{l+1}^\dagger (F_{ab})_l - \sum_{l=0}^\infty l \,\text{tr}(F_{ab})_{l}^\dagger (F_{ab})_{l-1} =0 \,,
\end{eqnarray}
where we apply the Gauss constraint $\mathbf{Q}_0=0$. 
The R-symmetry and automorphism action are easy to check since the part containing the $\mathbf{Q}^\dagger$ and $F_{ab}^\dagger$ blocks is symmetric in the $a= \lbrace 1,2 \rbrace $ flavour indices. 

The novel structure of the PSU$(1,1|2)$ Hamiltonian is given by the interactions between scalars and fermions with different flavours, \emph{i.e.} the single trace terms. 
To clarify its structure, we would like to first explain how the bosonic interaction term \eqref{eq:spinchainscalar-2} respects the global symmetry. 
The D-term and F-term expressions of \eqref{eq:spinchainscalar-2} provide two perspectives on the invariance under the SU$(1,1)$ action.  
The symmetry actions on the F-term
\eqref{eq:Hscalar} result in the following transformations
\begin{equation}
	(L_+)_D J_L^\dagger =(L+1) J_{L+1}^\dagger\,, \quad (L_+)_D J_L = -(L+1) J_{L-1}\,,
\end{equation}
where we should also notice that the action on $J_0$ identically vanishes:
\begin{equation}
(L_+)_D J_0 = 0 \, .
\end{equation}
This shows that $J_L^\dagger$ transforms under the $j=\frac{1}{2}$ representation of SU$(1,1)$ group. 
Therefore we find that the scalar interaction \eqref{eq:Hscalar} is invariant under the action of $L_+$ due to the telescopic structure in the summation:
\begin{equation}\label{eq:symmetry-Ftermscalar}
(L_+)_D  \sum_{L=0}^\infty \frac{1}{L+1} \text{tr}(J_L^\dagger J_L) = \sum_{L=0}^\infty \text{tr}(J_{L+1}^\dagger J_L) - 	\sum_{L=0}^\infty \text{tr}(J_{L}^\dagger J_{L-1}) =0 \, .
\end{equation}
On the contrary, the $L_+$ action on the $D$-term \eqref{eq:spinchainscalar-2} is more complicated. 
The $L_+$ action on each block is
\begin{align}
	\begin{split}
(L_+)_D [(\Phi_a^{\dagger})_m,(\Phi_b)_{m+l}] &= (m+1) [(\Phi^{\dagger}_a)_{m+1},(\Phi_b)_{m+l}] -(m+l) [(\Phi^{\dagger}_a)_m,(\Phi_b)_{m+l-1}] \, , \\
(L_+)_D [(\Phi^{\dagger}_a)_{n+l},(\Phi_b)_{n}] &= (n+l+1) [(\Phi_a^{\dagger})_{n+l+1},(\Phi_b)_{n}] - n [(\Phi^{\dagger}_a)_{n+l},(\Phi_b)_{n-1}] \, .
	\end{split}
\end{align}
The corresponding action on \eqref{eq:spinchainscalar-2} is thus
\begin{eqnarray}\label{eq:L+-action-scalar-2flavor-112}
&& (L_+)_D H_{\text{SMT}}^{\text{bos}} \\ \nonumber
&=& \sum_{m,n,l=0}^{\infty} \left( \frac{m}{m+n+l} - \frac{m+l}{m+n+l+1}\right) \text{tr}\Big([\Phi^{1\dagger}_{m},\Phi^1_{m+l-1}] [\Phi^{2\dagger}_{n+l},\Phi^2_n] 
+[\Phi^{1\dagger}_{m},\Phi^2_{m+l-1}] [\Phi^{2\dagger}_{n+l},\Phi^1_n] 
\Big) \\ \nonumber
&&+ \sum_{m,n,l=0}^{\infty} \left(\frac{n+l+1}{m+n+l+1} - \frac{n+1}{m+n+l+2}\right)  \text{tr}\Big( [\Phi^{1\dagger}_m,\Phi^1_{m+l}] [\Phi^{2\dagger}_{n+l+1},\Phi^2_n]
+[\Phi^{1\dagger}_m,\Phi^2_{m+l}] [\Phi^{2\dagger}_{n+l+1},\Phi^1_n]
\Big) \\ \nonumber
&& +\sum_{l=1}\sum_{m,n=0}^\infty \left( \frac{m}{m+n+l} - \frac{m+l}{m+n+l+1} \right) \text{tr}\Big( [\Phi^{2\dagger}_m,\Phi^2_{m+l-1}] [\Phi^{1\dagger}_{n+l},\Phi^1_n]
+ [\Phi^{2\dagger}_m,\Phi^1_{m+l-1}] [\Phi^{1\dagger}_{n+l},\Phi^2_n]
\Big) \\ \nonumber
&& + \sum_{l=1}\sum_{m,n=0}^\infty \left(\frac{n+l+1}{m+n+l+1} - \frac{n+1}{m+n+l+2}
\right)  
\text{tr} \Big([\Phi^{2\dagger}_m,\Phi^2_{m+l}] [\Phi^{1\dagger}_{n+l+1},\Phi^1_n] 
+[\Phi^{2\dagger}_m,\Phi^1_{m+l}] [\Phi^{1\dagger}_{n+l+1},\Phi^2_n] \, .
\Big)
\end{eqnarray}
Defining the function 
\begin{equation}
	f(l) = -l^2-l(m+n)+m \, ,
\end{equation}
the coefficients of \eqref{eq:L+-action-scalar-2flavor-112} become
\begin{align}
	\begin{split}
\frac{m}{m+n+l} - \frac{m+l}{m+n+l+1} &= \frac{f(l)}{(m+n+l)(m+n+l+1)} \\
\frac{n+l+1}{m+n+l+1} - \frac{n+1}{m+n+l+2} &= -\frac{f(l+1)}{(m+n+l+1)(m+n+l+2)} \, .
	\end{split}
\end{align}
This indicates that the summation over $l$ in \eqref{eq:L+-action-scalar-2flavor-112} is telescopic and only boundary terms of the summation over $l$ contribute. 
This then results in 
\begin{align}\label{eq:actionL+-secondtelesum}
	\begin{split}
& (L_+)_D H_{\text{SMT}}^{\text{bos}} \\
=& \sum_{m,n=0}^\infty	\frac{m}{(m+n)(m+n+1)} \text{tr}\Big([(\Phi^{\dagger}_1)_m,(\Phi_1)_{m-1}] [(\Phi^{\dagger}_2)_n,(\Phi_2)_n]+[(\Phi^{\dagger}_1)_m,(\Phi_2)_{m-1}] [(\Phi^{\dagger}_2)_n,(\Phi_1)_n]  \Big)\\
&- \frac{m+1}{(m+n+1)(m+n+2)}\text{tr} \Big( [(\Phi_2^{\dagger})_n,(\Phi_2)_n] [(\Phi_1^{\dagger})_{m+1},(\Phi_1)_m]+ [(\Phi_2^{\dagger})_n,(\Phi_1)_n] [(\Phi_1^{\dagger})_{m+1},(\Phi_2)_m]
\Big)
	\end{split}
\end{align}
The boundary term \eqref{eq:actionL+-secondtelesum} is still a telescopic sum over $m$, where only $m=0$ contributes. 
This identically vanishes, confirming the scalar interaction term is invariant under SU$(1,1)$ action.
Differently from the \eqref{eq:invariance-QQ-FF}, the invariance of $H_{\text{SMT}}^{\text{bos}}$ does not require extra constraints. 

The SU$(1,1)$ invariance of the purely fermionic interaction term in \eqref{eq:final_result_sphere_reduction}
is very similar to the symmetry action on the bosonic  D-terms \eqref{eq:L+-action-scalar-2flavor-112}-\eqref{eq:actionL+-secondtelesum}. 
The explicit calculation shows that the double telescopic sum structure (in terms of $m$ and $l$) will reappear when $L_+$ acts on the interaction term composed by fermions with different flavour. 
The symmetry action vanishes in the same way as \eqref{eq:actionL+-secondtelesum}: the boundary term simply vanishes. 
The whole procedure also applies to the mixed scalar-fermion block term in \eqref{eq:final_result_sphere_reduction}. 
As a result we conclude that the PSU$(1,1|2)$ Hamiltonian is invariant under the bosonic part of the spin group. 

Supersymmetry imposes a non-trivial relation between bosonic and fermionic interactions in~\eqref{eq:final_result_sphere_reduction}. 
The explicit calculation using the supercharges~\eqref{eq:def-supercharges} 
results in two types of structures which are found to vanish. 
The first kind of terms are made by single telescopic sums whose boundary term vanishes due to the Gauss constraint. 
The second kinds are double telescopic sums as \eqref{eq:actionL+-secondtelesum} and \eqref{eq:L+-action-scalar-2flavor-112} which are identically vanishing.

It is useful to reorganize the interaction in the following way. 
We combine the supersymmetry generators into
\begin{equation}
	\mathcal{Q}^\dagger= Q^\dagger +S^\dagger  \,.
\end{equation}
It is shown in \cite{Baiguera:2020jgy} that these are the supercharges generating the SU$(1,1)$ and SU$(2)$ R-symmetry. For example, $\{\mQ^\dagger,\mQ \} = 2L_0$.
The following combinations of $F$-blocks 
\begin{equation}
	\mathbf{F}^\dagger_l=(F^{\dagger}_{11})_l -(F^{\dagger}_{22})_l, \qquad 
	\mathbf{F}_l=(F_{11})_l -(F_{22})_l 
\end{equation}
will form an $\mN=1$ chiral multiplet with the $\mathbf{Q}$ block, as can be seen from the following $\mN=1$ supertransformation
\begin{align}\label{eq:mN=1-chiralmultiplet}
	\begin{split}
	\mQ^\dagger_D \mathbf{Q}_l^\dagger	&= -l \,\mathbf{F}_l^\dagger, \qquad
		\mQ^\dagger_D \mathbf{Q}_l  = l \,\mathbf{F}_{l-1} \\
		\mQ^\dagger_D \mathbf{F}_l^\dagger  &= -\mathbf{Q}^\dagger_{l+1}, \qquad
		\mQ^\dagger_D \mathbf{F}_l  = -\mathbf{Q}_l\,.
	\end{split}
\end{align}
The transformation \eqref{eq:mN=1-chiralmultiplet} then leads us to isolate the $\mQ^\dagger-$invariant part of Hamiltonian
\begin{equation}
	H_{\mN=1}= \sum_{l=1}^\infty \frac{1}{l} \text{tr} (\mathbf{Q}^\dagger_l \mathbf{Q}_l) + \sum_{l=0}^\infty \text{tr} (\mathbf{F}^\dagger_l \mathbf{F}_l)\,,
\end{equation}
whose invariance under $\mQ^\dagger$ requires to apply Gauss constraint. 
The overall PSU$(1,1|2)$ Hamiltonian \eqref{eq:final_result_sphere_reduction} can then be decomposed
\begin{equation}\label{eq:remainders}
	H_{\text{int}}= H_{\mN=1}+ \text{rest part} \,.
\end{equation}
The rest part of Hamiltonian under the $\mQ^\dagger$ action will exhibit the double telescopic sum structure as \eqref{eq:L+-action-scalar-2flavor-112}, therefore it will vanish identically.

The $\mN>1$ supermultiplets are usually constructed by relating multiple $\mN=1$ supermultiplets by additional supercharges. 
In \cite{Baiguera:2020mgk}, we discussed how the letters and blocks in SU$(1,2|2)$ subsector can be collected into an $\mN=2$ vector multiplet, which is made by an $\mN=1$ chiral multiplet and an $\mN=1$ vector multiplet. 
In this PSU$(1,1|2)$ subsector, the two scalar and fermionic letters naturally form the $\mN=2$ chiral multiplet. 
It is natural to expect that the Hamiltonian is made by an $\mN=2$ supermultiplet, where the $\mN=1$ chiral multiplet is formed by $(\mathbf{Q}_l,\mathbf{F}_l)$. 
Then we can expect the remaining terms in the Hamiltonian \eqref{eq:remainders} to be determined by another supermultiplet.  
The main obstruction of reorganizing the remaining terms into another chiral multiplet is the fermionic interaction interaction. 
We have shown in section \ref{sect-quantization_sphere_red} that the D-term form of the bosonic interaction can be nicely organized into an F-term whose symmetry properties are neatly shown in \eqref{eq:symmetry-Ftermscalar}. 
Such analogy in fermionic interactions cannot be simply formulated, as we will explain below.

\subsection*{Fermionic F-term}

The F-term simplification in the bosonic interaction term explained in section \ref{sect-quantization_sphere_red} makes the positivity structure manifest. 
However, the generalization of such analysis to the fermionic interaction term is less straightforward.
The difficulty originates from the complexity of structure constants of the SU$(N)$ gauge group. 
To explain this,
we will introduce following notations for the SU$(N)$ color group \cite{Haber:2019sgz}.
We denote as $T^A$ the generators of SU$(N)$ gauge group, where $A=1,...,N^2-1$.
The generators satisfy
\begin{equation}\label{eq:fdcouplings}
	T^A T^B =  \frac{1}{2N}\delta_{AB} \mathbbm{1} + \frac{1}{2} (d_{ABC}+i f_{ABC}) T^C \,.
\end{equation}
Then the D-term blocks of the fermions are
\begin{equation}
	\{\psi_1,\psi_2 \} =i \psi_1^A \psi_2^B f_{ABC} T_C\,.
\end{equation}
The $d_{ABC}$ term is called the $d$-type coupling. 
It appeared in studying the anomaly of QCD theory with SU$(N)$ color group \cite{Srednicki:2007qs}.

Let's start from the fermionic zero modes. 
By turning off all the higher level modes in the scalars and fermions, we only have the purely fermionic interaction term
\begin{equation}\label{eq:zeromode-fermionH}
	H = \text{tr}(	\{(\psi_1^{\dagger})_0,(\psi_1)_0 \} 	\{(\psi_2^{\dagger})_0,(\psi_2)_0 \} -	\{(\psi_1^{\dagger})_0,(\psi_2)_0 \} 	\{(\psi_2^{\dagger})_0,(\psi_1)_0 \} )\,.
\end{equation}
Naively the Jacobi identity cannot be applied to simplify \eqref{eq:zeromode-fermionH}, since the coefficients between two fermionic terms will result in exactly a $d$-type coupling due to Eq.~\eqref{eq:fdcouplings}. 
However, the Hamiltonian~\eqref{eq:zeromode-fermionH} is equivalent to the purely fermionic part of the Hamiltonian ~\eqref{eq:sphere_reduction_su23} in the SU$(2|3)$ sector. 
The difference between Eq.~\eqref{eq:zeromode-fermionH} and Eq.~\eqref{eq:sphere_reduction_su23} is precisely proportional to the Gauss constraint \eqref{eq:Gauss_law_su23}. 
Therefore \eqref{eq:zeromode-fermionH} is equivalent to
\begin{equation}
	H = \frac{1}{2}	\text{tr} \Big( \{(\psi_\alpha)_0,(\psi_\beta)_0 \} \{(\psi_\beta^\dagger)_0,(\psi_\alpha^\dagger)_0 \} \Big)
\end{equation}
which guarantees the positivity.

There is no Gauss constraint to help for higher levels of the fermionic modes. 
The full fermionic interaction term
\begin{equation}\label{eq:purefermion-interaction}
	H_{\text{SMT}}^{\text{ferm}}=	 \sum_{l=0}^{\infty} \sum_{m,n=0}^{\infty} \frac{\sqrt{(m+1)(n+1)}}{\sqrt{(m+l+1)(n+l+1)}} 
	\frac{ \tr \left(  \epsilon^{ac}  \epsilon^{bd} \lbrace (\psi_a^{\dagger})_m , (\psi_b)_{m+l} \rbrace \lbrace (\psi_c^{\dagger})_{n+l} , (\psi_d)_n \rbrace \right)  }{m+n+l+2}
\end{equation}
can be explicitly written in terms of the $L=m+n+l$ notation as \eqref{eq:L-notation-purescalar}
\begin{align}
	\begin{split}
H_1' &=\sum_{m,n=0}^L \sum_{m+n\le L} \frac{1}{L+2}\sqrt{\tfrac{(m+1)(n+1)}{(L-m+1)(L-n+1)}} (\psi^{\dagger}_1)^A_m (\psi_1)^B_{L-n} (\psi_2^{\dagger})^C_{L-m} (\psi_2)^D_n f_{ABE} f_{CDE} \\
H_2'&= \sum_{m,n=0}^L \sum_{m+n\le L} \frac{1}{L+2}\sqrt{\tfrac{(m+1)(n+1)}{(L-m+1)(L-n+1)}} (\psi^{\dagger}_1)^A_{L-m} (\psi_2)^D_n (\psi^{\dagger}_2)^C_m (\psi_1)^B_{L-n} f_{ADE}f_{CBE} \\
H_3'&= \sum_{m,n=0}^L \sum_{m+n < L} \frac{1}{L+2}\sqrt{\tfrac{(m+1)(n+1)}{(L-m+1)(L-n+1)}}
(\psi^{\dagger}_1)^A_{m} (\psi_2)^D_{L-n} (\psi_2^{\dagger})^C_{L-m} (\psi_1)^B_{n} f_{ADE}f_{CBE} \\
H_4'&= \sum_{m,n=0}^L \sum_{m+n< L} \frac{1}{L+2}\sqrt{\tfrac{(m+1)(n+1)}{(L-m+1)(L-n+1)}} (\psi^{\dagger}_1)^A_{L-m} (\psi_1)^B_n (\psi_2^{\dagger})^C_m (\psi_2)^D_{L-n} f_{ABE}f_{CDE}
	\end{split}
\end{align}
such that 
\begin{equation}
	H_{\text{SMT}}^{\text{ferm}} = H_1'+H_4'-H_2'-H_3' \,.
\end{equation}
Due to the normalization coefficients related to the levels of fermionic modes, the Jacobi identity 
\begin{equation}
	f_{ABE}  f_{CDE} - f_{ACE} f_{BDE} +f_{ADE} f_{BCE} =0
\end{equation}
is not enough to simplify the Hamiltonian. 
Instead, we also encounter the combinations $f_{ABE}  f_{CDE} -f_{ADE} f_{BCE}, $ which are generically related to $d$-type coefficients \cite{Haber:2019sgz}. For example:
\begin{equation}\label{eq:1}
	f_{ABE}f_{CDE} = \frac{2}{N} (\delta_{AC}\delta_{BD} - \delta_{AD}\delta_{BC}) +d_{ACE} d_{BDE} - d_{BCE} d_{ADE}\,.
\end{equation}
Simplifications will happen when $N$ is small. 
The $d_{ABC}$ are simply vanishing in the SU$(2)$ gauge group, while in the SU$(3)$ case, Eq.~\eqref{eq:1} can be simplified to \cite{Haber:2019sgz,macfarlane1968gell}
\begin{equation}
		f_{ABE} f_{CDE} -f_{ADE} f_{BCE} = 3d_{ACE} d_{BDE} -\delta_{AB} \delta_{CD} - \delta_{AD} \delta_{BC}+ \delta_{AC}\delta_{BD}
		\,.
\end{equation}  
However, this calculation for general SU$(N)$ group is difficult. 

Both the blocks in SU$(1,2|2)$ sector and the  $(\mathbf{Q}_l,\mathbf{F}_l), J_L$ block in the scalar interactions \eqref{eq:spinchainscalar-2} of the PSU$(1,1|2)$ sector have in common this pattern: the blocks can only be commutators of the bosonic letters, commutators of a bosonic and a fermionic letter and anticommutators of fermionic letters.
This pattern is equivalent to the fact that all the blocks in SMT transform in the adjoint representation of the SU$(N)$ gauge group. 
The appearance of $d_{ABC}$ coefficient in \eqref{eq:1} indicates that this pattern does not apply any more for interactions between fermions with different flavours. 
This can also be regarded as a non-trivial intertwining between the spin and colour groups.

%%%%%%%%%%%%%%%%%%%%%%%

\section{Discussion}
\label{sect-discussion}

In this paper, we have studied the Hamitonian of the Spin Matrix theories describing the effective degrees of freedom in the SU$(2|3)$ and PSU$(1,1|2)$ near-BPS limits.
We started from the results obtained quantizing the dilatation operator of $\mathcal{N}=4$ SYM theory and the action of the spin chain Hamiltonian on the excitations of the sector to derive the corresponding form of the quartic interactions.
Then we applied the spherical reduction method to classical $\mathcal{N}=4$ SYM over a three-sphere, and we gave a prescription to quantize the model.
The result is that the Hamiltonians obtained with these two techniques precisely match.
This shows that the application of the near-BPS limit and the quantization are commuting procedures.

We also analyzed the symmetry structure of the Hamiltonian, which is generically built from a set of fundamental blocks transforming in specific representations of the spin group.
This provides useful hints on how the spin group symmetry determines the Spin Matrix theory.
Although it is less clear how to organize the Hamiltonian of PSU$(1,1|2)$ sector into squares of blocks,
we used the enhanced psu$(1|1)^2$ symmetry to obtain $H_{\rm int}$ as the anticommutator of two conjugate cubic supercharges. 
This method not only provides a new technique to derive the effective interactions of the theory, but it also proves the positivity of the spectrum. 

The positivity property of the $\SU(2|3)$ near-BPS limit is well understood in terms of the F-term block structure which appears manifestly from sphere reduction.
A natural follow-up consists in considering the strongly-coupled regime of the $\SU(2|3)$ sector to find the analog of the giant graviton description employed for the $\SU(2)$ case \cite{Harmark:2016cjq}.
Such gravitational configurations describe the interactions between the exact $1/8$-BPS giant gravitons studied in \cite{Kinney:2005ej}. 
Similarly, the same construction can be generalized to other sectors. 
One of the interesting observations from \cite{Murthy:2020rbd} is that field theory observables like the superconformal index calculated in the free limit are interpolating between the giant graviton phase and the black hole phase. 
Understanding the overall landscape of SMT and its gravitational dual thus provides a tool to probe the dynamics and the phase transitions of these configurations.

On the other hand, a complete understandings of the supersymmetry structure of the $\mathrm{PSU}(1,1|2)$ subsector is still lacking.
The naive expectation would be to interpret the fundamental blocks as $\mN=2$ supermultiplets, which are used to construct the Hamiltonian.
Only the D-term structure, arising \emph{e.g.} from the $\mN=1$ chiral multiplets in Eq.~\eqref{eq:remainders}, is well-understood so far.
A possible tool to approach this problem resides in the representation theory of the PSU$(1,1|2)$ algebra \cite{Gotz:2005ka}.
A related question is how to simplify the fermionic interaction term \eqref{eq:purefermion-interaction} in an analog way as we did for the purely scalar interaction term \eqref{eq:spinchainscalar-2}, ending in an F-term structure.
If this task is achieved, one can identify the appropriate block which transforms in the representations of PSU$(1,1|2)$. 
As discussed in section \ref{sec:symmetry} and in reference \cite{Baiguera:2020mgk}, one of the benefits coming from the block structure is that the list of constraints satisfied by the ground states is given by setting the blocks to zero, \emph{i.e.} $B_l^a=0 .$ 
We may also get a better understanding of these blocks from an analysis similar to Section \ref{sect-positivity}, where the Hamiltonian is derived from the anticommutator of a simpler structure, \emph{i.e.} a cubic supercharge $\hat{\mathcal{Q}}.$
By imposing the supersymmetry invariance of the ground state $\hat{\mathcal{Q}} | 0 \rangle ,$ one can hope to identify the unknown block structure.

It was possible to find a semi-local formulation of the SMT describing the $\SU(1,1)$ near-BPS limits as field theories living on a circle \cite{Baiguera:2020jgy}.
In particular, we also found a superfield formulation of the action in the cases where supersymmetry was preserved in the spin group.
We believe that a similar structure in the $\mathrm{PSU}(1,1|2)$ case would be helpful to understand the block structure in terms of chiral superfields, which naturally provide an F-term interaction when integrating over half of superspace. 
While this analysis was started in \cite{Baiguera:2020jgy}, the new terms discovered in the present work and their non-trivial structure deserve further investigations.

The full PSU$(1,2|3)$ SMT can potentially provide alternative insights into studying the PSU$(1,1|2)$ subsector.  
This is the content of a project in progress \cite{Baiguera:2022}. 
However, there is a potential difference that we should alert. 
The fermions in PSU$(1,2|3)$ include the anti-chiral fermions transforming in triplet representation of SU$(3)_R$ and a chiral fermionic singlet due to the Dirac equation.
There is a priori no doublet structure between the chiral fermion and the anti-chiral fermion triplet.
The SU$(2)_A$ automorphism in PSU$(1,1|2)$ is an emergent symmetry transforming such fermion doublet.
Whether this structure can survive in the full PSU$(1,2|3)$ SMT is an open question, which could be crucial to determine the symmetry structure of the Hamiltonian.

The correspondence between gravity theory of AdS$_5\times S^5$ and $\mN=4$ SYM implies that the ground state of PSU$(1,2|3)$ SMT is dual to the $\frac{1}{16}-$BPS
states in the AdS$_5$ gravity, which include the BPS black hole solutions \cite{Gutowski:2004yv,Kunduri:2006ek}. 
It is a long term puzzle to find order $N^2$ entropy from the field theory computation \cite{Chang:2013fba,Kinney:2005ej}. 
This problem received many attentions recently by studying the superconformal index of $\mN=4$ SYM \cite{Cabo-Bizet:2018ehj,Benini:2018ywd,Hosseini:2017mds,Choi:2018hmj}.
While the $\frac{1}{16}-$BPS black holes are better understood, 
it was noticed in \cite{Choi:2018hmj} that in the $\frac{1}{8}-$BPS sector characterized by the PSU$(1,1|2)$ spin group, the calculation of superconformal index is predicting the existence of $\frac{1}{8}-$BPS black hole \footnote{There are other black hole solutions called EVH black holes in PSU$(1,1|2)$ subsector, as visited in \cite{Berkooz:2014uwa,Goldstein:2019gpz}. However, they are probably not the dominating saddle point in the large $N$ limit \cite{Choi:2018hmj}.}.
A gravitational solution of this kind was never found in the literature.
Studying the SMT in the PSU$(1,1|2)$ and PSU$(1,2|3)$ subsectors in pair might help us learn how to extract the dual gravitational information (especially the black holes) from the SMT. 

The study of PSU$(1,1|2)$ spin chain also appeared in other physics models like 2d CFT
\cite{OhlssonSax:2011ms}.
The holographic dual of these field theories are known to be string theory in AdS$_3\times S^3\times T^4$. 
In fact,  we reported in \cite{Baiguera:2020jgy} that the effective quantum field theory generating the PSU$(1,1|2)$ Hamiltonian is a 2d QFT after decompactifying the direction generating the infinite descendants.
It is interesting that the scalar fields and fermionic fields in the 2d QFT descriptions are all ghost--like.
This reminds us of the chiral CFT of 4D SCFT \cite{Beem:2013sza}, which is non-unitary. 
In fact, the decoupling condition of PSU$(1,1|2)$ subsector is equivalent to the Schur condition of chiral algebra. 
Besides, the spin group PSU$(1,1|2)$ defining the SMT is identically the chiral algebra of $\mN=4$ SYM \cite{Beem:2013sza,Bonetti:2016nma}. 
These non-trivial matches indicate a deep connection between these theories.

\section*{Acknowledgements}

S.B. acknowledges support from the Israel Science Foundation (grant No. 1417/21) and from the Kreitmann School of Advanced Graduate Studies. 
S.B. and T.H. are supported by the Independent Research Fund Denmark grant number DFF-6108-00340 “Towards a deeper understanding of black holes with non-relativistic holography”. 
Y.L. is supported by the UCAS program of special research associate and by the internal funds of the KITS, and by China Postdoctoral Science Foundation.

\appendix

\section{Conventions and details of the spherical reduction}
\label{app-notations_conventions}

In order to perform the sphere reduction procedure described in Section \ref{sect-effective_ham_from_sphere}, it is necessary to decompose the fields into spherical harmonics and introduce appropriate Clebsch-Gordan coefficients which describe the interactions between the modes.
We collect in this Appendix the fundamental material required to derive the results presented in the work, and we refer the reader to the following references for more details:
\begin{itemize}
\item Appendix A of \cite{Baiguera:2020jgy}: conventions on the decomposition of fields into spherical harmonics on the three-sphere for scalars, fermions and vector fields.
\item Appendix B of \cite{Baiguera:2020jgy}: expression of the free Hamiltonian of $\mathcal{N}=4$ SYM after sphere reduction, technical details on the treatment of fermionic modes, computation of the charges associated to rotations and R-symmetry, weights of the fields, interacting Hamiltonian.
\item Appendix A of \cite{Baiguera:2020mgk}: definition of Clebsch-Gordan coefficients and crossing relations between them when their momenta are saturated in a precise way. 
\end{itemize}
Details on the $\SU(2|3)$ and $\mathrm{PSU}(1,1|2)$ algebra and the letters of these sectors are given in Appendix \ref{app-algebra_SU23_PSU112}.

\subsection{Definition of the Clebsch-Gordan coefficients}
\label{app-definition_Clebsch}

We report for convenience the definitions of the Clebsch-Gordan coefficients, which are used in Section \ref{sect-effective_ham_from_sphere} to compute the effective Hamiltonian:
\beq
\mathcal{C}^{J_1 M_1}_{J_2 M_2; JM} = 
\sqrt{\frac{(2J+1)(2J_2+1)}{2J_1+1}}
C^{J_1 m_1}_{J_2 m_2; J m} C^{J_1 \tilde{m}_1}_{J_2 \tilde{m}_2; J \tilde{m}} \, ,
\label{eq:app_definitionClebsch_C}
\eeq
\beq
\begin{aligned}
\mathcal{D}^{J_1 M_1}_{J_2 M_2 \rho_2; JM \rho} & = 
(-1)^{\frac{\rho_2+\rho}{2} +1}
\sqrt{3(2J_2+1)(2J_2+2 \rho_2^2 +1)(2J+1)(2J+2 \rho^2 +1)} \\
& \times C^{J_1,m_1}_{Q_2,m_2; Q,m} C^{J_1, \tilde{m}_1}_{\tilde{Q}_2, \tilde{m}_2 ; \tilde{Q},\tilde{m}}
\begin{Bmatrix}
Q_2 & \tilde{Q}_2 & 1 \\
Q & \tilde{Q} & 1 \\
J_1 & J_1 & 0  
\end{Bmatrix} \, ,
\label{eq:app_definitionClebsch_D}
 \end{aligned}
\eeq
\beq
\begin{aligned}
 \mathcal{E}_{J_1 M_1 \rho_1; J_2 M_2 \rho_2; JM \rho} & =  \sqrt{6(2J_1+1)(2J_1 + 2 \rho_1^2 +1)(2J_2+1)(2J_2+2 \rho_2^2 +1)(2J+1)(2J+2 \rho^2 +1)} \\
& \times (-1)^{-\frac{\rho_1+\rho_2+\rho+1}{2}} 
\begin{Bmatrix}
Q_1 & \tilde{Q}_1 & 1 \\
Q_2 & \tilde{Q}_2 & 1 \\
Q & \tilde{Q} & 1 
\end{Bmatrix} 
\begin{pmatrix}
Q_1 & Q_2 & Q \\
m_1 & m_2 & m
\end{pmatrix}
\begin{pmatrix}
\tilde{Q}_1 & \tilde{Q}_2 & \tilde{Q} \\
\tilde{m}_1 & \tilde{m}_2 & \tilde{m}
\end{pmatrix} \, ,
\label{eq:app_definitionClebsch_E}
\end{aligned}
\eeq
\beq
\begin{aligned}
\mathcal{F}^{J_1 M_1 \kappa_1}_{J_2 M_2 \kappa_2; JM} = & (-1)^{\tilde{U}_1+U_2+J+\frac{1}{2}} \sqrt{(2J+1)(2J_2+1)(2J_2+2)}  \\
& \times C^{U_1, m_1}_{U_2, m_2; J,m} C^{\tilde{U}_1,\tilde{m}_1}_{\tilde{U}_2,\tilde{m}_2; J,\tilde{m}}
\begin{Bmatrix}
U_1 & \tilde{U}_1 & \frac{1}{2} \\
\tilde{U}_2 & U_2 & J
\end{Bmatrix} \, ,
\label{eq:app_definitionClebsch_F}
\end{aligned}
\eeq
\beq
\begin{aligned}
\mathcal{G}^{J_1 M_1 \kappa_1}_{J_2 M_2 \kappa_2; JM \rho} = & (-1)^{\frac{\rho}{2}} \sqrt{6(2J_2+1)(2J_2+2)(2J+1)(2J+2 \rho^2 +1)} \\
& \times C^{U_1,m_1}_{U_2,m_2; Q,m} C^{\tilde{U}_1, \tilde{m}_1}_{\tilde{U}_2, \tilde{m}_2 ; \tilde{Q},\tilde{m}}
\begin{Bmatrix}
U_1 & \tilde{U}_1 & \frac{1}{2} \\
U_2 & \tilde{U}_2 & \frac{1}{2} \\
Q & \tilde{Q} & 1 
\end{Bmatrix} \, ,
\label{eq:app_definitionClebsch_G}
\end{aligned}
\eeq
where we defined the quantities
\beq
U \equiv J + \frac{\kappa+1}{4} \spa
\tilde{U} \equiv J + \frac{1-\kappa}{4} \spa
Q \equiv J+ \frac{\rho(\rho+1)}{2} \spa
\tilde{Q} \equiv J + \frac{\rho(\rho-1)}{2} \, .
\label{eq:app_labels_harmonics}
\eeq

\subsection{Interacting Hamiltonian of $\mathcal{N}=4$ SYM after sphere reduction}

The full interacting Hamiltonian of $\mathcal{N}=4$ SYM theory after the decomposition of the fields into spherical harmonics is derived in \cite{Ishiki:2006rt}.
We report here the result using the following conventions for the fields:
\beq
(Z_a)_{JM}  \equiv \begin{pmatrix} (\Phi_1)_{JM} \\
 (-1)^{m-\tilde{m}} (\Phi^\dagger_2)_{J,-M} \\
  (-1)^{m-\tilde{m}} (\Phi^\dagger_3)_{J,-M} 
  \end{pmatrix} \, ,
  \label{eq:mapping_Z_scalars}
\eeq
\beq
(\Psi_A)_{J,M,\kappa=1} \equiv   (\psi^{\dagger}_A)_{J,-M,\kappa=1} \, , \qquad
(\Psi_A)_{J,M,\kappa=-1} \equiv (\psi_A)_{J,M, \kappa=-1} \, .
\label{eq:mapping_Psi_fermions}
\eeq 
These definitions have a technical origin to account for the different interpretation of scalars and fermions with respect to reference \cite{Ishiki:2006rt}. 

The result is:
\beq
\begin{aligned}
H_{\rm int}   = \sum_{J_i, M_i, \kappa_i, \rho_i} & \tr  \left\lbrace i g {\cal C}^{J_2M_2}_{J_1M_1;JM} \, \chi_{JM} \left([(Z_a^\dagger)_{J_2M_2},(\Pi^{(\Phi)\dagger}_a)_{J_1M_1}] + [Z^a_{J_1M_1},\Pi^{a(\Phi)}_{J_2M_2}]\right)  \right. \\
& \left. - 4g\sqrt{J_1(J_1+1)} {\cal D}^{J_2M_2}_{J_1M_1 0; JM\rho} \, A_{(\rho)}^{JM}
[Z^a_{J_1M_1},(Z_a^\dagger)_{J_2M_2}]  \right. \\
& \left. +g \mathcal{F}^{J_1 M_1 \kappa_1}_{J_2 M_2 \kappa_2; JM}  \, \chi_{JM} 
\lbrace (\Psi_A^{\dagger})_{J_1 M_1 \kappa_1} , \Psi^A_{J_2 M_2 \kappa_2} \rbrace \right. \\
& \left. + g \mathcal{G}^{J_1 M_1 \kappa_1}_{J_2 M_2 \kappa_2; JM \rho} \, A_{(\rho)}^{JM}
\lbrace (\Psi^{\dagger}_A)_{J_1 M_1 \kappa_1} , \Psi^A_{J_2 M_2 \kappa_2} \rbrace   
  \right. \\
  & \left. + \frac{g^2}{2} {\cal C}^{J_2 M_2}_{J_1 M_1;J M}{\cal C}^{J_3 M_3}_{J_4 M_4;J M}[Z^a_{J_1 M_1},(Z^{\dagger}_a)_{J_2 M_2}][Z^b_{J_3 M_3},(Z_b^{\dagger})_{J_4 M_4}] \right. \\
 & \left.   - \sqrt{2} ig (-1)^{-m_1+\tilde{m}_1+\frac{\kappa_1}{2}} \mathcal{F}^{J_1,-M_1,\kappa_1}_{J_2 M_2 \kappa_2; J M} 
 \Psi^4_{J_2 M_2 \kappa_2} [(Z_a)^{JM} , \Psi^a_{J_1 M_1 \kappa_1}]  \right. \\
 & \left.   + \sqrt{2}ig (-1)^{-m_1+\tilde{m}_1+\frac{\kappa_1}{2}} \mathcal{F}^{J_1,-M_1,\kappa_1}_{J_2 M_2 \kappa_2; J M} 
\, \epsilon_{abc} \Psi^a_{J_1 M_1 \kappa_1} [(Z_b^{\dagger})^{JM} , \Psi^c_{J_2 M_2 \kappa_2}]  \right. \\
  & \left.   + \sqrt{2}ig (-1)^{m_2-\tilde{m}_2+\frac{\kappa_2}{2}} \mathcal{F}^{J_1 M_1 \kappa_1}_{J_2, -M_2, \kappa_2; J M} 
 (\Psi^{\dagger}_4)_{J_2 M_2 \kappa_2} [(Z^{\dagger}_a)^{JM} , (\Psi^{\dagger}_a)_{J_1 M_1 \kappa_1}]  \right. \\
 & \left.   - \sqrt{2}ig (-1)^{m_2-\tilde{m}_2+\frac{\kappa_2}{2}} \mathcal{F}^{J_1 M_1 \kappa_1}_{J_2, -M_2, \kappa_2; J M} 
\, \epsilon_{abc} (\Psi^{\dagger}_a)_{J_1 M_1 \kappa_1} [(Z_b)^{JM} , (\Psi^{\dagger}_c)_{J_2 M_2 \kappa_2}]  \right. \\
& \left. + i g {\cal D}^{JM}_{J_1M_1\rho_1;J_2M_2\rho_2} \, \chi_{JM} [\Pi^{J_1M_1}_{(\rho_1)},A^{J_2M_2}_{(\rho_2)}] \right. \\
& \left. + g^2 \mathcal{C}^{JM}_{J_2 M_2; J_4, -M_4} \mathcal{D}_{JM; J_1 M_1 \rho_1; J_3 M_3 \rho_3} [A_{(\rho_1)}^{J_1 M_1} , Z^a_{J_2 M_2}] [A_{(\rho_3)}^{J_3 M_3} , (Z^{\dagger}_a)_{J_4 M_4}]   \right. \\
& \left. + 2 i g \rho_1(J_1 + 1) {\cal E}_{J_1M_1\rho_1;J_2M_2\rho_2;J_3M_3\rho_3}A^{J_1M_1}_{(\rho_1)}[A^{J_2M_2}_{(\rho_2)},A^{J_3M_3}_{(\rho_3)}] \right. \\
& \left. - \frac{g^2}{2} {\cal D}^{JM}_{J_1M_1\rho_1; J_3M_3\rho_3} {\cal D}_{JM;J_2M_2\rho_2; J_4 M_4 \rho_4}[A^{J_1M_1}_{(\rho_1)},A^{J_2M_2}_{(\rho_2)}][A^{J_3M_3}_{(\rho_3)},A^{J_4M_4}_{(\rho_4)}] \right. \\
& \left. -2g \sqrt{J_1(J_1+1)} \mathcal{D}_{J_2 M_2; J_1 M_1 0; J M \rho} \, \chi_{J_1 M_1} [\chi_{J_2 M_2} , A_{(\rho)}^{JM}] \right. \\
& \left. + \frac{g^2}{2} \mathcal{C}^{JM}_{J_1 M_1; J_3 M_3} \mathcal{D}_{JM; J_2 M_2 \rho_2; J_4 M_4 \rho_4} [\chi_{J_1 M_1} , A_{(\rho_2)}^{J_2 M_2}] [\chi_{J_3 M_3} , A_{(\rho_4)}^{J_4 M_4}]  \right. \\
& \left. + g^2 \mathcal{C}^{JM}_{J_1 M_1; J_2 M_2} \mathcal{C}_{JM; J_3 M_3; J_4 M_4} [\chi_{J_1 M_1} , Z^a_{J_2M_2}] [\chi_{J_3 M_3} , (Z_a^{\dagger})_{J_4 M_4}]
 \right\rbrace \, .
\end{aligned}
\label{eq:app_full_interacting_N=4SYM_Hamiltonian}
\eeq
We add few comments on the notation:
\begin{itemize}
\item The overall summation over all contracted indices involves momenta $(J,M),$ labels for fermions $(\kappa)$ and gauge fields $(\rho),$ and indices of all the fields under $\SU{(4)}$ R-symmetry. 
\item The fermions summed in the Yukawa term run over $a \in \lbrace 1,2,3  \rbrace$ and the corresponding Levi-Civita symbol is defined in such a way that $\epsilon_{123}=1.$
\item $\Pi^{\Phi}_a$ are the canonical momenta associated to the scalar fields $\Phi_a,$ while $\Pi_{(\rho)}$ is the symplectic partner of the gauge field $A_{(\rho)}.$
\item The Clebsch-Gordan coefficients were defined in Appendix \ref{app-definition_Clebsch}.
\end{itemize}
The terms involving the gauge fields, except for the terms contributing to the bosonic and fermionic currents, are not needed for the near-BPS limits included in this work, but have been put for completeness.
More simplifications of Eq.~\eqref{eq:app_full_interacting_N=4SYM_Hamiltonian} are directly reported in Section \ref{sect-effective_ham_from_sphere} when explicit sectors are considered.

\section{SU$(2|3)$ and PSU$(1,1|2)$ algebra}
\label{app-algebra_SU23_PSU112}

We will follow \cite{Harmark:2007px} for explaining the oscillator representations of the spin group. 
We define the oscillator ladder operators as
\begin{equation}
	[\mba^\alpha,\mba_\beta^\dagger] = \delta^\alpha_\beta, \qquad	[\mbb^{\dalpha},\mbb_{\dbeta}^\dagger] = \delta^{\dalpha}_{\dbeta}, \qquad 
	\{\mbc^a,\mbc_b^\dagger \} =\delta_b^a \,.
\end{equation}
The corresponding number operators are defined as 
\begin{equation}
	a^\alpha=\mba_\alpha^\dagger \mba^\alpha, \qquad b^{\dalpha} = \mbb_{\dalpha}^\dagger \mbb^{\dalpha}, \qquad  c^a =\mbc_a^\dagger \mbc^a\,.
\end{equation}

\subsection{Letters in SU$(2|3)$}
\label{app-review_SU23}

The SU$(2|3)$ subsector corresponds to the decoupling constraint
\begin{equation}
	D_0 = Q_1+Q_2+Q_3 \, ,
\end{equation}
which results in 
\begin{equation}
	b^1=b^2=0,\qquad c^4=1\,.
\end{equation}
There are three scalars $Z,X,W$ and two fermions $\psi_\pm$ in this subsector, whose oscillator representations are 
\begin{equation}
	\ket{Z} = \mbc_3^\dagger \mbc_4^\dagger \ket{0},\, 	\ket{X} = \mbc_2^\dagger \mbc_4^\dagger \ket{0},\, \ket{W} = \mbc_1^\dagger \mbc_4^\dagger \ket{0}, \quad  \ket{\psi_+} = \mba_1^\dagger \mbc_4^\dagger \ket{0},\,  \ket{\psi_-} = \mba_2^\dagger \mbc_4^\dagger \ket{0}\,.
\end{equation}
For compactness of notation, we are using $\ket{\Phi_a}$ to represent the scalar triplet with the map
\begin{equation}
	\Phi_1 \equiv Z, \quad \Phi_2 \equiv X, \quad \Phi_3 \equiv W \,.
\end{equation}
Notice that the scalar label is not equal to the oscillator label, but instead $\ket{\Phi_a}= \mbc_{4-a}^\dagger \mbc_4^\dagger \ket{0}$, $a=1,2,3$. The supercharges are generically of the form 
$Q_{a\alpha}\equiv \mba_\alpha\mbc_{4-a}^\dagger$. This way of labelling the letters will be useful for writing the transformation rules in a compact way. 

With all these tools, we can then rewrite the symmetry generators in terms of the fields. The supercharges are
\begin{equation}
	Q_{a\alpha} = \text{tr} (\Phi_a^\dagger \psi_{\alpha}), \quad a=1,2,3;\quad \alpha=\pm \, .
\end{equation}
The SU$(2)$ part of algebra only acts on the fermion doublet, which is 
\begin{equation}
	I_+ = \text{tr} (\psi_+^\dagger\psi_-), \quad I_- = \text{tr} (\psi_-^\dagger\psi_+), \quad I_0 =\frac{1}{2} \text{tr} \Big(\psi_+^\dagger\psi_+ -\psi_-^\dagger\psi_-  \Big) \,.
\end{equation}
Similarly, the SU$(3)$ R-symmetry acting on the scalars are
\begin{align}\label{eq:SU3algebra}
	\begin{split}
		R_{12} & = \text{tr} (X^\dagger W) , \quad R_{13} =\text{tr} (Z^\dagger W), \quad R_{23} = \text{tr} (Z^\dagger X) \, , \\
		R_{21} &= \text{tr} (W^\dagger X), \quad R_{31} = \text{tr} (W^\dagger Z), \quad R_{32} = \text{tr} (X^\dagger Z) \, , \\
		R_0 &= \frac{1}{2} \text{tr}(Z^\dagger Z - X^\dagger X), \qquad \mathbf{R}_0 = \frac{1}{2} \text{tr}(Z^\dagger Z+ X^\dagger X -2 W^\dagger W) \,.
	\end{split}
\end{align}
The representations of SU$(2)$ are parametrized by the spin $j$, and the corresponding dimension of the spin-$j$ representation is $2j+1$. 
The states in the spin-$j$ representation are parametrized by quantum number $ |m|\le j$. 
The representations of SU$(3)$ are parametrized by two labels $(n,m)$. Such theories are widely used in phenomenological studies of QCD. In terms of the symmetric tensor language, $n$ labels the number of upper indices of the tensor while $m$ labels the number of lower indices. 
This can be already understood by means of the Young Tableaux. 
The dimension of $(n,m)$ representation is 
\begin{equation}
	d(n,m) = \frac{1}{2}(n+1)(m+1) (n+m+2) \,.
\end{equation}
The representations are thus sometimes written in terms of the dimension rather than $(n,m)$ labels. 
The most commonly used representations are the fundamental  $\mathbf{3}=(1,0)$, the anti-fundamental representation $\bar{\mathbf{3}} = (0,1)$ and adjoint representation $\mathbf{8}=(1,1)$.
The fermions $\psi_\alpha$ transform as a doublet of SU$(2)$, which is the $j=\frac{1}{2}$ representation. 
The scalars $Z^\dagger,X^\dagger,W^\dagger$ transform in the fundamental representation. 
The weights of the operators are summarized in table \ref{table:dimension-su23-letter}.
\begin{table}
	\centering
	\begin{tabular}{|c|c|c|c|c|c|}
		\hline
		& $Z^\dagger$ & $X^\dagger$ & $W^\dagger$  & $\psi^\dagger_+$  & $\psi_-^\dagger$  \\
		\hline
		$(R_0,\mathbf{R}_0)$		&  $(\frac{1}{2},\frac{1}{2})$ & $(-\frac{1}{2},\frac{1}{2})$  & $(0,1)$ & $(0,0)$ & $(0,0)$ \\
		\hline
		$(j;m)$		& $0$ & $0$ & $0$ & $(\frac{1}{2},\frac{1}{2})$ & $(\frac{1}{2},-\frac{1}{2})$  \\
		\hline
	\end{tabular}
	\caption{The charges and weight of the letters in the $\SU(2|3)$ sector. One can also calculate these data from \cite{Harmark:2007px}. We identify $\Phi_a^\dagger$ with $Z^\dagger,X^\dagger,W^\dagger$ respectively. Thus $a=1,2,3$ will be mapped to the corresponding values of $(R_0,\mathbf{R}_0)$.
	}
	\label{table:dimension-su23-letter}
\end{table}

\subsection{Letters in PSU$(1,1|2)$}
\label{app-review_PSU112}

The near-BPS limit of the PSU$(1,1|2)$ sector reads
\begin{equation}
	D_0 =S_1+Q_1+Q_2 \, ,
\end{equation}
which results in \begin{equation}
	a^2=b^2=0, \quad c^1=0,\quad c^4=1 \,.
\end{equation}
The allowed letters are $\Phi_{1,2},\psi_1$ and $\bar{\chi}_7\equiv \psi_2$. Here we rename the fermion $\bar{\chi}_7$ from the convention in \cite{Harmark:2007px}. 
The set of letters in this subsector have already been shown in \cite{Baiguera:2020jgy} as 
\begin{eqnarray}\nonumber
&& \ket{d_1^n Z} = \ket{\Phi_n^1}=\frac{1}{n!} (\mba_1^\dagger \mbb_1^\dagger)^n \mbc_3^\dagger \mbc_4^\dagger \ket{0}, \quad  \ket{d_1^n \psi_1} =\ket{\psi_n^1} = \frac{1}{\sqrt{n!(n+1)!}}
 (\mba_1^\dagger \mbb_1^\dagger)^n \mba_1^\dagger \mbc_4^\dagger \ket{0}  \, , \\  \nonumber
&& \ket{d_1^n X} = \ket{\Phi_n^2}=\frac{1}{n!} (\mba_1^\dagger \mbb_1^\dagger)^n \mbc_2^\dagger \mbc_4^\dagger \ket{0}, \quad 
 \ket{d_1^n \psi_2} = -\ket{\psi_n^2} = \frac{1}{\sqrt{n!(n+1)!}}
(\mba_1^\dagger \mbb_1^\dagger)^n \mbb_1^\dagger \mbc_2^\dagger\mbc_3^\dagger\mbc_4^\dagger \ket{0}  \,.
\end{eqnarray}
The PSU$(1,1|2)$ symmetry generators include the SU$(2)$ R-symmetry which acts on the scalar doublet
\begin{equation}
	R_+ =\mbc_3^\dagger \mbc^2, \quad R_- =\mbc_2^\dagger \mbc^3, \quad R_0 = \frac{1}{2} (\mbc_3^\dagger \mbc_3-\mbc_2^\dagger \mbc_2) \, ,
\end{equation}
the SU$(1,1)$ subalgebra 
\begin{equation}
L_0 = \frac{1}{2}(1+\mba_1^\dagger \mba_1+\mbb_1^\dagger \mbb_1), \quad L_+ =\mba_1^\dagger \mbb_1^\dagger,\quad L_- =\mba_1 \mbb_1 \, ,
\end{equation}
and the supercharges: 
\begin{align}
	\begin{split}
& Q =\mba_1 \mbc_3^\dagger, \quad Q^\dagger =\mba_1^\dagger \mbc_3, \quad S=\mbb_1 \mbc_3, \quad S^\dagger= \mbb_1^\dagger \mbc_3^\dagger \, , \\
& \tilde{Q}  =\mba_1 \mbc_2^\dagger, \quad \tilde{Q}^\dagger= \mba_1^\dagger \mbc_2, \quad \tilde{S}= \mbb_1 \mbc_2, \quad \tilde{S}^\dagger = \mbb_1^\dagger \mbc_2^\dagger \,.
		\end{split}
\end{align}
We are ready to rewrite the symmetry generators in terms of the modes of the fields. The supercharges are 
\begin{align}\label{eq:def-supercharges}
	\begin{split}
		Q^\dagger &= \sum_{n=0}^\infty \sqrt{n+1} \,\text{tr}\, \Big( (\psi_1^{\dagger})_n (\Phi_1)_n \Big) + \sum_{n=0}^\infty \sqrt{n+1} \,\text{tr}\,\Big((\Phi_{2}^{\dagger})_{n+1} (\psi_2)_n \Big) \, , \\
		\tilde{Q}^\dagger &= \sum_{n=0}^\infty \sqrt{n+1} \,\text{tr}\,\Big( (\psi_1^{\dagger})_n (\Phi_2)_n \Big) - \sum_{n=0}^\infty \sqrt{n+1} \,\text{tr}\,\Big( (\Phi_{1}^{\dagger})_{n+1} (\psi_2)_n \Big) \, , \\
		S^\dagger &= \sum_{n=0}^\infty \sqrt{n+1} \,\text{tr}\,\Big( (\Phi_{1}^{\dagger})_{n+1} (\psi_1)_n \Big) +\sum_{n=0}^\infty \sqrt{n+1} \,\text{tr}\,\Big((\psi_2^{\dagger})_n (\Phi_{2})_n \Big) \, , \\
		\tilde{S}^\dagger &=- \sum_{n=0}^\infty \sqrt{n+1} \,\text{tr}\,\Big((\psi_2^{\dagger})_n (\Phi_1)_n \Big) + \sum_{n=0}^\infty \sqrt{n+1}\,\text{tr}\,\Big( (\Phi_{2}^{\dagger})_{n+1} (\psi_1)_n \Big) \,.
	\end{split}
\end{align}
The R-symmetry generators are
\begin{equation}
	R_0 = \frac{1}{2} \sum_{n=0}^\infty \text{tr} \left((\Phi_2^{\dagger})_n (\Phi_2)_n -(\Phi_1^{\dagger})_n (\Phi_1)_n \right), \quad R_+ =R_-^\dagger =\sum_{n=0}^\infty \text{tr} \Big( (\Phi_2^{\dagger})_n  (\Phi_1)_n  \Big) \,.
\end{equation}
Notice that the R-symmetry only acts on scalar doublets while fermions transform trivially under this symmetry. However, as the basic ingredients of the $\mN=2$ supersymmetric field theory, the two scalars and the two fermions in the PSU$(1,1|2)$ are supposed to organize into two $\mN=1$ chiral multiplet to form an $\mN=2$ hypermultiplet. 
The fermions should also transform as a doublet.
The corresponding symmetry is called the SU$(2)$ automorphism which is not part of global PSU$(2,2|4)$ symmetry of $\mN=4$ SYM, but is an emergent symmetry in the PSU$(1,1|2)$ subsector \cite{Beisert:2007sk}. 
The symmetry generators can be written as 
\footnote{We should note that the two fermions in PSU$(1,1|2)$ subsector are not the same fermions in SU$(2|3)$ subsector. In terms of the notation of \cite{Harmark:2007px}, $\psi^2\equiv \bar{\chi}_7$ while $\psi_1\equiv \psi_+ \equiv\chi_1$ are the same fermions in both sectors. }
\begin{equation}\label{eq:SU2-automorphism}
		\mathfrak{R}_0  = \frac{1}{2} \sum_{n=0}^\infty \text{tr} \Big((\psi_2^{\dagger})_n (\psi_2)_n -(\psi_{1}^{\dagger})_n (\psi_1)_n \Big), \quad \mathfrak{R}_+ =  \sum_{n=0}^\infty  \text{tr}\Big((\psi_2^{\dagger})_n (\psi_1)_n \Big) \,.
\end{equation}
As the SU$(2)$ automorphism is also a global symmetry of the PSU$(1,1|2)$ subsector, the final Hamiltonian must be also invariant under this action.

\addcontentsline{toc}{section}{References}

\bibliography{newbib}
\bibliographystyle{newutphys}

\end{document}